%% file: main.tex
\documentclass{article}
\input{shorcuts}

\title{Robust and efficient estimation for the Generalized Extreme-Value distribution with application to flood frequency analysis in the UK}
\author{Nathan Huet$^1$, Ilaria Prosdocimi$^1$ \\
$^1$ {\small Ca’ Foscari University of Venice, Venice, Italy}}

\begin{document}

\maketitle

\begin{abstract}
    A common approach for modeling extremes, such as peak flow or high temperatures, is the three-parameter Generalized Extreme-Value distribution. 
    This is typically fit to extreme observations, here defined as maxima over disjoint blocks. 
    This results in limited sample sizes and consequently, the use of classic estimators, such as the maximum likelihood estimator, may be inappropriate, as they are highly sensitive to outliers.
To address these limitations, we propose a novel robust estimator based on the minimization of the density power divergence, controlled by a tuning parameter $\alpha$ that balances robustness and efficiency. When $\alpha = 0$, our estimator coincides with the maximum likelihood estimator; when $\alpha = 1$, it corresponds to the $L^2$ estimator, known for its robustness. We establish convenient theoretical properties of the proposed estimator, including its asymptotic normality and the boundedness of its influence function for $\alpha > 0$. The practical efficiency of the method is demonstrated through empirical comparisons with the maximum likelihood estimator and other robust alternatives. Finally, we illustrate its relevance in a case study on flood frequency analysis in the UK and provide some general conclusions in Section \ref{sec:conclude}.
\end{abstract}
\section{Introduction}

\input{intro}

\section{Background} \label{sec:background}



This section begins with a brief overview of extreme value theory and the GEV distribution. It then introduces the density power divergence along with its key properties, following the development in \cite{basu1998robust}. For more comprehensive coverage of extreme value theory, the reader is referred to \cite{coles2001introduction, resnick2008extreme}, while foundational concepts in robust statistics can be found in \cite{hampel86,huber2011robust}.

\subsection{Generalized extreme value distribution}
\input{evd}

\subsection{Minimum density power divergence estimators}
\input{mdpde}

\section{MDPD estimator for GEV distributions} \label{sec:GEVforMDPDE}
\input{mdpdeGEV}


\section{Simulation study}\label{sec:simulation_study}

\input{expes}

\section{Application: flood frequency analysis in the UK}\label{sec:application}
\input{appli}

\section{Conclusion} \label{sec:conclude}
In this paper, we extend a classic robust and efficient estimation method, the MDPD estimator, to the case of the GEV distribution. We establish its consistency and asymptotic normality, and prove the boundedness of its influence function, thereby confirming its strong robustness properties. Through simulations on both contaminated and uncontaminated datasets, we demonstrate the reliability of the MDPD estimator and its advantages over existing methods in the literature. Compared to other robust methods, the MDPD estimator stands out for its high efficiency in low-contamination settings, while maintaining strong robustness under heavy contamination. Another advantage of our approach lies in its ease of implementation for both positive and negative shape parameters, with a low rate of non-convergence or implausible convergence outputs. Finally, we also present a case study in which the MDPD estimator offers a compelling alternative to classical approaches in flood frequency analysis in the presence of PILFs. In this study, the MDPD estimator is shown to be an elegant approach in which no outlier detection preliminary step is necessary to yield satisfactory results. 

Beyond this application, the method could be applied to other contexts where outliers, in the left or right tail, may contaminate extreme value data. 
For example, it is often the case that annual maxima flows routinely collected by measuring authorities are representative of differ flood generating processes \citep[possibly leading to PILFs appearing the data,][]{Barth2017mixed}: if the proportion of data generated by one of the mechanism is small robust approaches such as those presented in this work could lead to a correct estimation of the main process without the need of specifying the different sub-populations but without ignoring the potential issues connected to the analysis of non-identically distributed records \citep{Singh2005nonid}. 
Furthermore, several extensions of the current MDPD framework should be possible. A particularly valuable direction would be to extend the framework to allow the parameters to depend on covariates (e.g., time), thereby capturing more complex data structures and improving the fit for certain processes. As highlighted by \cite{juarez2004robust}, one strength of the MDPD method is that such an extension to nonstationary settings can be achieved relatively easily. In such cases, however, theoretical results under misspecified settings (see Remark~\ref{rem:misspecified}) would still need to be established. Another interesting direction would be to study the convergence of the MDPD estimator under second-order properties. Indeed, it is known that the ML estimator of the shape index of the GEV or GP distribution can exhibit significant asymptotic bias \citep{dombry2019maximum}. Hence, it would be interesting to compare the asymptotic biases of the ML and MDPD estimators, which would likely provide further evidence supporting the MDPD estimator as a viable alternative.
Finally, another promising line of research lies in the development of a systematic method 
for selecting the hyper-parameter $\alpha$ in a suitably “optimal” way for GEV distributions.



\clearpage
\section*{Reproducibility}
All data and code required to reproduce the analyses and figures in this paper are publicly available at \url{https://github.com/HuetNathan/robustGEV}.

\section*{Acknowledgments}
This work was supported by the DoE 2023-2027 (MUR, AIS.DIP.ECCELLENZA2023\_27.FF) project. IP acknowledges that this study was carried out within the RISE project and received funding from the European Union Next-GenerationEU - National Recovery and Resilience Plan (NRRP) – MISSION 4 COMPONENT 2, INVESTIMENT 1.1 Fondo per il Programma Nazionale di Ricerca e Progetti di Rilevante Interesse Nazionale (PRIN) – CUP N.H53D23002010006. This publication reflects only the authors’ views and opinions; neither the European Union nor the European Commission can be considered responsible for them. The British River peak flow data was downloaded from the National River Flow Archive (\url{https://nrfa.ceh.ac.uk/}).
The authors also wish to thank Sergio F. Juárez for providing his PhD manuscript.

\clearpage
\appendix
\input{appendix}

\clearpage
\input{appendix_simulation}

\clearpage
\input{appendix_appli}

\clearpage
\bibliographystyle{apalike} 
\bibliography{biblio}
\end{document}

%% file: shorcuts.tex
\usepackage{diagbox}
\usepackage{babel}
\usepackage{amsfonts}
\usepackage{amssymb,amsmath}
\usepackage[T1]{fontenc}  
\usepackage{hhline}
\usepackage{fancyhdr}
\usepackage{lmodern}
\usepackage{mathabx}
\usepackage{mathrsfs}
\usepackage{calrsfs}
\usepackage{tcolorbox}
\usepackage{calc}
\usepackage{dsfont}
\usepackage{color}
\usepackage{amsthm}
\usepackage{mathtools}
\usepackage{enumitem}
\usepackage{hyperref}
\usepackage{xcolor}
\usepackage{geometry}
\usepackage{natbib}
\usepackage{booktabs}
\usepackage{multirow}
\usepackage{makecell}

\hypersetup{
  colorlinks   = true, 
  urlcolor     = blue, 
  linkcolor    = blue, 
  citecolor   =  violet 
}

\newcommand{\mb}{\mathbb}

\newcommand{\s}{\geq}
\newcommand{\m}{\leq}

\newcommand{\1}{\mathds{1}}

\newcommand{\ie}{i.e.}

\newtheorem{theorem}{Theorem}[section]
\newtheorem{proposition}{Proposition}[section]
\newtheorem{corollary}{Corollary}[section]

\theoremstyle{definition}

\newtheorem{remark}{Remark}[section]

%% file: intro.tex
The \textit{Generalized Extreme-Value} (GEV) distribution, introduced by \cite{jenkinson1955frequency}, plays a central role in modeling rare events across a wide range of fields, including flood frequency analysis \citep{morrison2002stochastic,castellarin2012review,prosdocimi2015detection}, temperature extremes \citep{stein2017should,AuldTemp}, heavy precipitation \citep{feng2007modeling,papalexiou2013battle,nerantzaki2022assessing,gaetan2025precip}, droughts \citep{piwowar2023drought}, and extreme wind prediction \citep{friederichs2012forecast,soukissian2015effect}. This extensive use of the GEV distribution is justified by a fundamental result in \textit{Extreme Value Theory} (EVT), namely the Fisher–Tippett–Gnedenko theorem \citep{fisher1928limiting,gnedenko1943distribution}, which states that the limiting distribution of rescaled maxima of i.i.d. random variables, if non-degenerate, is a GEV distribution (see Section~\ref{sec:background}). More importantly, this result ensures that rescaled maxima over blocks of observations (typically over fixed time intervals such as a day, month, or year) are well modeled by a GEV distribution \citep[see][]{coles2001introduction}. This modeling framework, known as the \textit{Block Maxima} (BM) approach, defines extremes as the maxima within blocks of observations. Another possible approach is to define as extremes the observations which exceed a high threshold: this leads to the \textit{Peaks-over-Threshold} (PoT) approach, another widely used EVT framework. In the PoT setting, the limit distribution of exceedances is found to be a \textit{Generalized Pareto} (GP) distribution. The reader is referred to \cite{coles2001introduction} and \cite{resnick2008extreme} for a comprehensive exposition of EVT.

Over the years, numerous estimation methods have been developed for extreme-value distributions (see \cite{de2010parameter} and \cite{de2010parameter2} for detailed overviews of inference methods for the GP distribution). The most commonly used approach remains the classical \textit{Maximum Likelihood} (ML) estimation. The asymptotic behavior of ML estimators for the GEV and GP distributions has been extensively studied; in both cases, asymptotic normality has been established under the condition that the shape parameter exceeds $-1/2$ \citep[see][]{bucher2017maximum,dombryGEV,smith1985maximum}. Alternative popular approaches include methods of moments estimators, notably the \textit{Probability Weighted Moments} (PWM) estimator \citep{hosking1985estimation,hosking1987parameter}. Bayesian inference has also been successfully applied to extremes \citep[see, e.g.,][]{yoon2010full,stephenson2016bayesian}. In both the BM and the PoT frameworks, a bias–variance trade-off arises: choosing a large block size (or a high threshold) increases the likelihood that the resulting maxima (or exceedances) fall within the extreme-value regime and are approximately independent. However, this typically reduces the effective sample size, which increases statistical uncertainty. 
Furthermore, due to the nature of the data itself, sample extremes are prone to containing abnormal observations, further affecting the performance of traditional estimators such as ML and and PWM estimators. 

To address this issue,  robust estimators have been developed in recent decades. For example, \cite{peng2001robust} proposed a median-based estimator for the GP distribution that uses the empirical median rather than the moments, enhancing resistance to outliers. However, this approach lacks efficiency and relies on the GP’s two-parameter structure; therefore, no known extension exists for the three-parameter GEV distribution. Another robust method applicable to both the GEV and GP distributions is the \textit{Optimal Bias-Robust Estimator} (OBRE) introduced by \cite{dupuis1998robust}. The OBRE belongs to the family of M-estimators \citep[see, e.g.,][]{hampel86,huber2011robust} and minimizes a rescaled likelihood by assigning a weight to each observation: the more an observation deviates from the bulk of the data, the smaller its weight. More recently, \cite{lin2024multi} introduced the multi-quantile estimator, which expresses the three GEV parameters in terms of three quantiles, then estimates them by solving the resulting system of empirical equations.

In this paper, we propose a new estimator for the parameters of the GEV distribution that combines high efficiency and high robustness. Like the OBRE, it offers a tuning parameter to balance the trade-off between these two objectives. This estimator is the \textit{Minimum Density Power Divergence} (MDPD) estimator, originally proposed by \cite{basu1998robust}. As its name suggests, the MDPD estimator minimizes a divergence between two densities: in particular the divergence which is minimized lies between the Kullback–Leibler divergence and the squared $L^2$ distance. When one of the two densities used  by the divergence is the empirical distribution, minimizing the Kullback–Leibler divergence yields the ML estimator, which is known for its optimal asymptotic efficiency, while minimizing the squared $L^2$ distance yields an estimator with strong robustness. Thus, the MDPD estimator can be viewed as a compromise, combining the efficiency of ML with the robustness of $L^2$ methods, making it well-suited for the GEV framework. Indeed, \cite{juarez2004robust} have already demonstrated the strong theoretical and empirical performance of the MDPD estimator in the GP case.

Influential points can have a large impact on the estimation of extreme value distribution parameters in practice. This is clearly showcased in Section~\ref{sec:application} in which we carry out a flood frequency analysis, focusing on peak flow datasets characterized by the presence of left-tail outliers caused by years of zero or very low river flows. In this type of situation, it is sometimes recommended to identify and possibly remove the so-called \textit{Potentially Influential Low Floods} \citep[PILFs,][]{england2018guidelines}, thus limiting their influence in the estimation. Our proposed approach avoids the removal of these points allowing for the robust estimation of the extreme value distribution parameters without the need of identifying influential observations. 


This paper is structured as follows. Section~\ref{sec:background} provides foundational concepts in EVT and introduces the GEV distribution, followed by a general introduction to the density power divergence framework. Section~\ref{sec:GEVforMDPDE} contains the core theoretical contributions of the paper, including the definition of the MDPD estimator in the GEV setting and a proof of its asymptotic normality in Theorem~\ref{th:asympt_norm}. Section~\ref{sec:simulation_study} presents an extensive simulation study to empirically assess the performance of the MDPD estimator for GEV distributions. Finally, in Section~\ref{sec:application}, we apply the MDPD estimator to peak flow datasets in the UK.


%% file: evd.tex
A key result in univariate extreme value theory is the Fisher–Tippett–Gnedenko theorem \citep{fisher1928limiting,gnedenko1943distribution}. Let $\{Z_1, \dots, Z_n\}$ be a sample of i.i.d. random variables. Assume there exist two sequences $(a_n)_n, (b_n)_n \in \mathbb{R}^{\mathbb{N}}$, with $a_n > 0$, such that
\begin{equation*}
    \mb{P}\Big(\frac{\max\{Z_1,...,Z_n\} - b_n}{a_n} \m z\Big) \rightarrow F(z),
\end{equation*}
as $n \rightarrow +\infty$, where $F$ is a non-degenerate distribution function. Then, the limiting distribution $F$ necessarily belongs to the \textit{Generalized Extreme-Value} (GEV) family, which is given by
\begin{equation*}
    F(x;\mu,\sigma,\xi) = \exp\Big(-\Big(1+\xi\Big(\frac{x-\mu}{\sigma}\Big)\Big)^{-1/\xi}\Big)\1\{x \in D_{\mu,\sigma,\xi}\},
\end{equation*}
where $\mu,\sigma,\xi \in \mb{R}\times ]0,+\infty[\times \mb{R}$ are the location, scale and shape parameters, respectively, and where the support of the distribution $D_{\mu,\sigma,\xi}$ is defined as
  \begin{equation*}
    D_{\mu,\sigma,\xi} =
    \begin{cases*}
      [\mu - \sigma/\xi,+\infty[, & if $\xi >0$; \\
      \mb{R}, & if $\xi =0$; \\
      ]-\infty,\mu - \sigma/\xi],   & if $\xi <0$.
    \end{cases*}
    \end{equation*}
The corresponding probability density function of the GEV distribution is given by:
\begin{equation*}
    f(x;\mu,\sigma,\xi) = \frac{1}{\sigma}\Big(1+\xi\Big(\frac{x-\mu}{\sigma}\Big)\Big)^{-(\xi+1)/\xi}\exp\Big(-\Big(1+\xi\Big(\frac{x-\mu}{\sigma}\Big)\Big)^{-1/\xi}\Big)\1\{x \in D_{\mu,\sigma,\xi}\}.
\end{equation*}


%% file: mdpde.tex
In this section, we present the basics of the density power divergence introduced in \cite{basu1998robust}. Let $f$ and $g$ be two density functions defined on the same space $\mathscr{X}$. The density power divergence between $f$ and $g$ is defined as
\begin{equation}\label{eq:dpd}
d_\alpha(g, f)=\int_{\mathscr{X}}\left(f^{1+\alpha}(x)-\left(1+\frac{1}{\alpha}\right) g(x) f^\alpha(x)+\frac{1}{\alpha} g^{1+\alpha}(x)\right) d x
\end{equation}
for $\alpha > 0$, and as
\begin{equation}\label{eq:hellinger}
d_0(g, f)=\int_{\mathscr{X}} g(x) \log \left(\frac{g(x)}{f(x)}\right) d x
\end{equation}
for $\alpha = 0$, so that $d_0(g, f) = \lim_{\alpha \rightarrow 0} d_\alpha(g, f)$.

Let $\{X_1, \dots, X_n\}$ be i.i.d. random elements defined on $\mathscr{X}$, and denote by $g_n$ their empirical density function. Let $\mathscr{F} = \{f(x ; \theta) : x \in \mathscr{X}, \theta \in \Theta\}$ be a parametric family of density functions. A \textit{Minimum Density Power Divergence} (MDPD) estimator $\hat{\theta}_{\alpha,n} \in \Theta$ is defined as the element of $\Theta$ for which the corresponding model density is closest to the empirical density $g_n$ in terms of $d_\alpha$, that is:
\begin{equation*}
\hat{\theta}_{\alpha,n} \in \arg\min_{\theta \in \Theta} d_\alpha\big(g_n, f(\, \cdot\,;\theta)\big).
\end{equation*}
Rewriting Equation~\eqref{eq:dpd}, the MDPD estimator is equivalently obtained by minimizing the empirical criterion
\begin{equation}\label{eq:H_alpha}
H_\alpha(\theta)=\int_{\mathscr{X}} f^{1+\alpha}(x ; \theta) d x - \left(1+\frac{1}{\alpha}\right) \frac{1}{n} \sum_{i=1}^n f^\alpha\left(X_i ; \theta\right).
\end{equation}

Regarding Equations~\eqref{eq:dpd} and~\eqref{eq:hellinger}, note that the density power divergence reduces to the squared $L^2$ distance when $\alpha = 1$, and tends to the Kullback-Leiber divergence as $\alpha \rightarrow 0$. Therefore, the MDPD method offers a trade-off between efficiency and robustness depending on the choice of $\alpha$: when $\alpha = 0$ the MDPD estimator coincides with the ML estimator, which is known for its efficiency, and when $\alpha = 1$ it coincides with the $L^2$ estimator, which is known for its robustness \citep[see][for a detailed investigation in the case of the GP distribution]{vandewalle2007robust}.

In \cite{basu1998robust}, the authors establish several theoretical properties of the MDPD estimator, under standard regularity and identifiability assumptions, such as consistency and asymptotic normality of the estimator. They also analyze the influence function of the MDPD estimator, providing a closed-form expression based on $M$-estimation theory and proving its boundedness for $\alpha > 0$ in particular cases 
\citep[see Section~2 in][for additional details on the influence function]{vandewalle2007robust}. They also prove that the MDPD estimator is equivariant w.r.t. reparameterization but not, in general, under transformation of the data.



%% file: mdpdeGEV.tex
Let $\{X_1,...,X_n\}$ a sample of i.i.d. random variables from a GEV distribution, and denote by $g_n$ their empirical distribution. The parametric model is the family of GEV densities $\mathscr{F}_{GEV} = \{f(x ; \mu,\sigma,\xi) : x \in D_{\mu,\sigma,\xi}, (\mu,\sigma,\xi) \in \mb{R}\times\mb{R}_{>0}\times\mb{R}\}$. In this context, the function $H_\alpha$ \eqref{eq:H_alpha} can be rewritten as
\begin{equation}
    H_\alpha(\mu,\sigma,\xi) = \frac{1}{\sigma^\alpha}\Big(\frac{1}{1+\alpha}\Big)^{\alpha(\xi +1)+1}\Gamma\big(\alpha(\xi +1)+1\big)- \Big(1+\frac{1}{\alpha}\Big)\frac{1}{n}\sum_{i=1}^n f^\alpha(X_i;\mu,\sigma,\xi), \label{eq:Halpha}
\end{equation}
over $(\mu,\sigma,\xi) \in \mb{R}\times\mb{R}_{>0}\times]-(1+\alpha)/\alpha,+\infty[$ and where $\Gamma$ is the classic Gamma function. The condition $\xi > -(1+\alpha)/\alpha$ ensures to restrict the study over a domain on which the $\Gamma$-function is well-defined and smooth. Finally, a MDPD estimator $(\hat{\mu}_{\alpha,n},\hat{\sigma}_{\alpha,n},\hat{\xi}_{\alpha,n})$ is defined as 
\begin{equation*}
   (\hat{\mu}_{\alpha,n},\hat{\sigma}_{\alpha,n},\hat{\xi}_{\alpha,n}) \in \arg\min_{(\mu,\sigma,\xi)}H_\alpha(\mu,\sigma,\xi),
\end{equation*}
    over $\mb{R}\times\mb{R}_{>0}\times]-(1+\alpha)/\alpha,+\infty[$.

Denote the score function and the information of the GEV distribution by $S(x;\mu,\sigma,\xi)$, and $i(x;\mu,\sigma,\xi)$, respectively (see Section~\ref{appendix:score_and_inf} in the Appendix for explicit formulas). Define the $3\times3$ matrices $K_\alpha$ and $J_\alpha$ as

\begin{align*}
J_\alpha(\mu,\sigma,\xi)&=  \int_{D_{\mu,\sigma,\xi}} S(x ; \mu,\sigma,\xi) S^\top(x ; \mu,\sigma,\xi) f^{1+\alpha}(x ; \mu,\sigma,\xi) d x 
\end{align*}
and
\begin{equation*}
    K_\alpha(\mu,\sigma,\xi)=\int_{D_{\mu,\sigma,\xi}} S(x ; \mu,\sigma,\xi) S^\top(x ; \mu,\sigma,\xi) f^{1+2 \alpha}(x ; \mu,\sigma,\xi) d x-U_\alpha(\mu,\sigma,\xi) U_\alpha^\top(\mu,\sigma,\xi),
\end{equation*}
where $U_\alpha(\mu,\sigma,\xi) \in \mb{R}^3$ is given by
\begin{equation*}
    U_\alpha(\mu,\sigma,\xi) = \int_{D_{\mu,\sigma,\xi}} S(x ; \mu,\sigma,\xi) f^{1+\alpha}\big(x ; \mu,\sigma,\xi) d x.
\end{equation*}

These quantities appear in the expression of the asymptotic variance of the estimator’s limiting normal distribution. The asymptotic normality of the MDPD estimator for the GEV distribution is stated in the following theorem.

\begin{theorem}[Consistency and asymptotic normality]\label{th:asympt_norm}
    Suppose $g$ is a GEV density and let $(\mu_0,\sigma_0,\xi_0)$ be the target parameters, \ie, $g = f(\cdot;\mu_0,\sigma_0,\xi_0)$. Suppose $\xi_0 > -(1+\alpha)/(2+\alpha)$, for fixed $\alpha>0$. Then, there exists a sequence of MDPD estimators $\big((\hat{\mu}_{\alpha,n},\hat{\sigma}_{\alpha,n},\hat{\xi}_{\alpha,n})\big)_{n \s 1}$. In addition, this sequence is consistent for $(\mu_0,\sigma_0,\xi_0)$, as $n \rightarrow +\infty$, and  
    \begin{equation*}
        \sqrt{n}(\hat{\mu}_{\alpha,n}-\mu_0,\hat{\sigma}_{\alpha,n}-\sigma_0,\hat{\xi}_{\alpha,n}-\xi_0)^\top \overset{d}{\longrightarrow} \mathcal{N}\big(0,J^{-1}_\alpha(\mu_0,\sigma_0,\xi_0)K_\alpha(\mu_0,\sigma_0,\xi_0)J^{-1}_\alpha(\mu_0,\sigma_0,\xi_0) \big),
    \end{equation*}
    as $n \rightarrow + \infty$.
\end{theorem}

The proof of the theorem is provided in Section~\ref{appendix:proof_asympt_norm} of the Appendix. The same arguments used in the proof of asymptotic normality for the GP distribution in \cite{juarez2004robust} are applied to establish Theorem~\ref{th:asympt_norm}. The proof consists of verifying that the GEV distribution satisfies the conditions of Corollary A.3.4 in \cite{phdjuarez} (stated as Corollary~\ref{cor:juarez} in Section~\ref{appendix:proof_asympt_norm}), namely certain regularity conditions on the score and information functions. However, particular care is required for the GEV distribution, primarily due to the complex forms of these functions (see Section~\ref{appendix:score_and_inf}).



The restrictions $\xi_0 > -(1+\alpha)/\alpha$ in the definition of $H_\alpha$ \eqref{eq:Halpha} and $\xi_0 > -(1+\alpha)/(2+\alpha)$ in Theorem \ref{th:asympt_norm} are identical to those in the GP case. Consequently, the same observations hold for the GEV distribution: as $\alpha \rightarrow 0$, we recover the classical restriction $\xi_0 > -1/2$ under which the asymptotic normality of the ML estimator is guaranteed \citep[see][]{bucher2017maximum}; for $\alpha > 0$, the region in which asymptotic normality holds is enlarged compared to the ML estimator.

\begin{figure}[h!]
\centering
  \includegraphics[width=.3\textwidth]{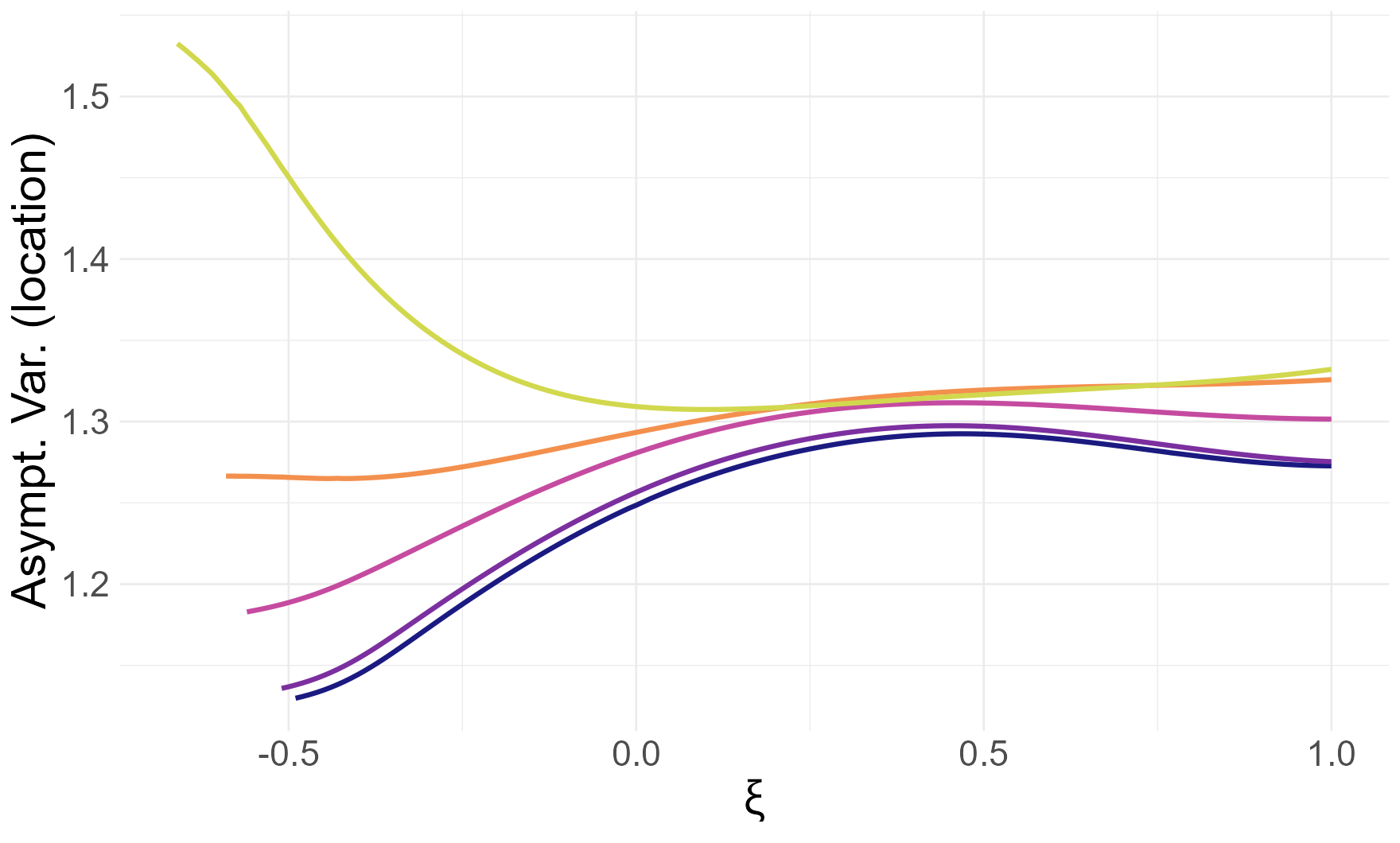}
  \hspace{0.4cm}
  \includegraphics[width=.3\textwidth]{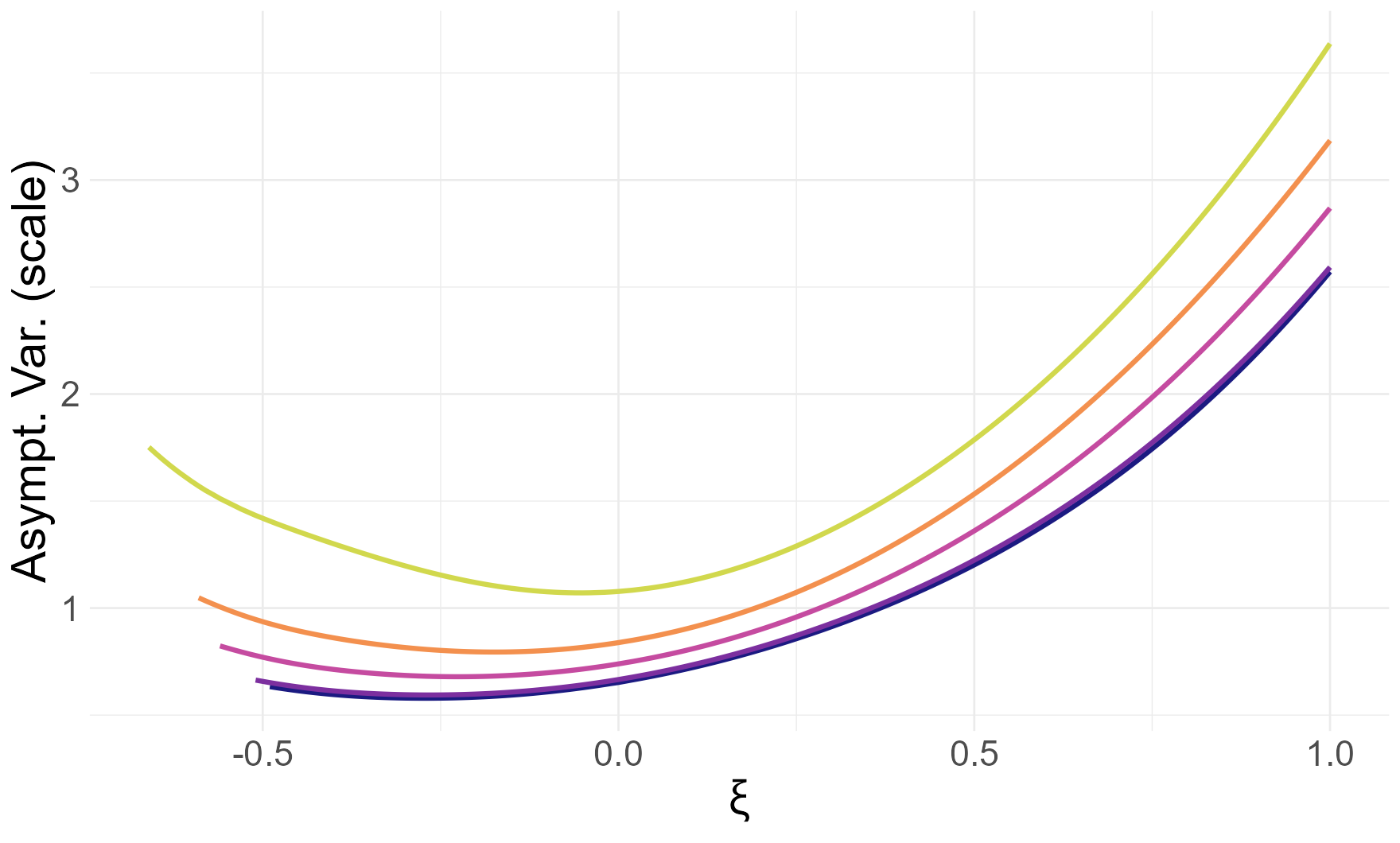}
  \hspace{0.4cm}
  \includegraphics[width=.3\textwidth]{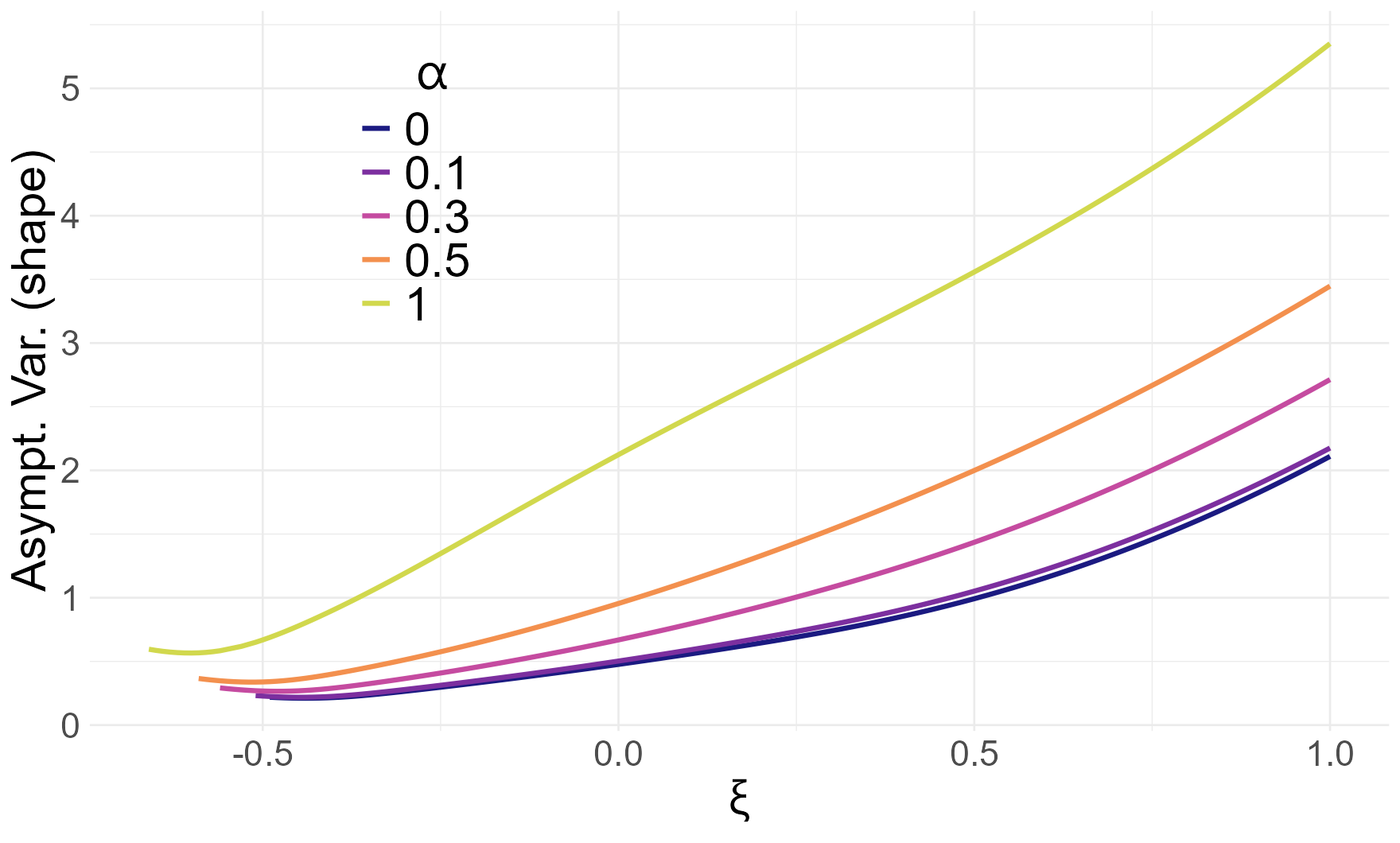}
  \caption{Asymptotic variance of the MDPD estimator involved in Theorem~\ref{th:asympt_norm} for different values of $\alpha$ and of the ML estimator ($\alpha =0$) as a function of the shape parameter $\xi_0$, for the location parameter (left), the scale parameter (middle), and the shape parameter (right). The true parameter values are $\mu_0 = 0$ and $\sigma_0 = 1$.\label{fig:asymptotic_variance}}
\end{figure}

The asymptotic variance in Theorem~\ref{th:asympt_norm} can be used to construct confidence intervals for the parameter estimators (see Section~\ref{sec:application}). It is therefore important to understand what is lost when using the MDPD estimator instead of the ML estimator, which is known for its optimal asymptotic variance. Figure~\ref{fig:asymptotic_variance} displays the asymptotic variance of the MDPD estimator for different values of $\alpha$ and of the ML estimator ($\alpha = 0$) as a function of the shape parameter. We observe that the asymptotic variance of the MDPD estimator for a reasonable value of $\alpha$, such as $\alpha = 0.1$, is very close to the optimal variance of the ML estimator, while still offering highly satisfactory efficiency and robustness performance (see Section~\ref{sec:simulation_study}). This figure also illustrates that the region of asymptotic normality becomes broader as $\alpha$ increases.

\begin{remark}[Model misspecification.]\label{rem:misspecified} In this article, we only prove asymptotic guarantees under the assumption of a correctly specified model, \ie, the true density $g$ belongs to the parametric family $\mathcal{F}$. At the cost of additional technicalities, similar asymptotic results could be established in the misspecified setting, as shown in the case of the GP model in \cite{juarez2004robust}.

Assuming that the true density $g$ follows a GEV distribution is supported by the Fisher–Tippett–Gnedenko theorem, which provides asymptotic justification for modeling block maxima with a GEV distribution \citep[see Chapter~3 in][]{coles2001introduction}. However, despite this theoretical guarantee, the assumption is often questionable in practice. Real-world data typically exhibit some degree of dependence (usually temporal or spatial) and often deviate from the fixed-parameters GEV model. Another reason  for this model deviation also lies in that fact that the "true extreme regime" in which the GEV assumption fully holds is never truly observed in finite samples. As a result, there is no assurance that the data-generating process falls within the assumed GEV family, or that the parameters remain fixed. For instance, many studies dealing with nonstationary GEV models assume a time-invariant shape parameter which is a debatable assumption \citep[see, e.g.,][]{jayaweera2025evidence}. In the PoT framework, it also remains unclear whether the threshold for defining extremes should vary with time \citep{eastoe2009modelling}. An interesting direction for future work would be to explore the situation where the true density $g$ belongs to one GEV family, but the inference is performed under a different GEV model. This line of research lies however beyond the scope of the present paper.

\end{remark}

An important concept in robust statistics is the influence function of an estimator \citep{hampel1974influence}. It measures the sensitivity of an estimator to a single observation and, in particular, quantifies how severely an outlier can affect it. It is well known that maximum likelihood estimators are generally non-robust to outliers, characterized by the unboundedness of their influence functions. 

Following \cite{juarez2004robust}, the influence function of the MDPD estimator for the GEV distribution admits a closed form \citep[see][]{basu1998robust} and is bounded, as stated in the following result.
\begin{proposition}[Influence function] In the settings of Theorem~\ref{th:asympt_norm}, for $\alpha \s 0$, the influence function $IF_\alpha$ of the MDPD estimator is given by
\begin{equation*}
    IF_\alpha\left(x; \theta_0\right)=J_\alpha^{-1}(\theta_0)\left[S(x ; \theta_0) f^\alpha(x ; \theta_0)-U_\alpha(\theta_0)\right],
\end{equation*}
    and is bounded for $\alpha>0$, for all $x \in D_{\mu_0,\sigma_0,\xi_0}$ and where $\theta_0=(\mu_0,\sigma_0,\xi_0)$.
\end{proposition}
The closed-form expression follows from the general results on the MDPD established in \cite{basu1998robust}, and the boundedness follows from the definition of the GEV density $f$ \eqref{appendix_eq:gev_density} and the score function $S$ (see Section~\ref{appendix:score_and_inf} in the Appendix). The boundedness of the influence function implies that the effect of an outlier on the parameter estimates is limited. This property provides a theoretical justification for the robustness of the MDPD estimator and represents an important advantage over the ML estimator, whose influence function is unbounded, meaning that a single outlier can severely impact the inference. The influence function of the MDPD estimator for the GEV distribution decomposes into three components
\begin{equation}\label{eq:compo_IF}
    IF_\alpha\left(x; \theta_0\right)
    = \big(IF_{\alpha,\mu_0}\left(x; \theta_0\right),\,
           IF_{\alpha,\sigma_0}\left(x; \theta_0\right),\,
           IF_{\alpha,\xi_0}\left(x; \theta_0\right)\big),
\end{equation}
where each component represents the influence that an observation $x$ has on the MDPD estimator of one specific parameter of the GEV distribution. It is worth noting that, for the ML estimator, not only is the global influence function $IF_0$ unbounded, but each marginal influence function $IF_{0,\mu_0}$, $IF_{0,\sigma_0}$, and $IF_{0,\xi_0}$ is also unbounded. This implies that an outlier can have an uncontrolled negative impact on the ML-estimation of every parameter of the GEV distribution.

\begin{figure}[h!]
\centering
  \includegraphics[width=.3\textwidth]{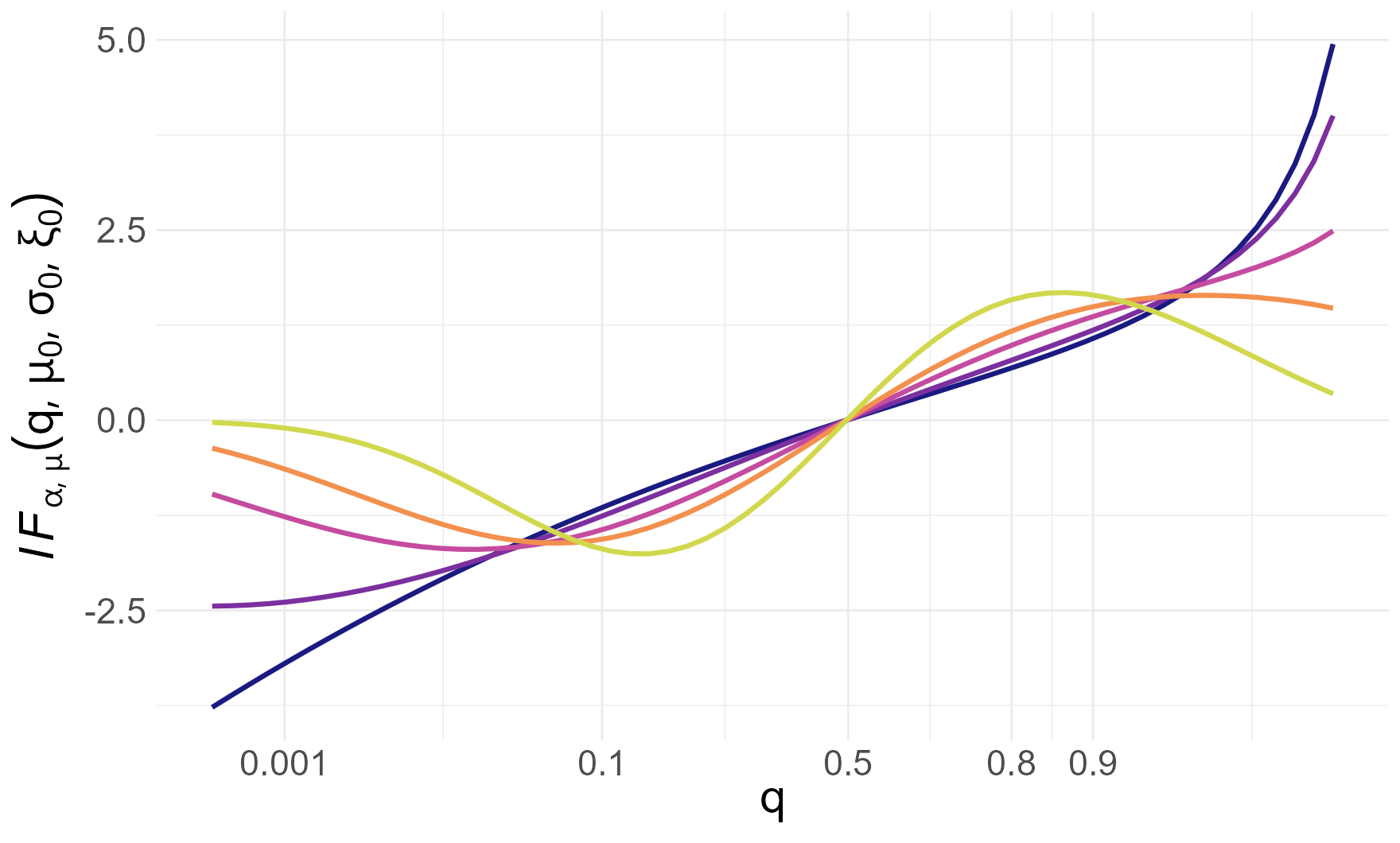}
  \hspace{0.4cm}
  \includegraphics[width=.3\textwidth]{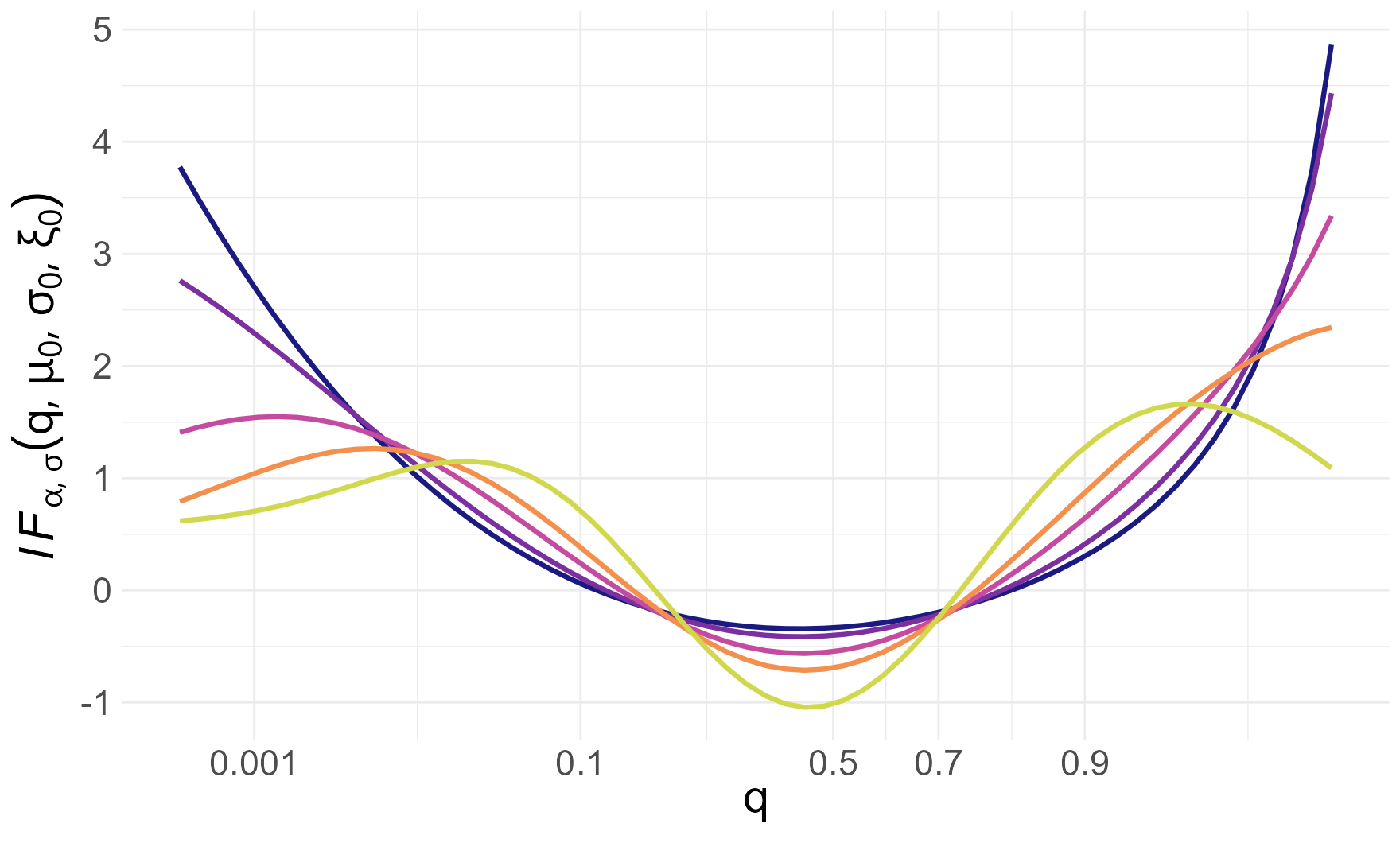}
  \hspace{0.4cm}
  \includegraphics[width=.3\textwidth]{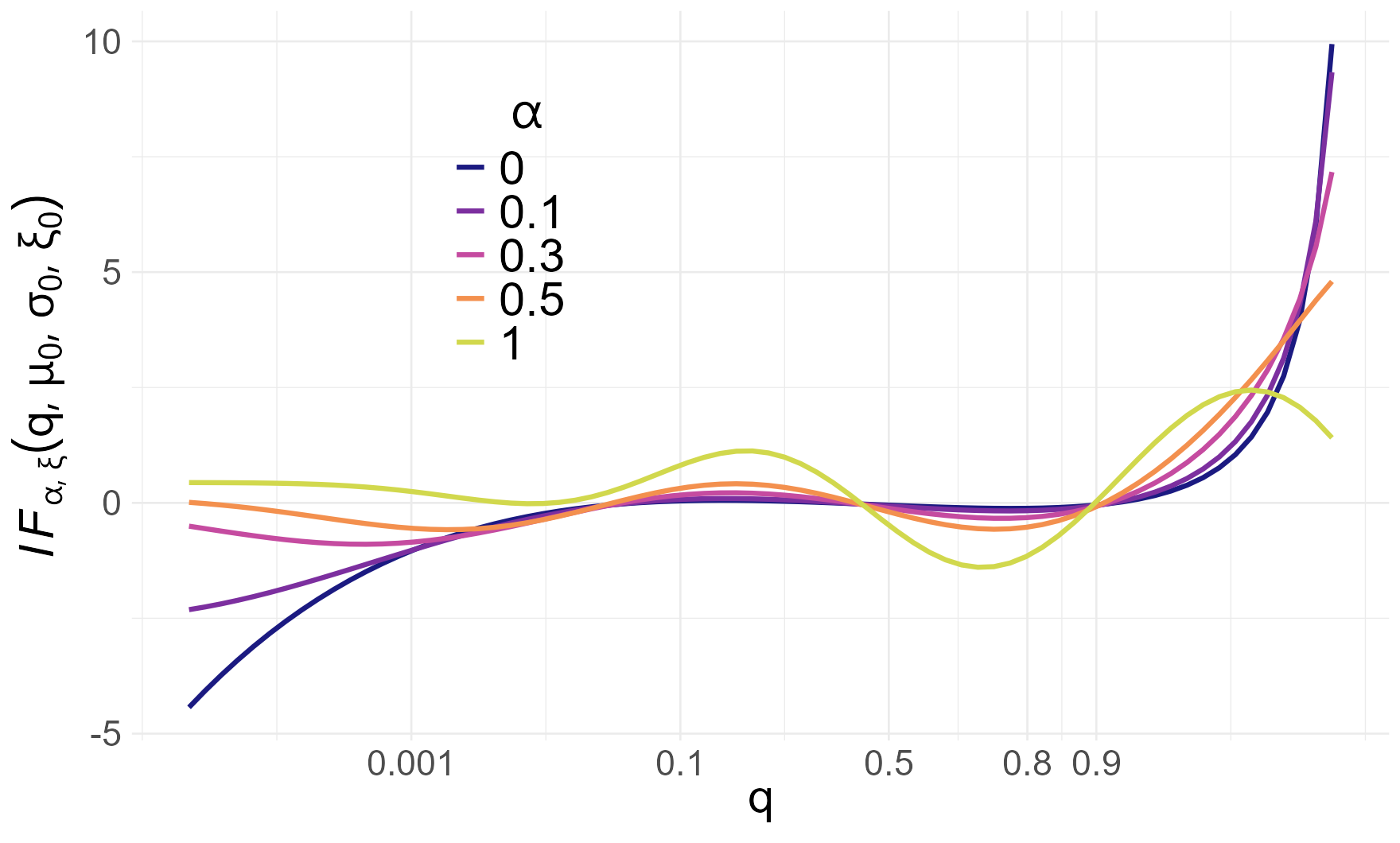}
  \caption{Influence functions of the MDPD estimators \eqref{eq:compo_IF} for different values of $\alpha$ and of the ML estimator ($\alpha = 0$), for the location parameter (left), the scale parameter (middle), and the shape parameter (right). The true parameter values are $\mu_0 = 0$, $\sigma_0 = 1$, and $\xi_0 = -0.3$ corresponding to a upper bound for the domain of approximately 3.3. The x-axis represents the quantile level at which the influence functions are evaluated. \label{fig:if_xi_negative}}
\end{figure}

\begin{figure}[h!]
\centering
  \includegraphics[width=.3\textwidth]{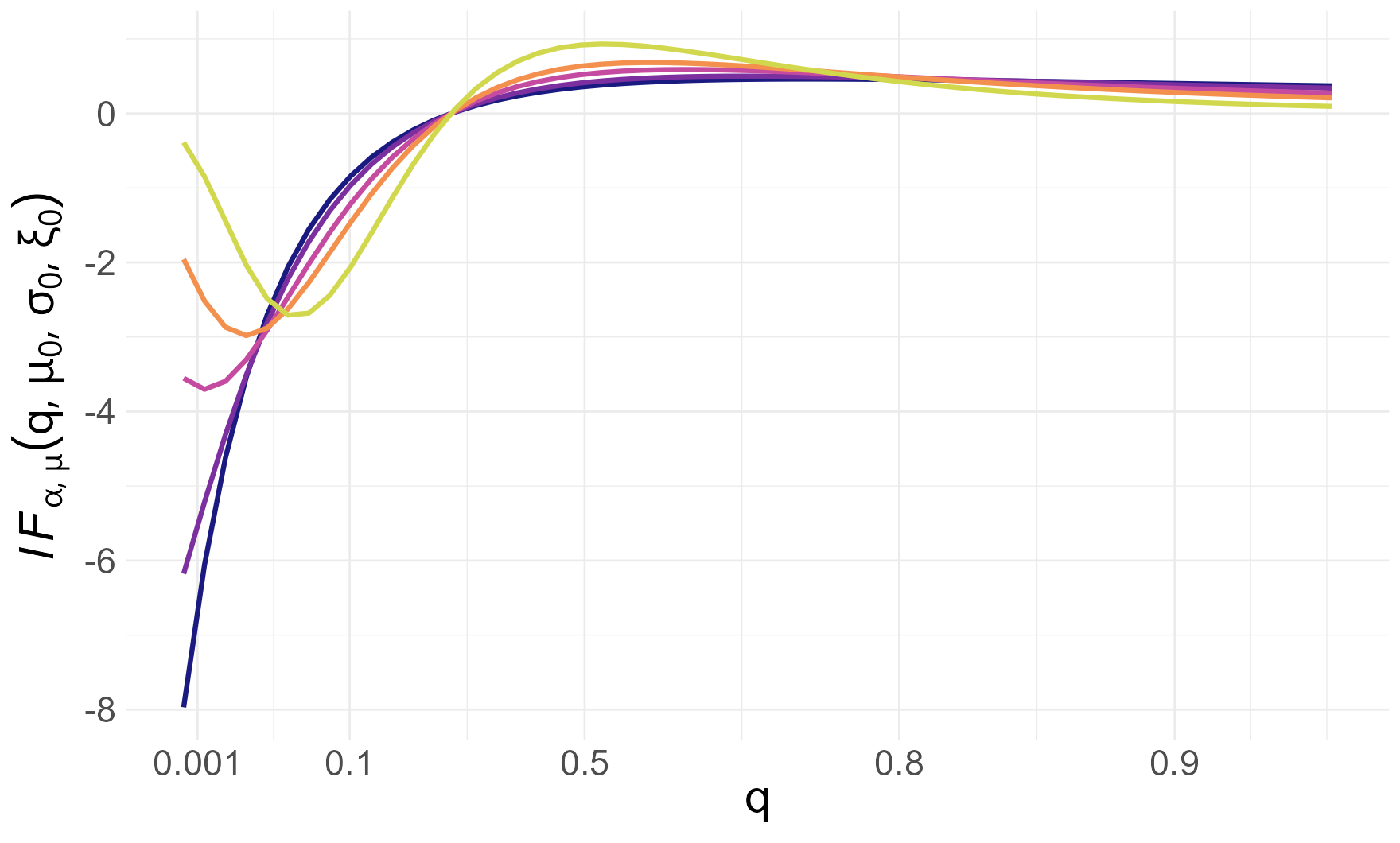}
  \hspace{0.4cm}
  \includegraphics[width=.3\textwidth]{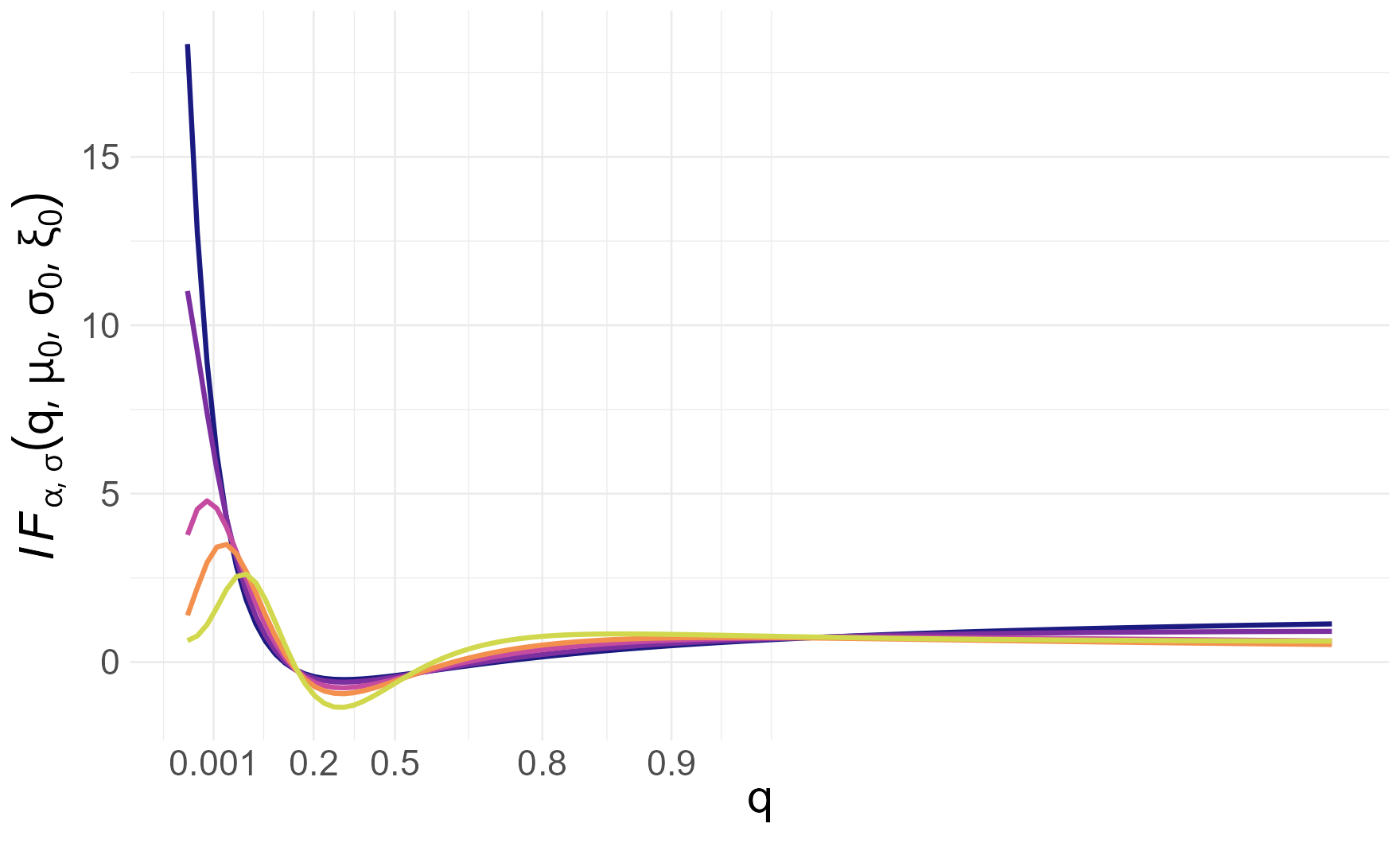}
  \hspace{0.4cm}
  \includegraphics[width=.3\textwidth]{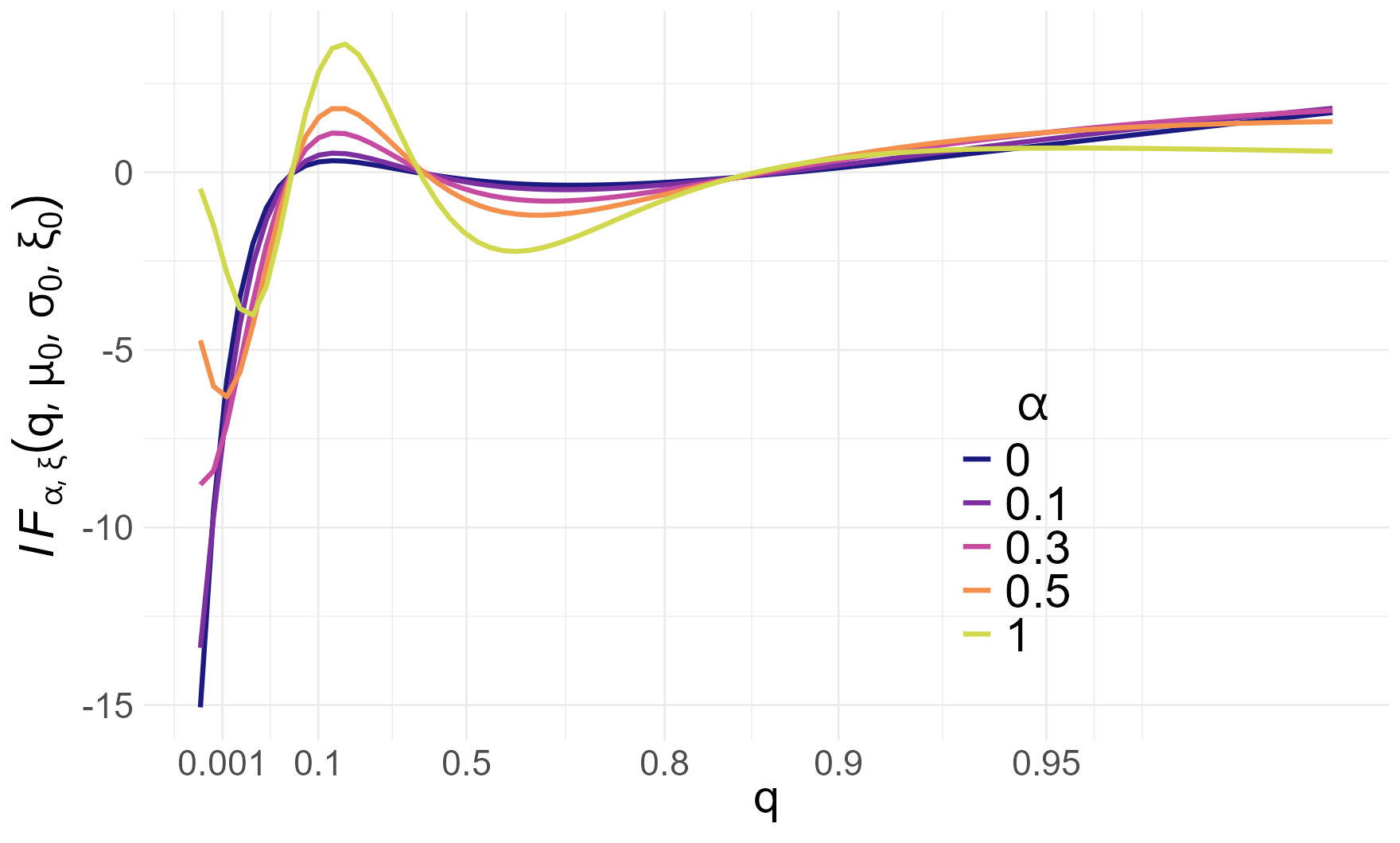}
  \caption{Influence function of the MDPD estimators \eqref{eq:compo_IF} for different values of $\alpha$ and of the ML estimator ($\alpha = 0$), for the location parameter (left), the scale parameter (middle), and the shape parameter (right). The true parameter values are $\mu_0 = 0$, $\sigma_0 = 1$, and $\xi_0 = 0.3$ corresponding to a lower bound for the domain of approximately -3.3. The x-axis represents the quantile level at which the influence functions are evaluated. 
  \label{fig:if_xi_positive}}
\end{figure}

Figures~\ref{fig:if_xi_negative} and~\ref{fig:if_xi_positive} show the influence functions of the MDPD estimator for different values of $\alpha$, as well as for the ML estimator ($\alpha = 0$), for each parameter under both positive and negative values of $\xi_0$, respectively. These figures illustrate the boundedness of the influence functions of the MDPD estimator and the unboundedness in some direction of those of the ML estimator. As $\alpha$ increases, the influence functions stabilize more rapidly, highlighting the increasing robustness of the MDPD estimator with larger $\alpha$ values. Interestingly, for a negative shape parameter, an observation shows approximately the same absolute influence whether it approaches the upper bound (around 3.3) or the infinite lower-bound of the distribution. In contrast, for a positive shape parameter, an observation approaching the lower bound (around $-3.3$) has an exploding influence, whereas one approaching the infinite upper-bound has a relatively-mild impact.

%% file: expes.tex
In this section, we assess the performance of the MDPD estimator from several perspectives, particularly its efficiency and robustness, in comparison to classical estimators, such as the ML estimator. The extreme value methods from the \texttt{R} package \texttt{mev} are employed, following the guidelines of \cite{belzile2023modeler}. The code used to run our experiments is available at \url{https://github.com/HuetNathan/robustGEV}.

A recurrent question in power divergence methods is the choice of the tuning parameter $\alpha$. Ideally, this choice should be data-driven, adapting to the proportion and nature of outliers in the data to balance robustness and efficiency, as well as the bias–variance trade-off. Some existing methods rely on minimizing an asymptotic variance of the estimator \citep[][]{hong2001automatic}, an empirical version derived from the limiting distribution in Theorem~\ref{th:asympt_norm}, or an asymptotic mean square error criterion \citep[][]{warwick2005choosing}. However, these approaches require selecting a pilot estimator beforehand to compute the different measure quantities, and this choice can significantly influence the selected value of $\alpha$ and in fine the resulting MDPD estimator. To address this dependency, \cite{basak2021optimal} propose to iterate the method of \cite{warwick2005choosing} by using at each step the estimator from the previous iteration as the pilot for the next. Alternatively, \cite{sugasawa2021selection} avoid asymptotic criteria and the need for a pilot estimator by minimizing a scoring function directly with respect to $\alpha$. In this study, following the approach of \cite{juarez2004robust}, we adopt a naive but practical strategy: selecting a small, fixed value for $\alpha$, typically $0.05$ or $0.1$, which has been shown empirically to provide a reasonable compromise between robustness and efficiency.

To obtain a global measure of performance, we compute the error using the Wasserstein distance of order~1 \citep[as in, e.g.,][]{alonso2014comparing,negahban2025framework}, defined as
\begin{equation}\label{eq:wasserstein}
W_1(F_1,F_2) =\int_{\mb{R}} |F_1(x)-F_2(x)|dx,
\end{equation}
where $F_1$ and $F_2$ are cumulative distribution functions \citep[see][for more details about the Wasserstein distance]{panaretos2019statistical}.
In our context, we take $F_1$ to be the true underlying distribution and $F_2$ to be the distribution obtained using the parameters estimated by the different methods under comparison.

Table~\ref{table:relative_wd} illustrates the efficiency of the MDPD estimator for various values of $\alpha$, which, we recall, controls the trade-off between efficiency and robustness (the closer $\alpha$ is to zero, the more efficient the estimator; the closer it is to one, the more robust the estimator). The table reports the ratios of Wasserstein distances between distributions estimated using ML (numerator) and MDPD (denominator). The models are trained on different uncontaminated samples drawn from GEV distributions with fixed location and scale parameters $\mu_0 = 0$ and $\sigma_0 = 1$, and varying shape parameters $\xi_0 \in \{-0.4, -0.2, ..., 0.8\}$. A ratio below 1 indicates that the MDPD estimator performs worse than the ML estimator; a ratio equal to 1 indicates comparable performance; and a ratio above 1 would indicate that the MDPD estimator outperforms the ML estimator. Since the ML estimator is known for its optimal efficiency in uncontaminated settings, as expected, none of the ratios exceed 1.
For moderate shape parameters varying from $-0.4$ to $0.4$, which is the typical range in practical applications, the ratios remain very close to 1: from 
0.98 to 1.00 for $\alpha=0.05$, and from 0.96 to 1.00 for $\alpha =0.1$. These results confirm that a naive choice of $\alpha = 0.05$ or $0.1$ should yield MDPD estimators with high efficiency, and the drop in efficiency is relevant only when using fairly large values of $\alpha$. 
The loss of efficiency of the MDPD estimator is more pronounced for high positive values of the shape parameter $\xi_0$. 
Note that throughout this section the shape parameters are always kept strictly below one in order to theoretically satisfy the required moment assumption for using the Wasserstein distance of order 1 (existence of the first moment).

\begin{table}[ht!]
\caption{
Ratio of average Wasserstein distances between ML- and MDPD-estimated distributions, for several values of $\alpha$ (column-wise) and several $\xi_0$-varying uncontaminated models (row-wise). The averages are computed over $d=200$ replications of samples of size $n=100$. \label{table:relative_wd}}
\vspace{0.2cm}

\begin{center}
\begin{small}
\begin{tabular}{c|ccccccccccc}
    \diagbox{$\xi_0$}{$\alpha$} & 0.02 &0.05&0.1&0.15&0.2&0.3&0.4&0.5&0.6&0.7&0.8 \\ \hline
    -0.4 & 0.98 & 0.98 & 0.99 & 0.98 & 0.97 & 0.96 & 0.95 & 0.94 & 0.90 & 0.86 & 0.85 \\
    -0.2 &0.99 & 1.00 & 0.99 & 1.00 & 0.98 & 0.97 & 0.95 & 0.91 & 0.90 & 0.85 & 0.85 \\
    0 & 1.00 & 1.00 & 0.99 & 0.99 & 0.99 & 0.97 & 0.94 & 0.91 & 0.87 & 0.87 & 0.88 \\
    0.2 &  1.00 & 1.00 & 1.00 & 0.96 & 0.98 & 0.91 & 0.85 & 0.79 & 0.73 & 0.61 & 0.60\\
    0.4 & 1.00 & 0.99 &0.96 & 0.94 & 0.90 & 0.84 & 0.72 & 0.69 & 0.67 & 0.58 & 0.47  \\
    0.6 & 1.00 & 0.98 & 0.96 & 0.87 & 0.86 & 0.79 & 0.64 & 0.63 & 0.57 & 0.50 & 0.50 \\
    0.8 &  0.94 & 0.84 & 0.81 & 0.83 & 0.71 & 0.64 & 0.49 & 0.53 & 0.36 & 0.32 & 0.23
\end{tabular}
\end{small}
\end{center}
\end{table}

To assess the performance of our estimator, we compare the MDPD estimator with the classical ML estimator and with a recently proposed robust method, the \emph{Multi-Quantile} (MQ) estimator, introduced in \cite{lin2024multi}. 
Performance is assessed on both uncontaminated and contaminated GEV samples, in order to evaluate efficiency and robustness, respectively. The data are generated according to the following mixture contamination model:
\begin{equation*}
(1-\varepsilon)\mathrm{GEV}(\mu_0,\sigma_0,\xi_0) + \varepsilon\mathrm{GEV}(\mu_1,\sigma_1,\xi_1).
\end{equation*}
For each configuration, the three parameters are estimated over $d = 200$ samples of size $n = 100$. We compute the error using the Wasserstein distance of order 1 in Equation \eqref{eq:wasserstein}. In our setting, $F_1$ is the cdf of a GEV distribution with parameters $(\mu_0, \sigma_0, \xi_0)$, and $F_2$ is the cdf of a GEV distribution with estimated parameters $(\hat{\mu}_0, \hat{\sigma}_0, \hat{\xi}_0)$ obtained from the different methods.

We consider three scenarios: one with a positive shape parameter ($\xi_0 = 0.1$), one with a negative shape parameter ($\xi_0 = -0.1$), and one with a zero shape ($\xi_0 = 0$). In all cases, the location and scale parameters are fixed at $\mu_0 = 0$ and $\sigma_0 = 1$, and the contamination proportion is set to $\varepsilon = 0.1$. We vary either the contaminating shape parameter $\xi_1$ or the contaminating scale parameter $\sigma_1$—but never both simultaneously. For the MQ procedure, the triplet of quantile levels must be specified in advance. Based on Figure~5 in the Appendix of \cite{lin2024multi}, we use the triplet $(q_1, q_2, q_3) = (0.1, 0.3, 0.9)$ for $\xi_0 > 0$, $(0.1, 0.7, 0.9)$ for $\xi_0 < 0$, and $(0.1, 0.4, 0.9)$ for $\xi_0 = 0$. 

\begin{figure}[ht!]
\centering
  \includegraphics[width=.47\textwidth]{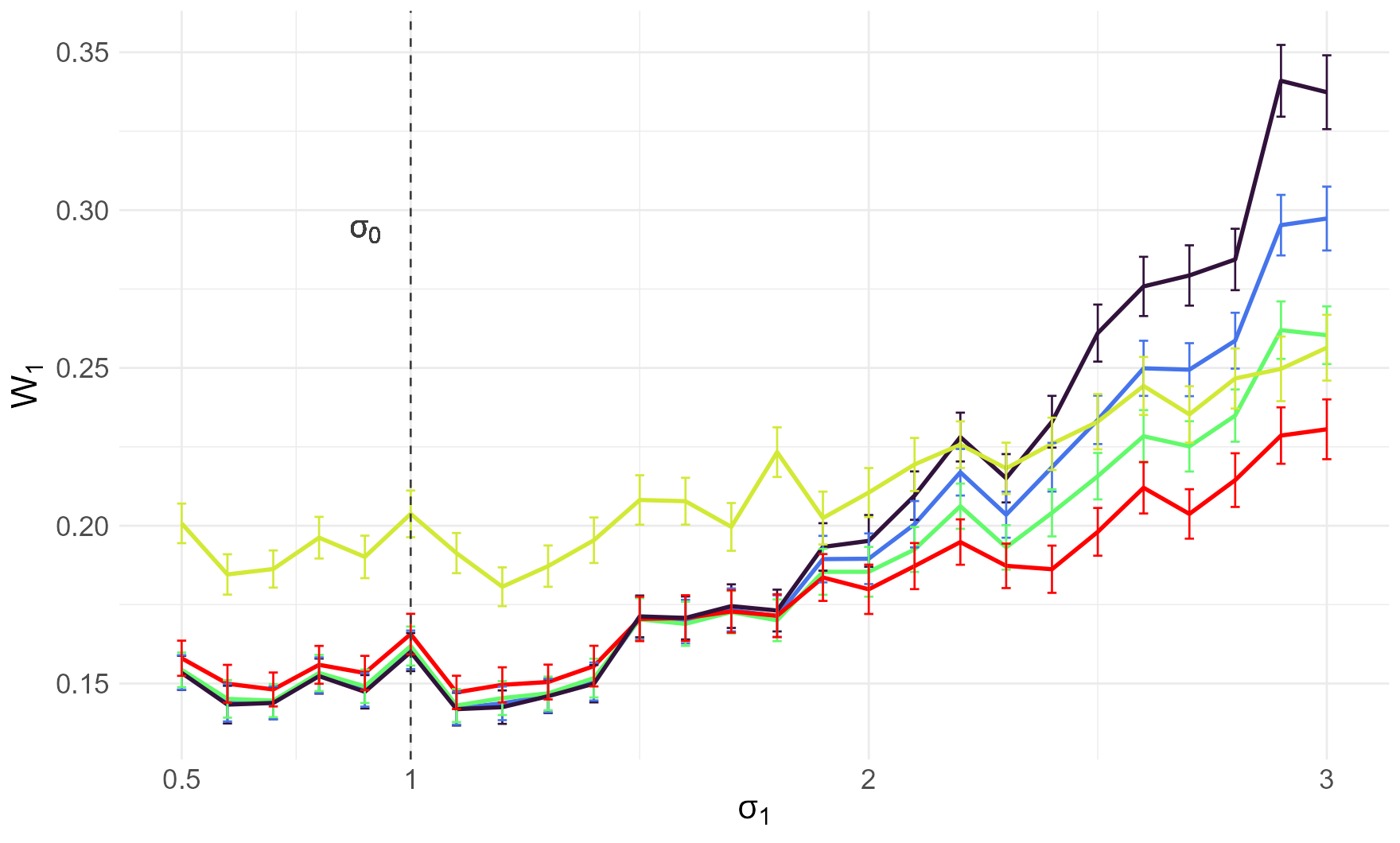}
  \hspace{0.7cm}
  \includegraphics[width=.47\textwidth]{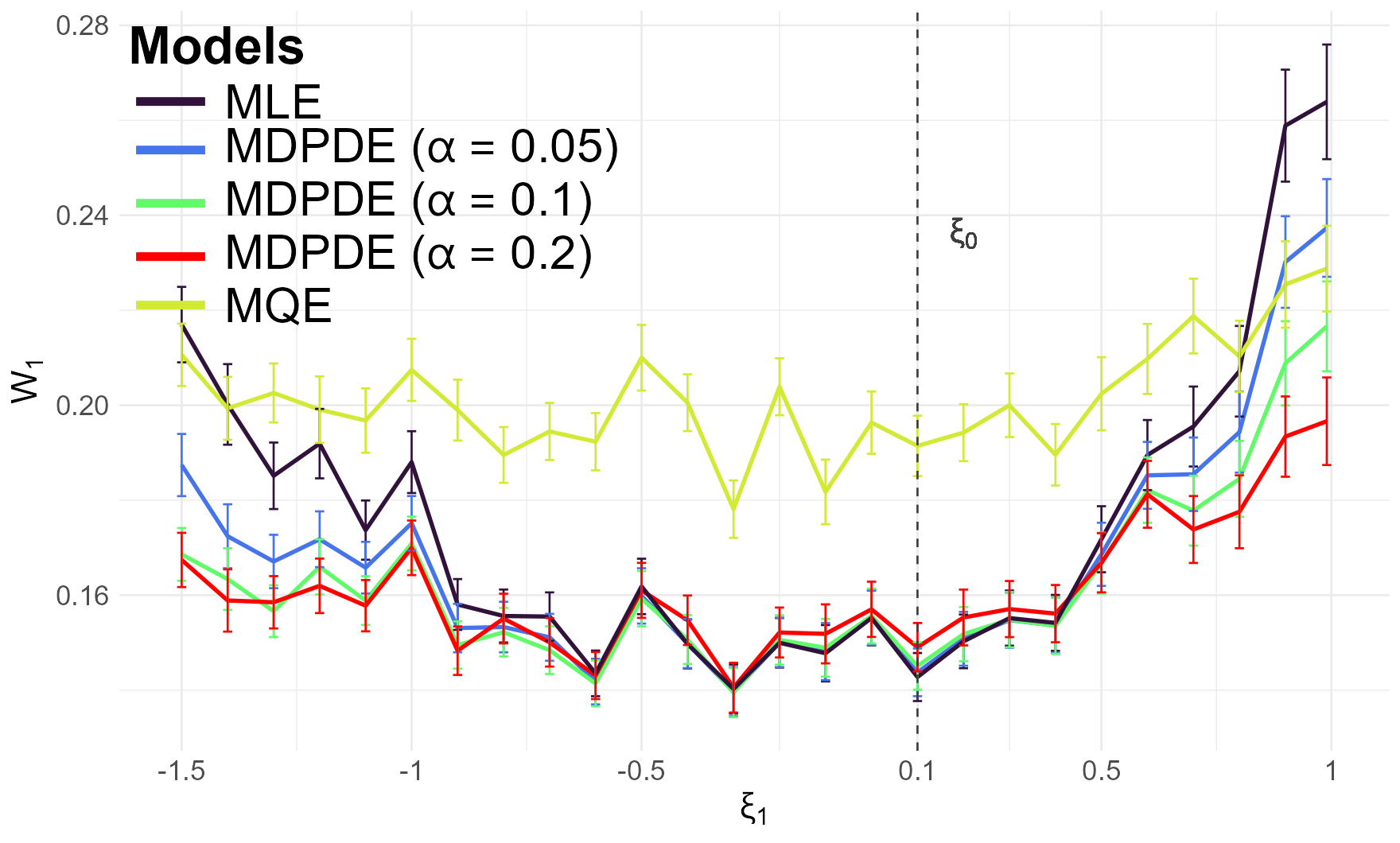}
  \caption{Average Wasserstein distance over 200 replications (with standard errors) across various contaminated models. In the left panel, the shape parameter $\xi_1$ varies while the 
  location and the scale parameters are fixed to the true model parameters ($\mu_1 = \mu_0$ and $\sigma_1=\sigma_0$). In the right panel, the scale parameter $\sigma_1$ varies while the location and the shape parameters are fixed to the true model parameters ($\mu_1 = \mu_0$ and $\xi_1=\xi_0$). Each sample has size $n = 100$, with contamination proportion $\varepsilon = 0.1$. The true model parameters are $\mu_0 = 0$, $\sigma_0 = 1$, and $\xi_0 = 0.1$. \label{fig:wd_positive_xi}}
\end{figure}

\begin{figure}[ht!]
\centering
  \includegraphics[width=.47\textwidth]{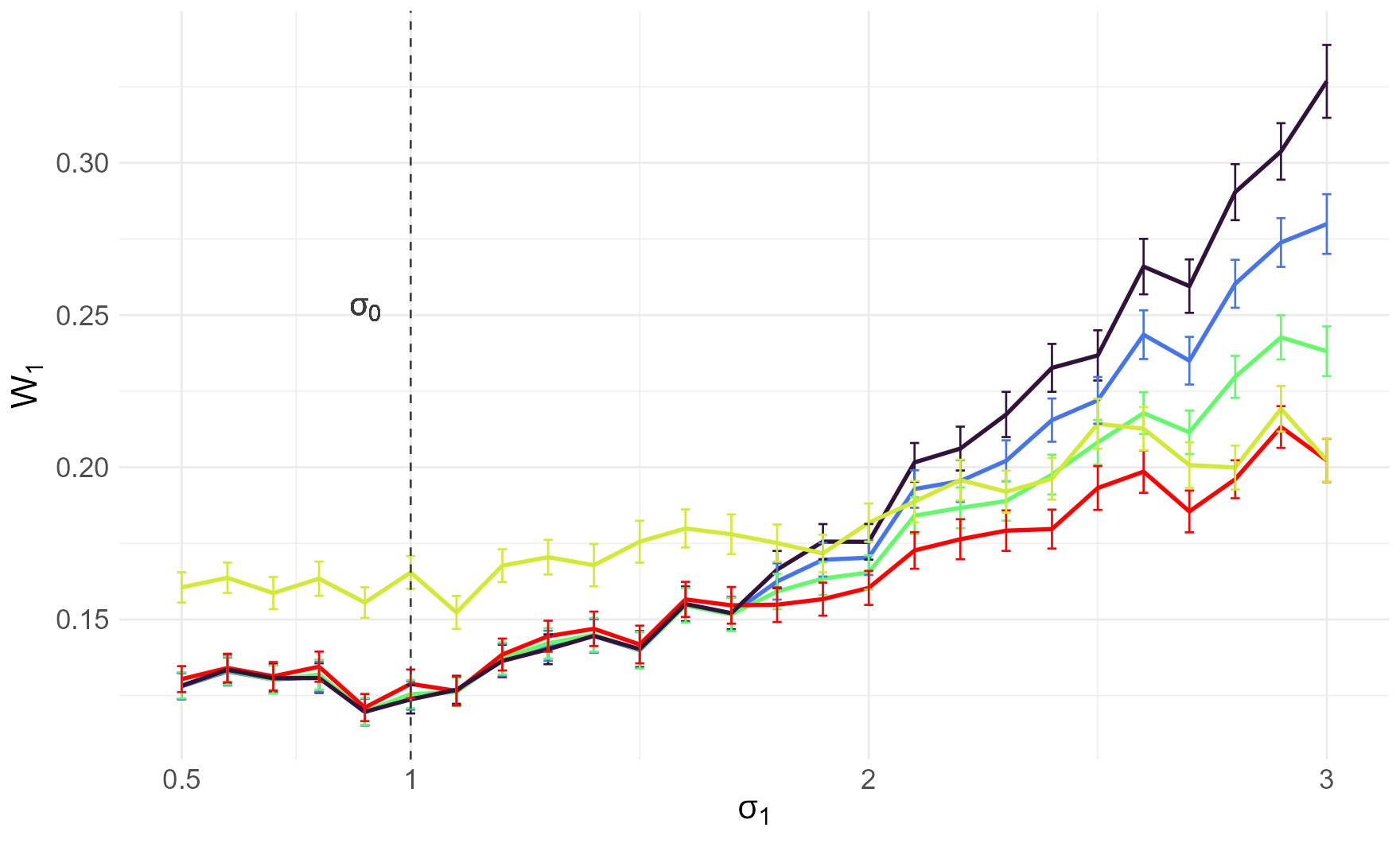}
  \hspace{0.7cm}
  \includegraphics[width=.47\textwidth]{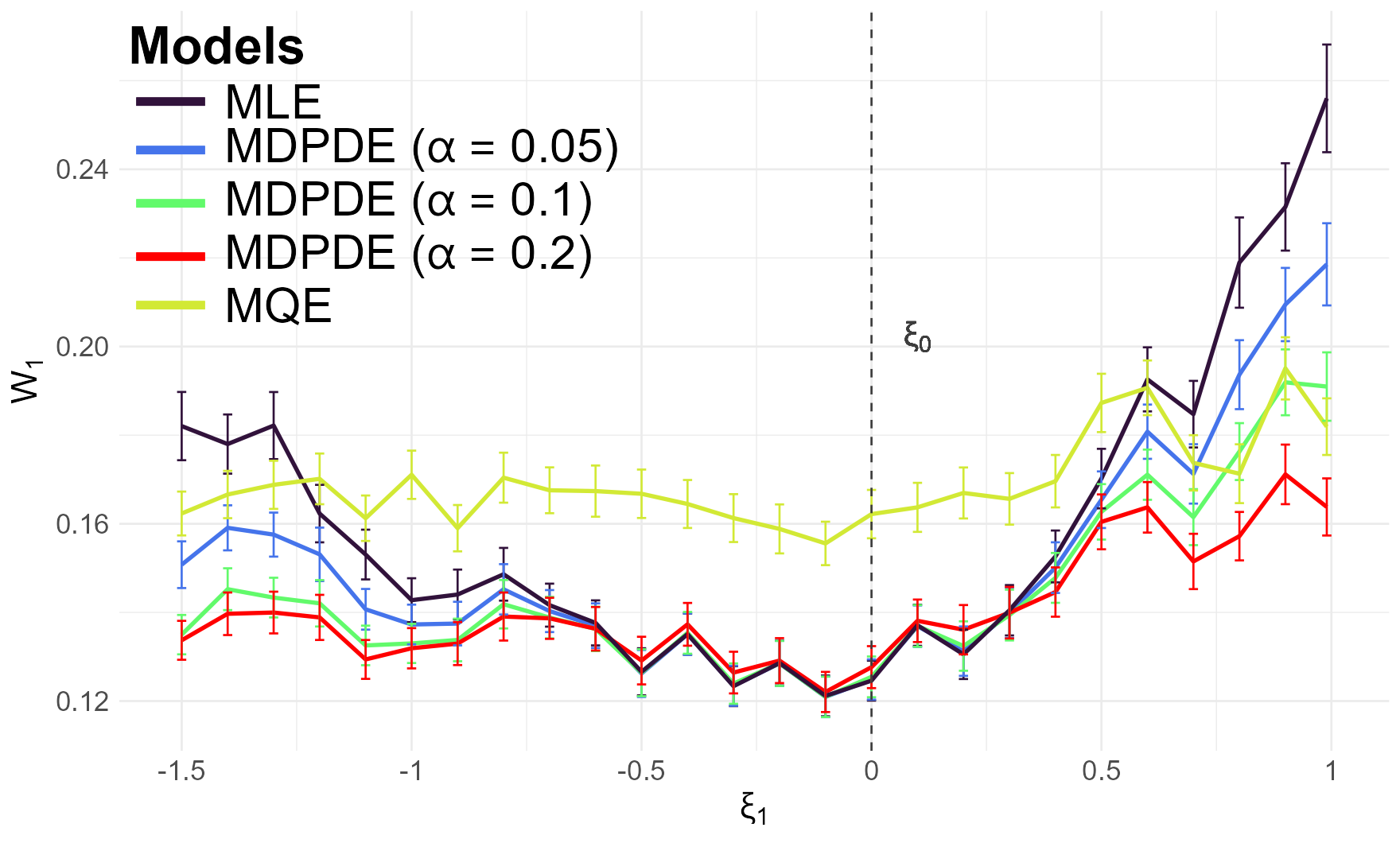}
  \caption{Average Wasserstein distance over 200 replications (with standard errors) across various contaminated models. In the left panel, the shape parameter $\xi_1$ varies while the 
  location and the scale parameters are fixed to the true model parameters ($\mu_1 = \mu_0$ and $\sigma_1=\sigma_0$).
  In the right panel, the scale parameter $\sigma_1$ varies while the location and the shape parameters are fixed to the true model parameters ($\mu_1 = \mu_0$ and $\xi_1=\xi_0$). Each sample has size $n = 100$, with contamination proportion $\varepsilon = 0.1$. The true model parameters are $\mu_0 = 0$, $\sigma_0 = 1$, and $\xi_0 = 0$.\label{fig:wd_gumbel}}
\end{figure}

\begin{figure}[ht!]
\centering
  \includegraphics[width=.47\textwidth]{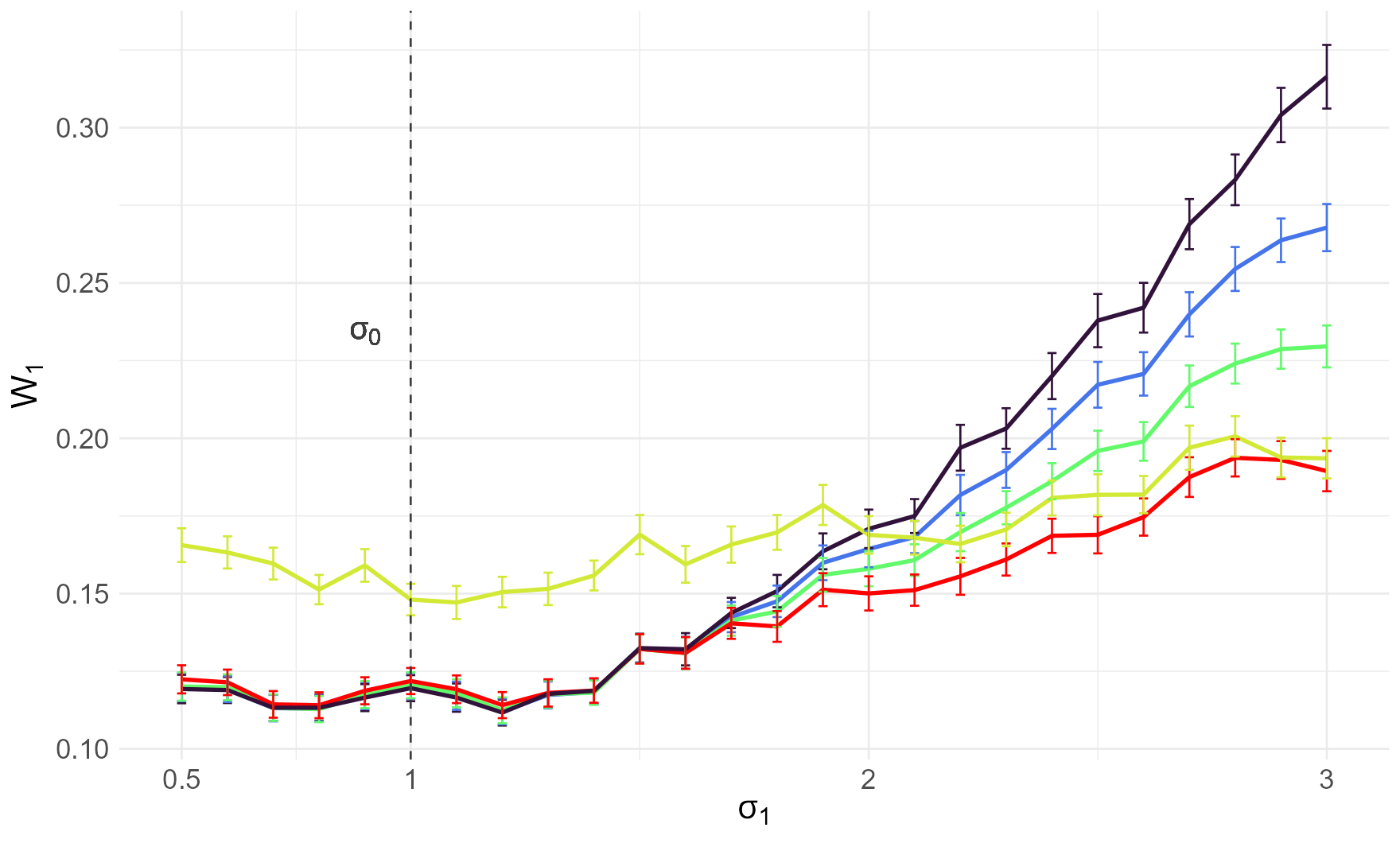}
  \hspace{0.7cm}
  \includegraphics[width=.47\textwidth]{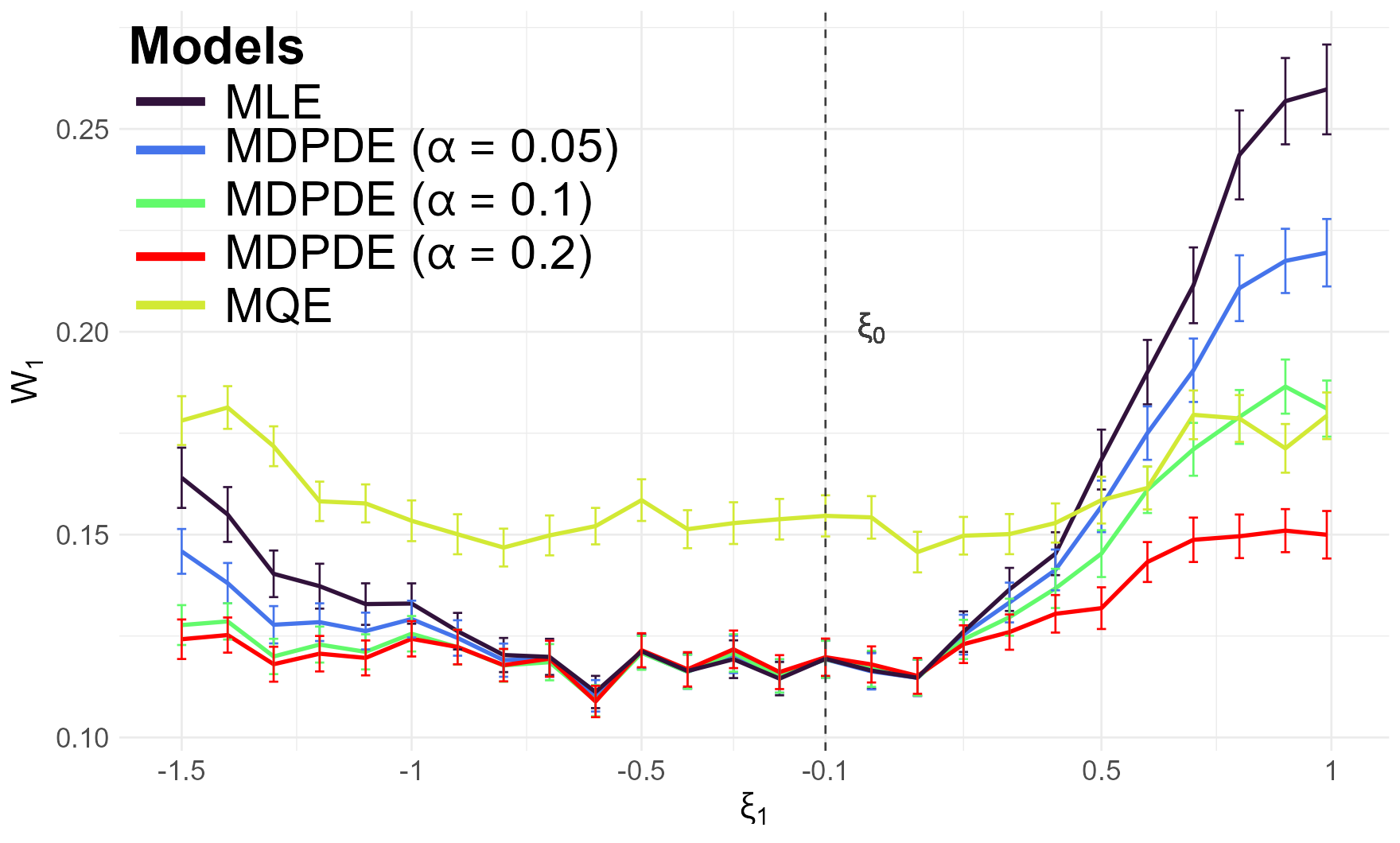}
  \caption{Average Wasserstein distance over 200 replications (with standard errors) across various contaminated models. In the left panel, the shape parameter $\xi_1$ varies while the location and the scale parameters are fixed to the true model parameters ($\mu_1 = \mu_0$ and $\sigma_1=\sigma_0$) In the right panel, the scale parameter $\sigma_1$ varies while the location and the shape parameters are fixed to the true model parameters ($\mu_1 = \mu_0$ and $\xi_1=\xi_0$). Each sample has size $n = 100$, with contamination proportion $\varepsilon = 0.1$. The true model parameters are $\mu_0 = 0$, $\sigma_0 = 1$, and $\xi_0 = -0.1$.\label{fig:wd_negative_xi}}
\end{figure}

The results are presented in Figures~\ref{fig:wd_positive_xi}, \ref{fig:wd_gumbel}, and~\ref{fig:wd_negative_xi}. Additional experiments for this setting, as well as for cases with $\varepsilon = 0.2$, $\varepsilon = 0.05$, or with a smaller sample size $n = 50$, are reported in Section~\ref{sec:simu_appendix} of the Appendix. Further comparisons with classical robust estimators, namely the \emph{radius-minimax estimator} and the \emph{optimal MSE estimator}, are also provided for the positive shape parameter case 
\citep[as they are only implemented in this setting in the R packages \texttt{ROptEst} and \texttt{RobExtremes};][]{horbenko2018package,kohl2019package} in Section~\ref{sec:simu_appendix}.



A first general remark concerns our choice of the true model: we set the true shape parameter $\xi_0$ to $-0.1$, $0$, or $0.1$, values typically observed in environmental applications. These choices ensure that both the mean and variance exist, and that they fall within the “stability domain” of the ML estimator, allowing for a fair comparison with it in a setting where it is expected to perform optimally.

Regarding the plots, the results depicted by Figures~\ref{fig:wd_positive_xi}, \ref{fig:wd_gumbel}, and~\ref{fig:wd_negative_xi} are unanimous. In uncontaminated or lightly contaminated scenarios, the ML estimator slightly outperforms the others. When the contaminating parameters are smaller than the true ones, the performances of the ML and MDPD estimators are generally comparable, though the MDPD often shows a slight advantage in shape contaminated settings. However, with a contamination with larger parameters than the true ones, the performance of the ML estimator deteriorates significantly, while the MDPD estimator maintains greater stability and accuracy, confirming its higher robustness. As expected, the more severe the contamination, the better the performance of the MDPD estimator with $\alpha = 0.2$ compared to its versions with $\alpha = 0.1$ or $\alpha = 0.05$. In all cases, the uncertainty levels associated with both estimators are of the same order of magnitude.  Another advantage of the MDPD estimator lies in its more reliable convergence during numerical optimization. Empirically, the minimization procedure converges more consistently than that of the ML estimator, possibly due to a slightly wider “stability region”, \ie, the region where asymptotic normality holds. However, it is worth noting that although larger values of $\alpha$ are generally expected to reduce convergence failures, this trend is not observed in Table~\ref{tab:missing_value} in the Appendix. Regarding the MQ estimator, it exhibits satisfying robustness to heavy-tailed contamination in scale or shape compared to the other estimators (see Figures~\ref{fig:wd_gumbel} and~\ref{fig:wd_negative_xi}). However, in other settings, and particularly in the standard uncontaminated case, the MQ estimator performs much worse than both the ML and MDPD estimators, highlighting its low efficiency. In fact, the performance of the MQ estimator appears unaffected by contamination: its MSE remains stable across levels of contamination, yielding acceptable values under heavy contamination but excessively large ones for uncontaminated or lightly contaminated cases. In conclusion, the MDPD estimator empirically provides a reasonable compromise between robustness and efficiency, making it a compelling choice across a variety of scenarios. Using the MDPD estimator when the data originates from an uncontaminated distribution does not lead to a relevant loss in performance in the estimation, but when the true distribution is contaminated using the more robust procedure ensures that the parameters of the main distribution are still mostly correctly characterized.

%% file: appli.tex
\begin{figure}[ht!]
\centering
\includegraphics[width=\textwidth]{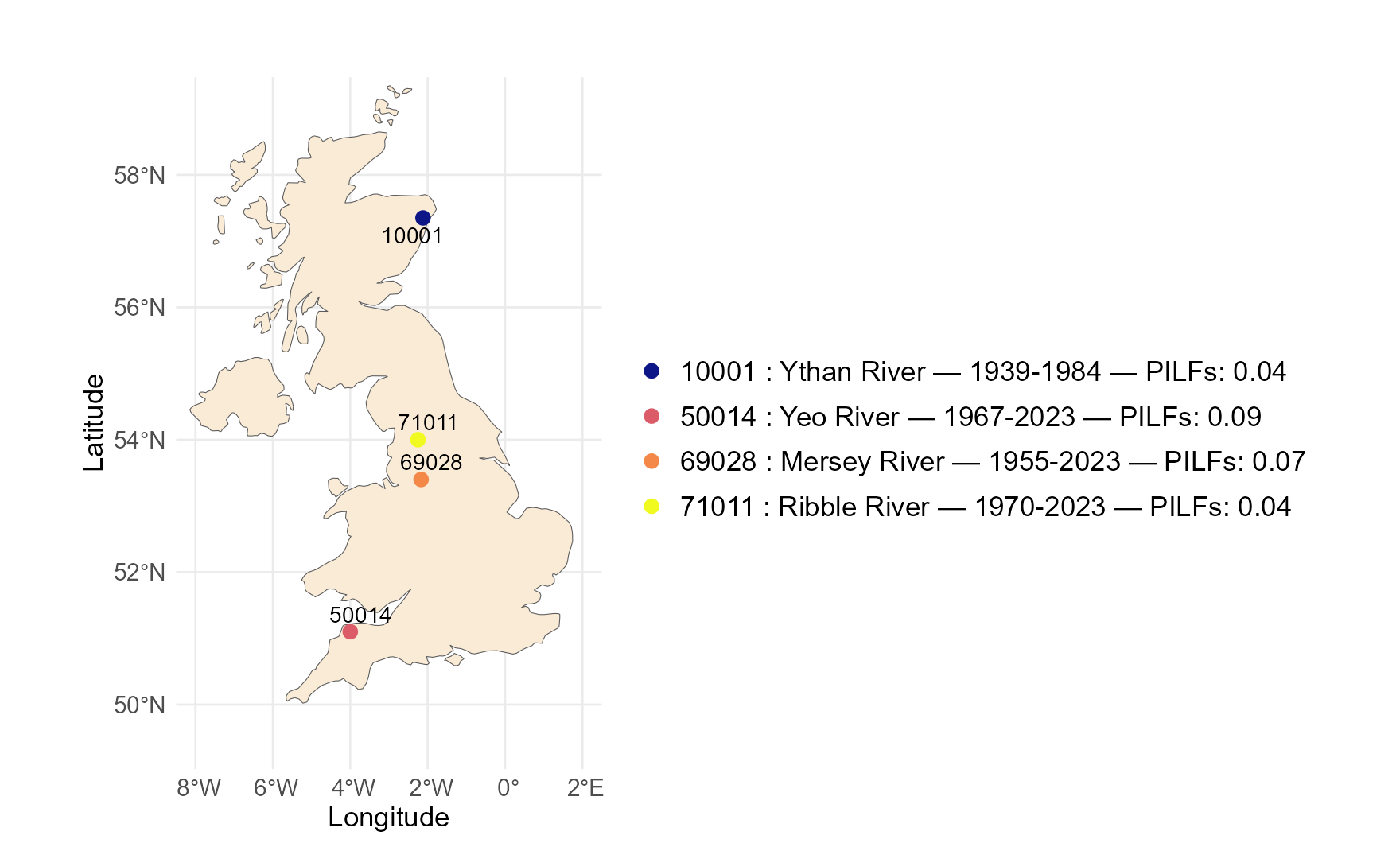}

  \caption{Location of the four gauging stations within the UK. The legend indicates their respective NRFA number, the river on which the catchment is located, the time range of the data considered, and the proportion of PILFs in the dataset. \label{fig:map_uk}}
\end{figure}


In addition to experiments on simulated datasets, we evaluate the performance of our new estimator for the GEV parameters using real-world data and provide insights into its practical application. Specifically, we analyze peak flow datasets from the United Kingdom (UK), provided by the National River Flow Archive \citep{nrfa}. These data consist of annual maximum river flows recorded across the UK. As noted in the introduction, they contain left-tail outliers, called Potentially Influential Low Floods (PILFs). PILFs are often considered to arise from a different generative process than the rest of the annual maxima and are typically removed to focus the analysis on more extreme observations. For a comprehensive discussion of PILFs, we refer the interested reader to \cite{england2018guidelines}. Finding those influential points is a practical challenge in which our estimator provides a valuable alternative. A common approach to detecting PILFs in a series of annual peak flows is to use the Grubbs–Beck test \citep{grubbs1972extension} to assess whether the lowest observation comes from the same distribution as the others. This test can be applied iteratively to detect several PILFs \citep{cohn2013generalized,lamontagne2016robust}.
Our analysis shows that the MDPD estimator naturally downweights the influence of PILFs, providing a robust alternative for modeling peak flow time series without the need to explicitly remove these observations.

\begin{figure}[ht!]
\begin{minipage}{0.8\textwidth}
    \centering
\includegraphics[width=.45\textwidth]{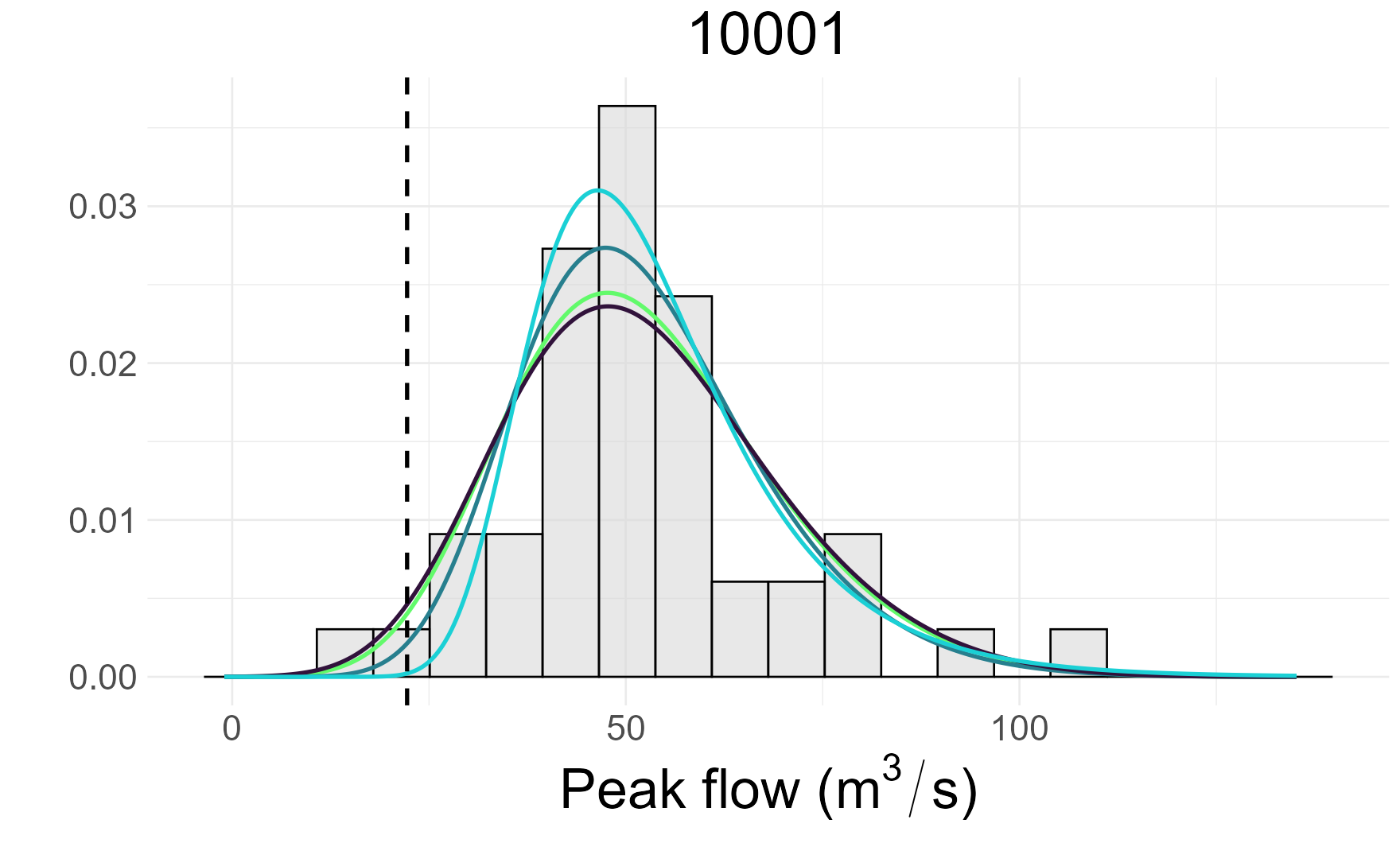}
\hspace{0.6cm}
\includegraphics[width=.45\textwidth]{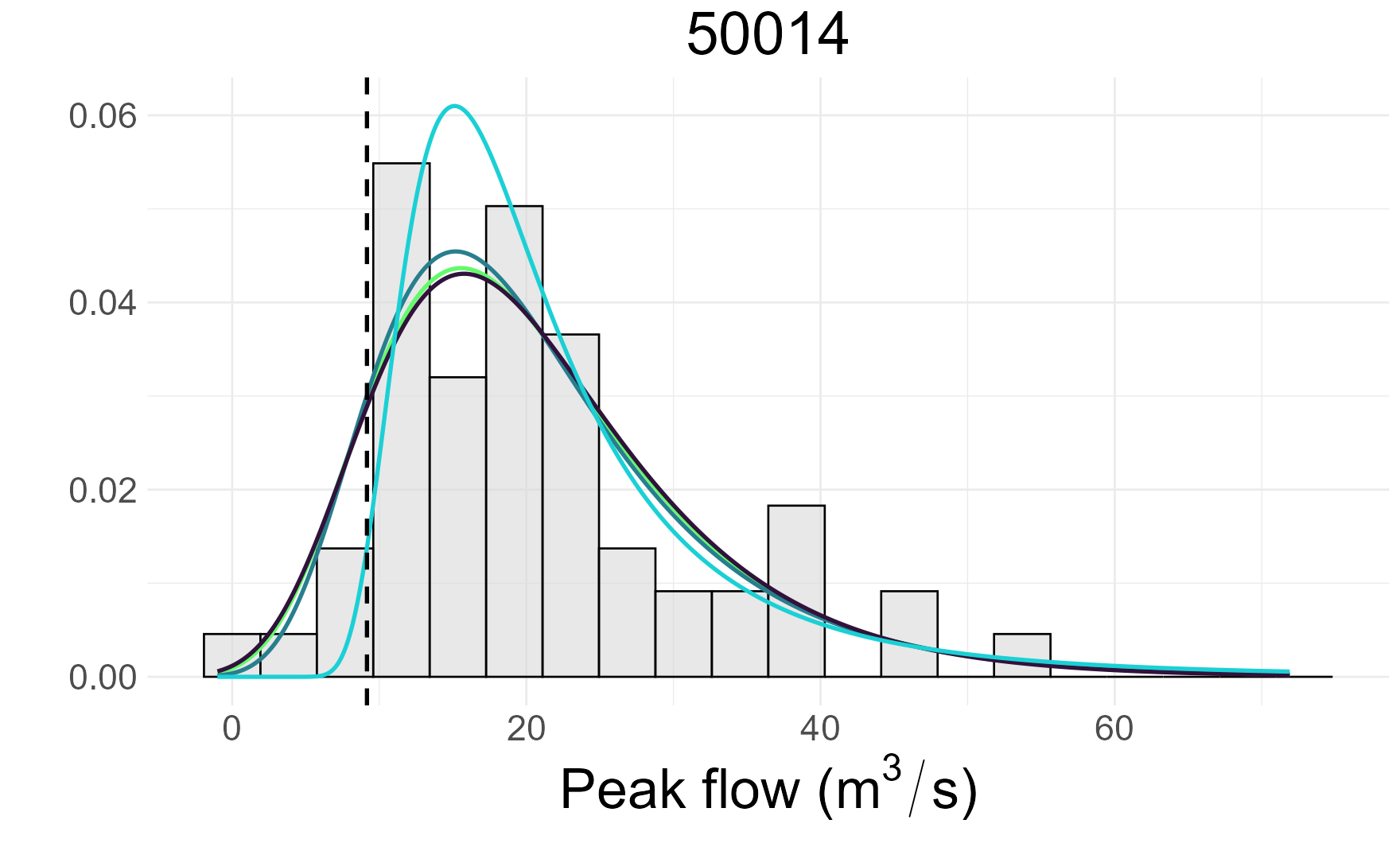}

\vspace{0.6cm} 

\includegraphics[width=.45\textwidth]{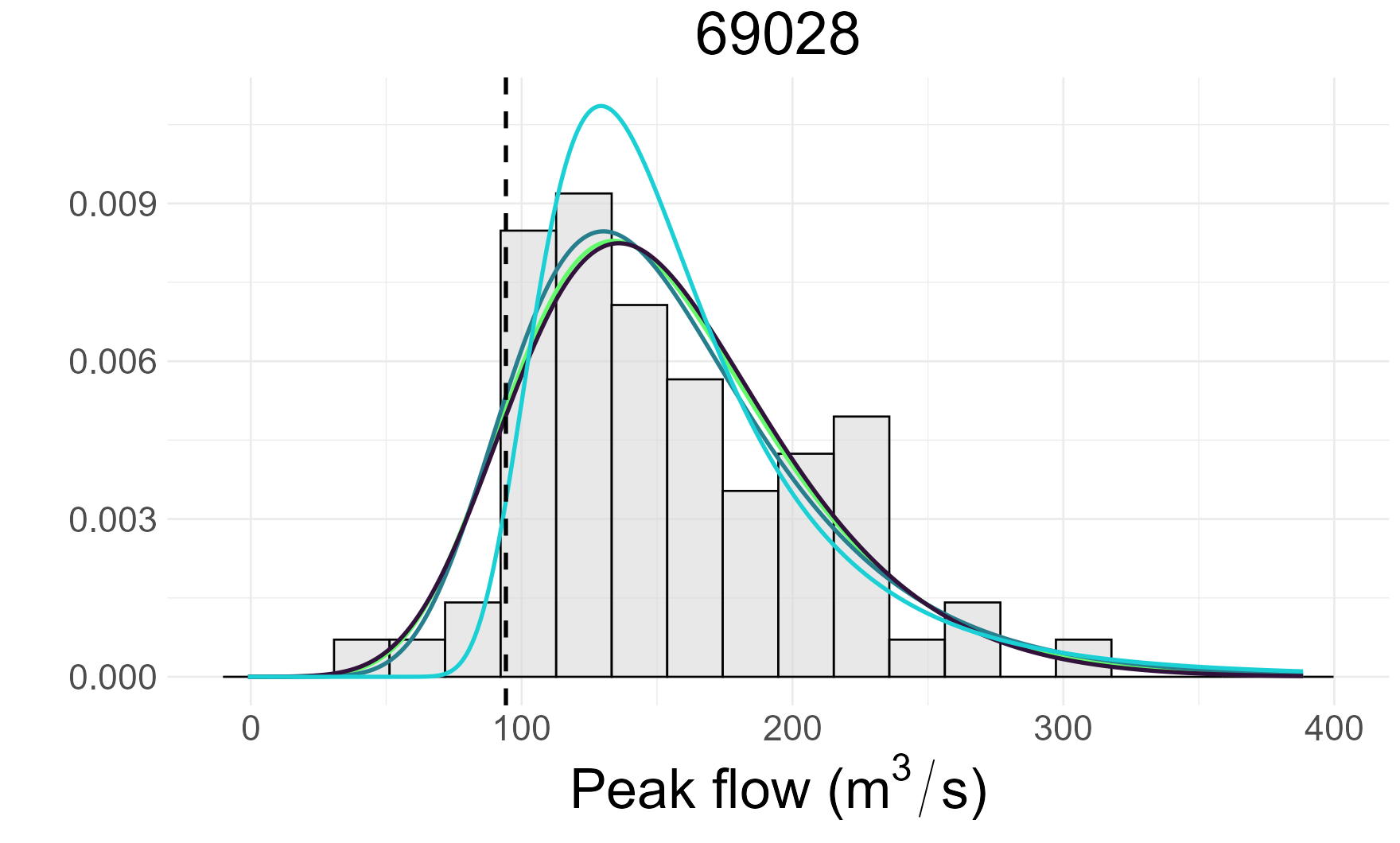}
\hspace{0.6cm}
\includegraphics[width=.45\textwidth]{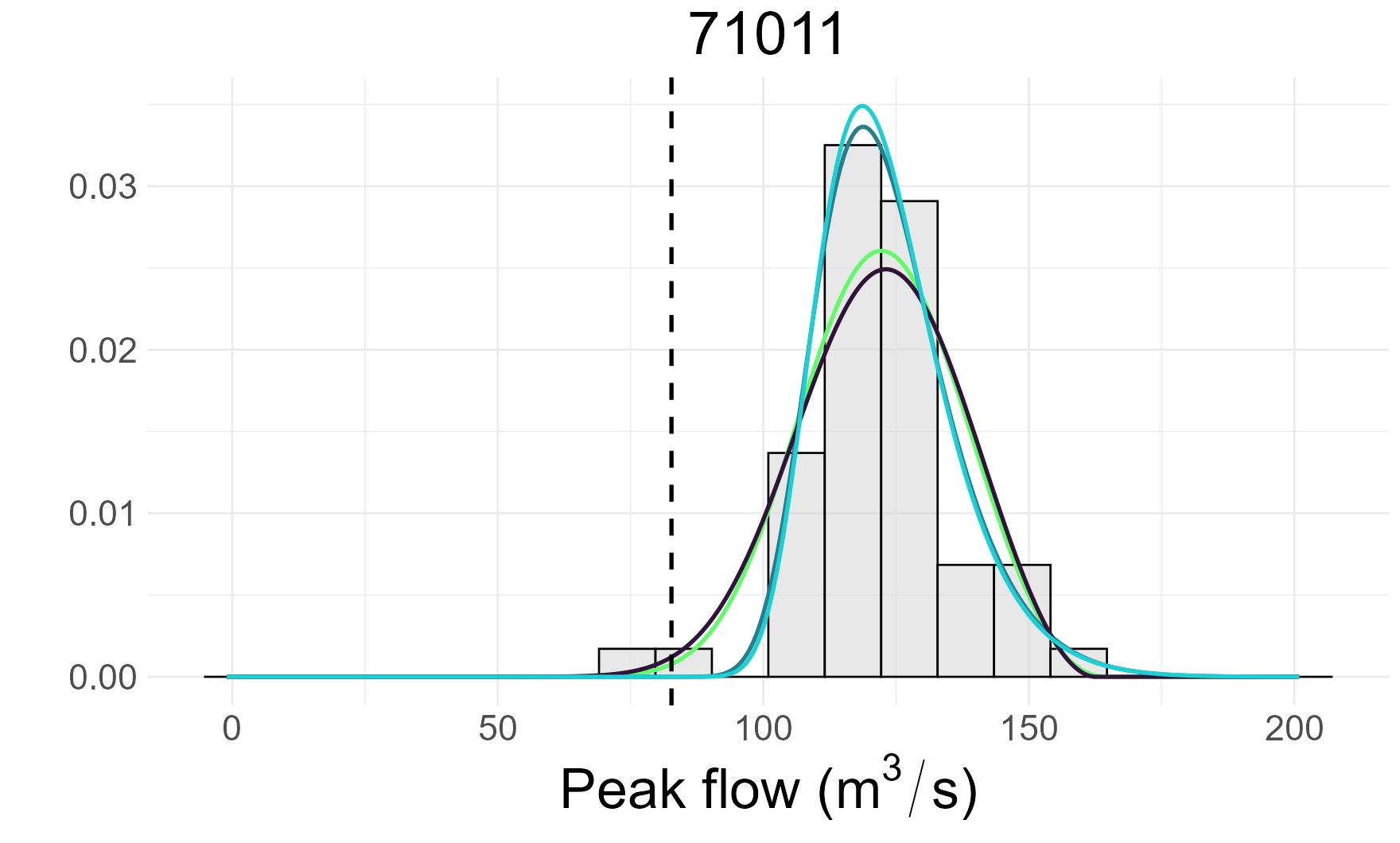}
\end{minipage}
\hfill
\begin{minipage}{0.18\textwidth}
\includegraphics[width=\textwidth]{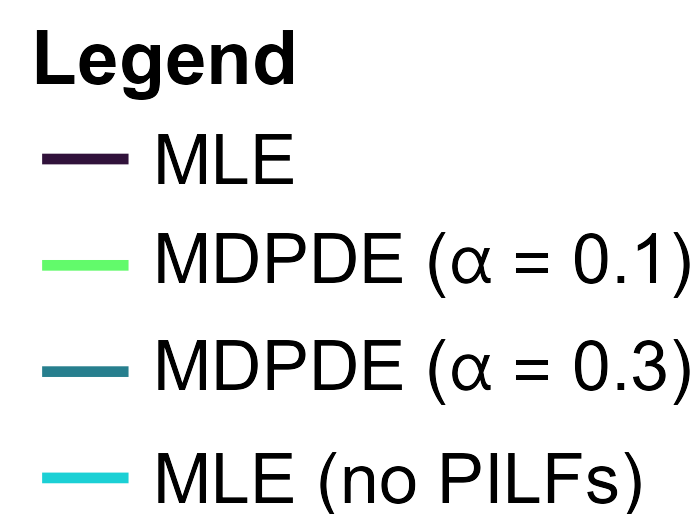}
\end{minipage}
  \caption{Histograms of annual peak flows at the five gauging stations: 
  The curves correspond to the fitted GEV densities from the four different models. The dotted vertical grey lines indicate the largest PILF, \ie, the largest observation excluded in the ML model without PILFs.\label{fig:appli_density}}
\end{figure}

Four gauging stations across the UK are considered in the analysis.
Figure~\ref{fig:map_uk}  shows their respective locations, the time range of the data, and the proportion of PILFs. The PILFs were identified manually by examining the $p$-values resulting from the Grubbs–Beck test procedure, implemented via the \texttt{MGBT} function from the \texttt{MGBT} package \citep{MGBTpackage}. Specifically, observations associated with $p$-values preceding a noticeable gap in the ordered $p$-value magnitudes were flagged as PILFs. Sample sizes for the stations range from $n=46$ to $n=69$ peak flow values, and the proportion of identified PILFs varies from 0.04 to 0.09. These values are consistent with the simulation settings considered in Sections~\ref{sec:n50e01} and~\ref{sec:n50e005} of the Appendix, where the sample size is fixed at $n=50$ and the contamination proportion $\varepsilon$ is set to 0.05 or 0.1. In those simulations, the robustness of the MDPD estimator has already been empirically demonstrated.

\begin{figure}[h!]
\begin{minipage}{0.8\textwidth}
    \centering
\includegraphics[width=.45\textwidth]{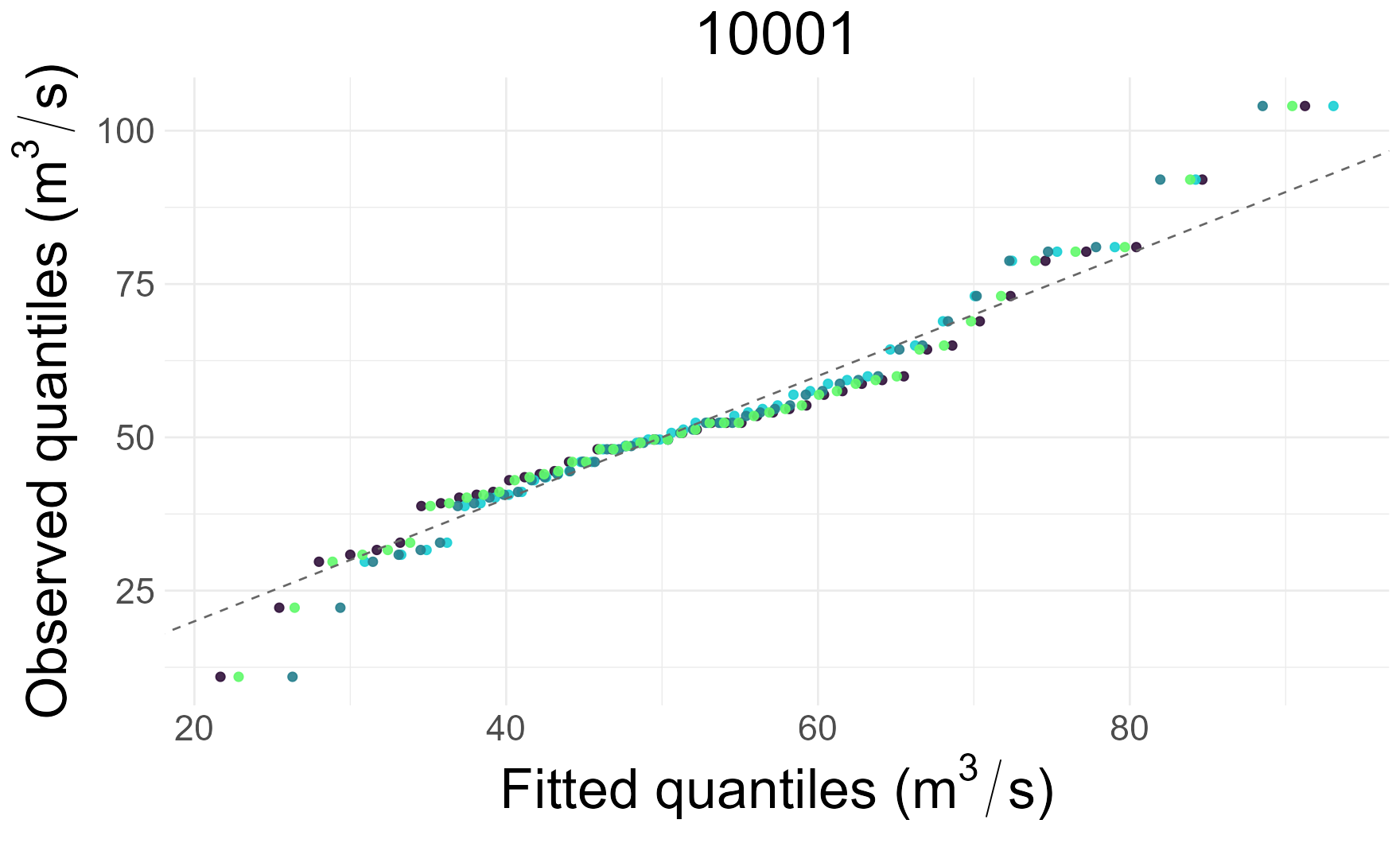}
\hspace{0.6cm}
\includegraphics[width=.45\textwidth]{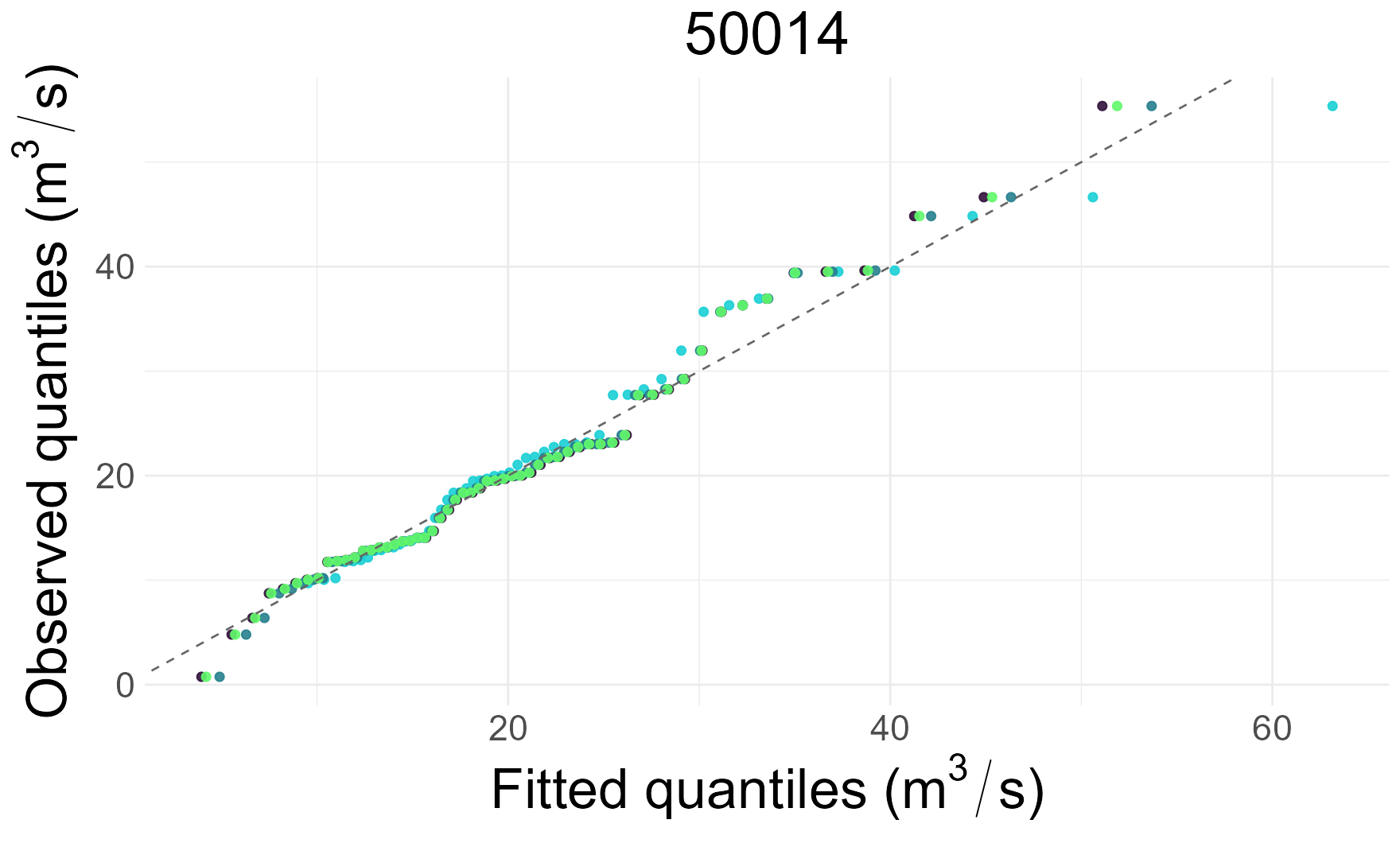}

\vspace{0.6cm} 

\includegraphics[width=.45\textwidth]{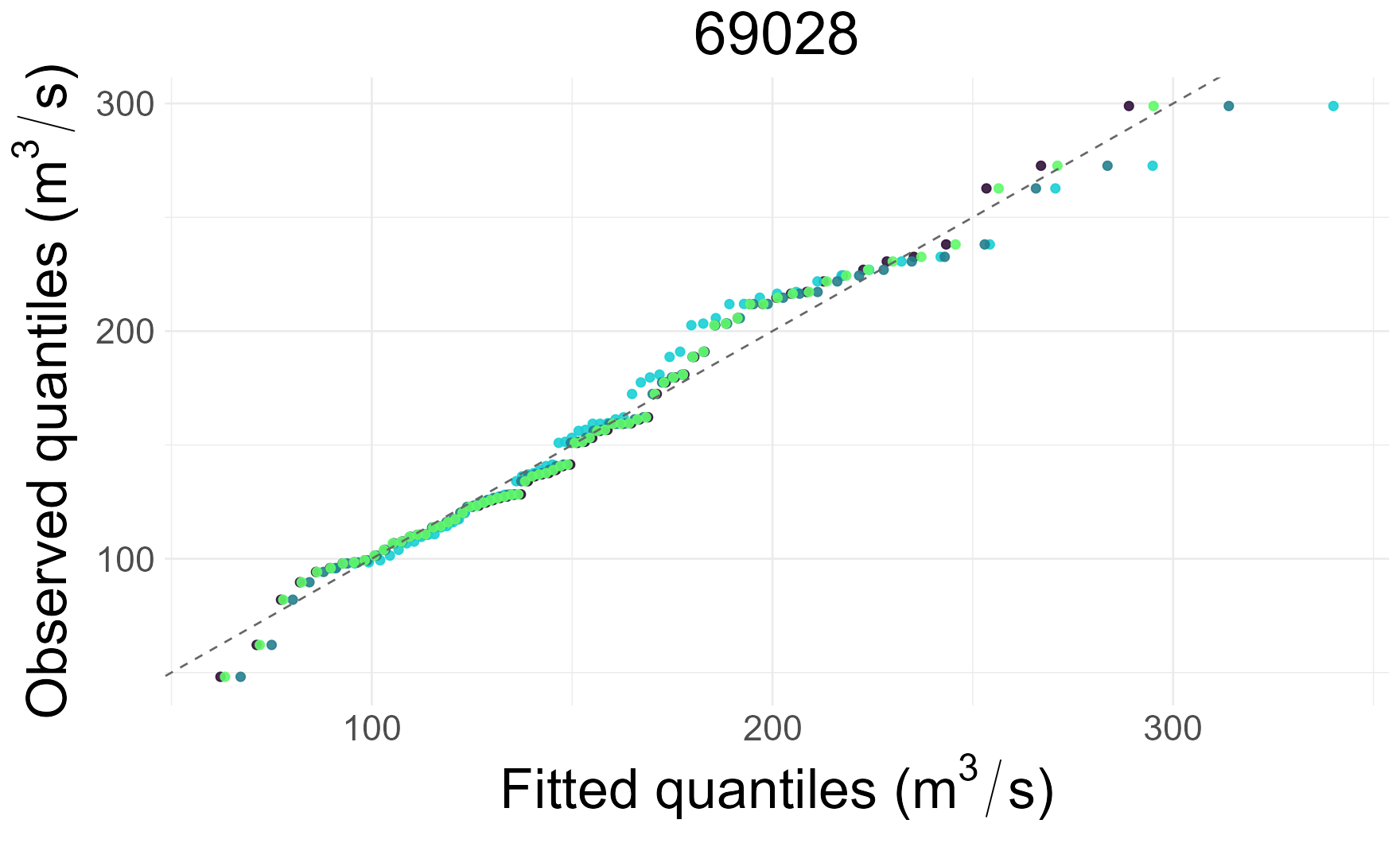}
\hspace{0.6cm}
\includegraphics[width=.45\textwidth]{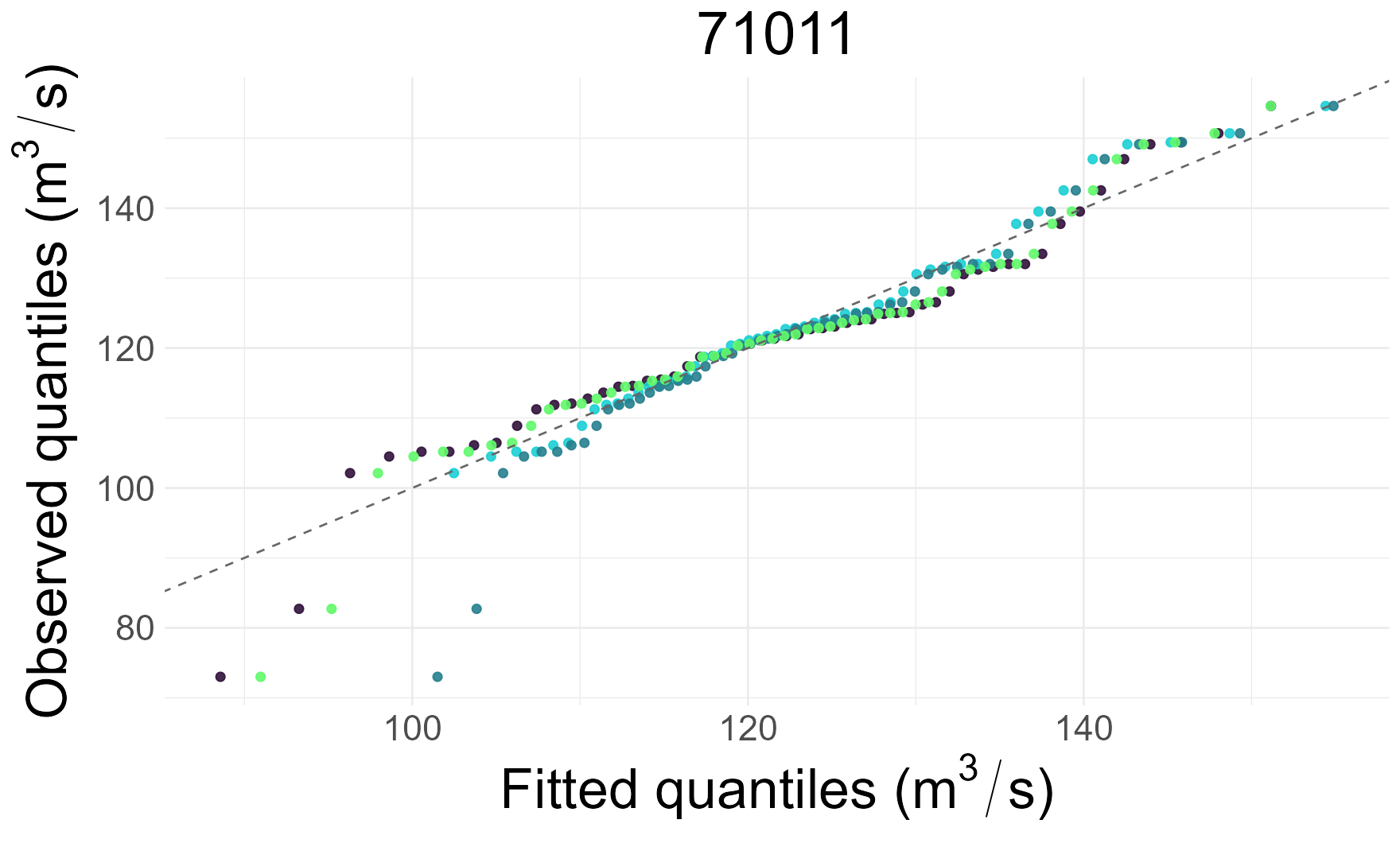}
\end{minipage}
\hfill
\begin{minipage}{0.18\textwidth}
\includegraphics[width=\textwidth]{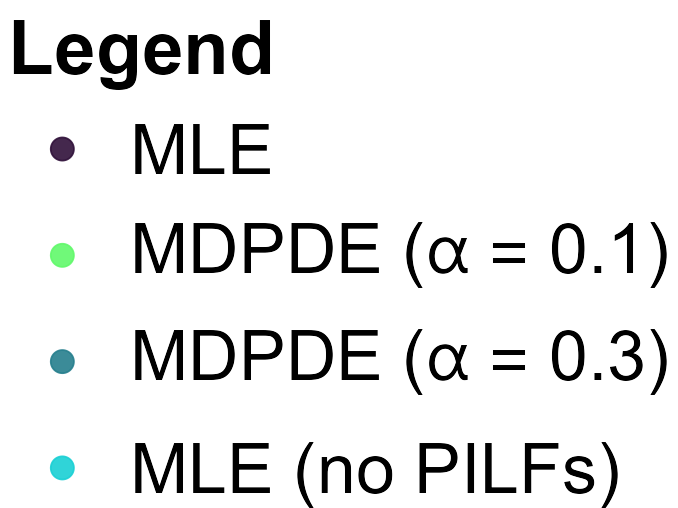}
\end{minipage}
  \caption{QQ-plots comparing observed quantiles (y-axis) to fitted quantiles (x-axis) at the four gauging stations
  for the four different models.\label{fig:appli_qqplot}}
\end{figure}

\begin{figure}[h!]
\begin{minipage}{0.7\textwidth}
    \centering
\includegraphics[width=.45\textwidth]{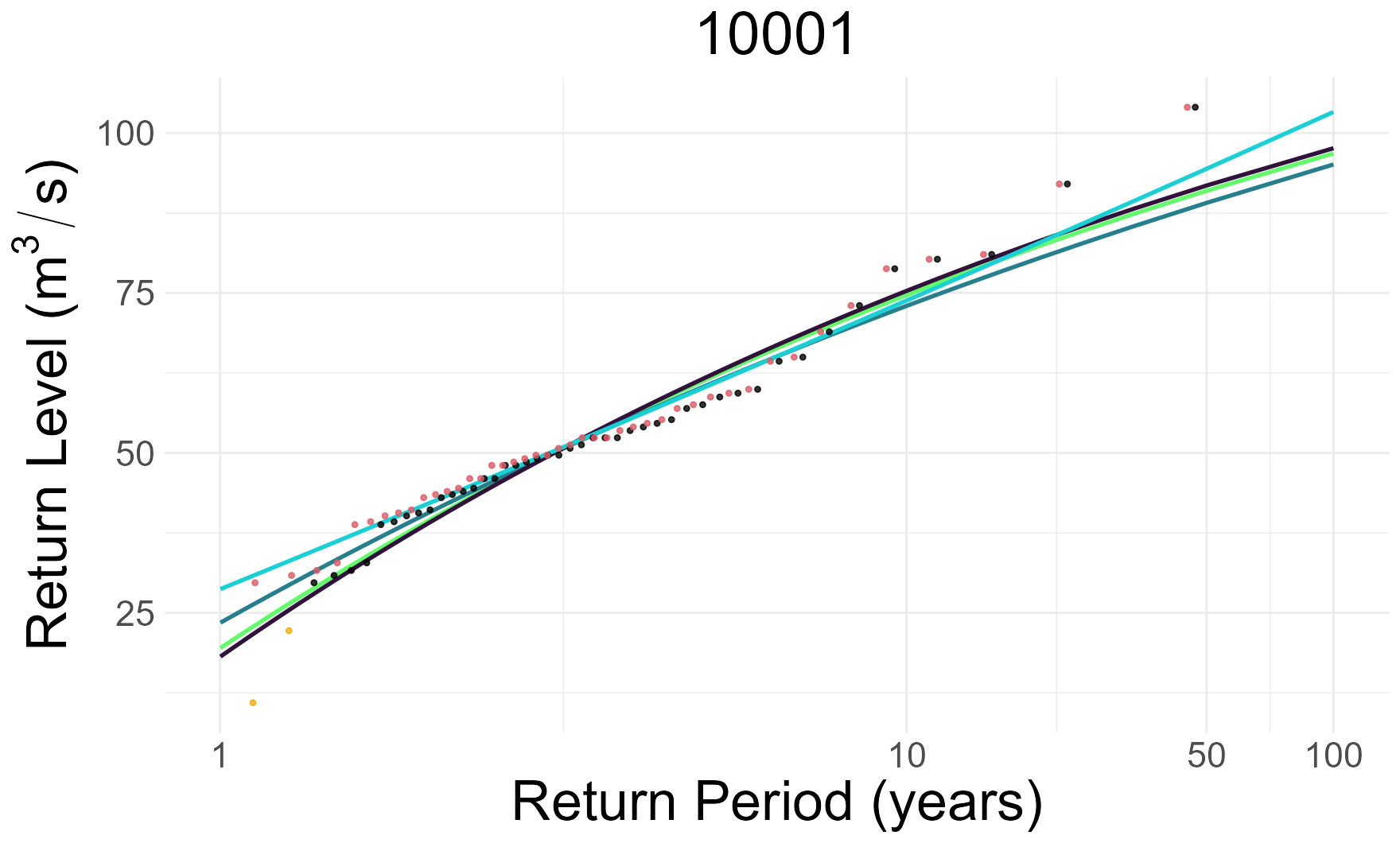}
\hspace{0.6cm}
\includegraphics[width=.45\textwidth]{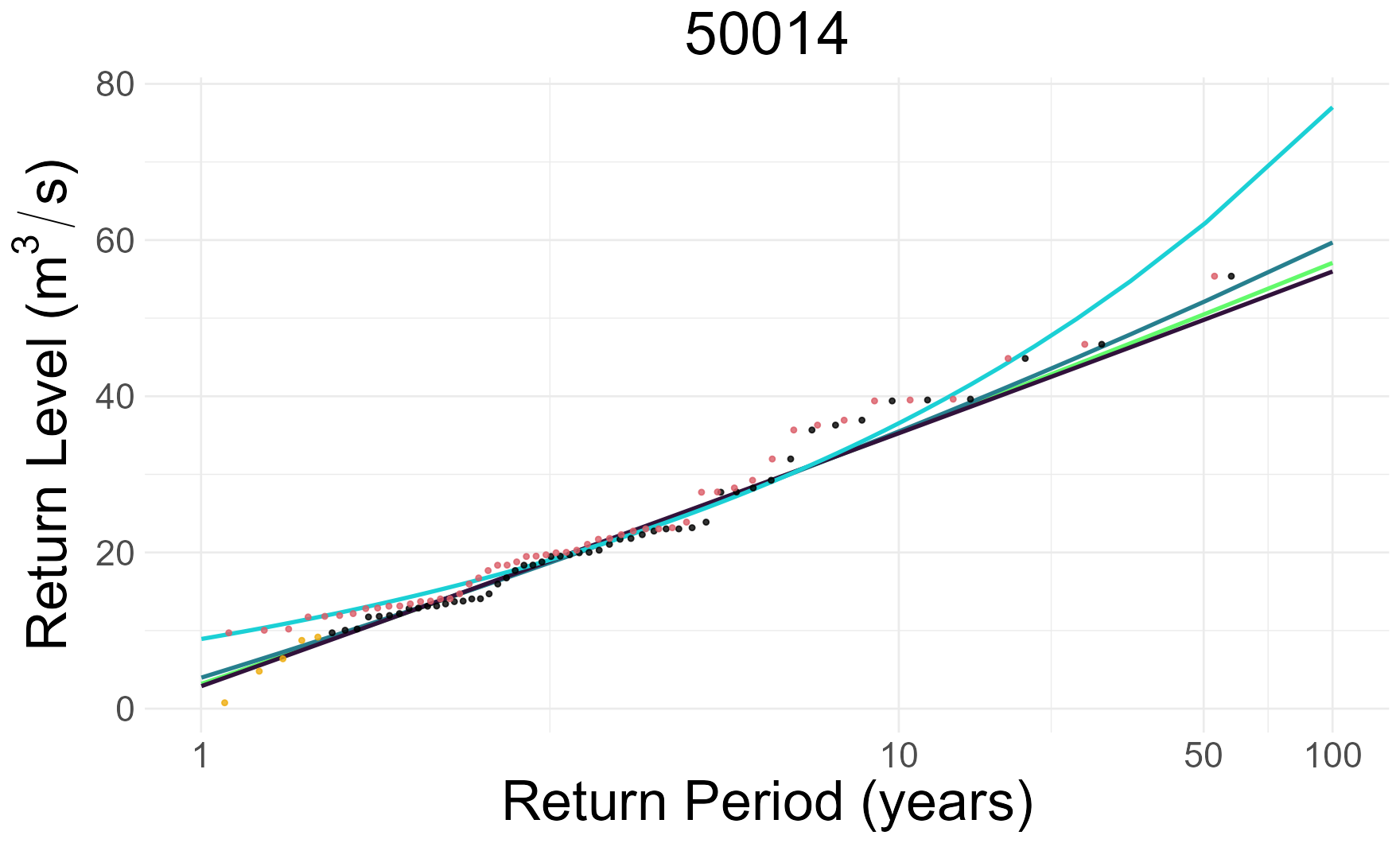}

\vspace{0.6cm} 

\includegraphics[width=.45\textwidth]{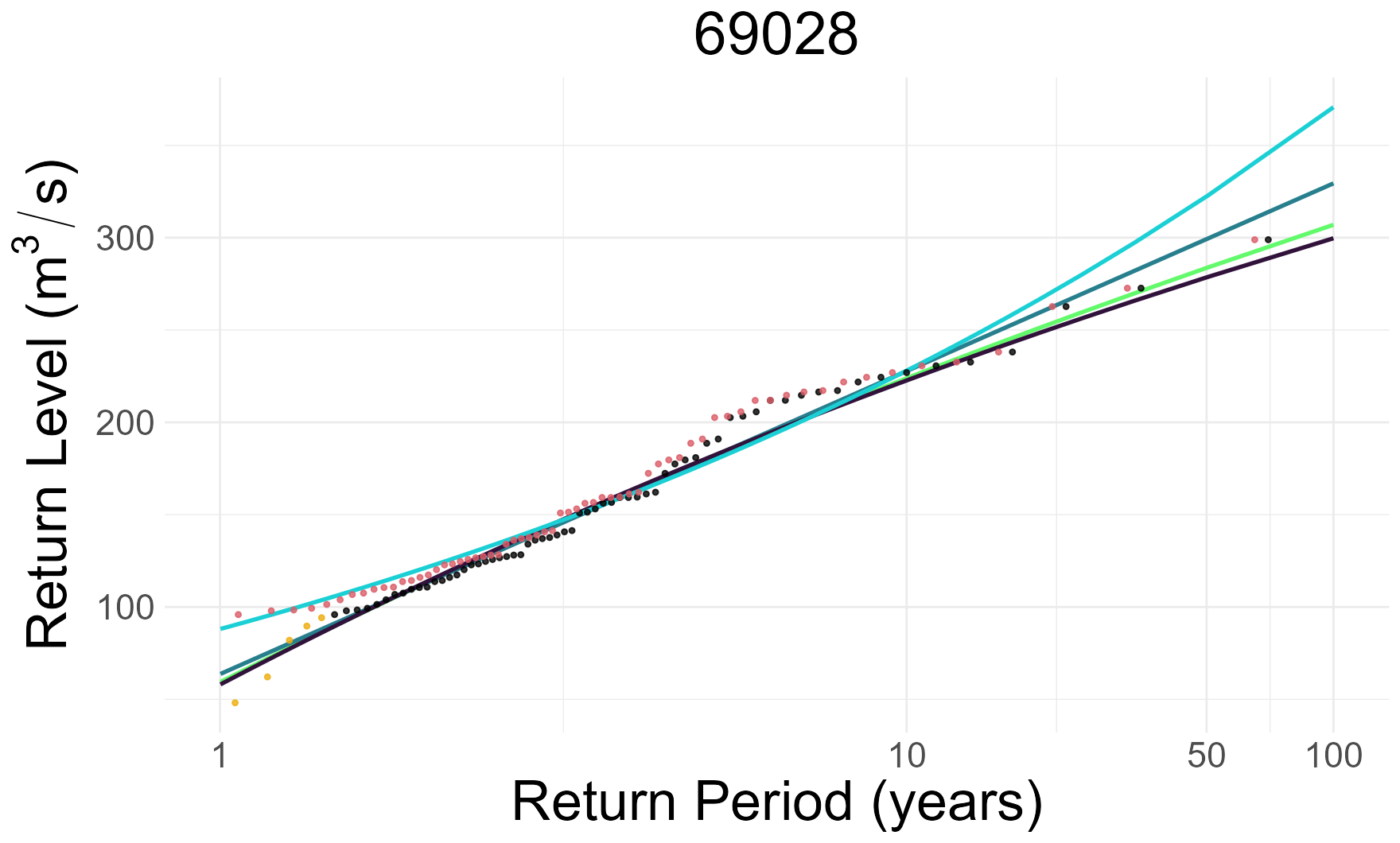}
\hspace{0.6cm}
\includegraphics[width=.45\textwidth]{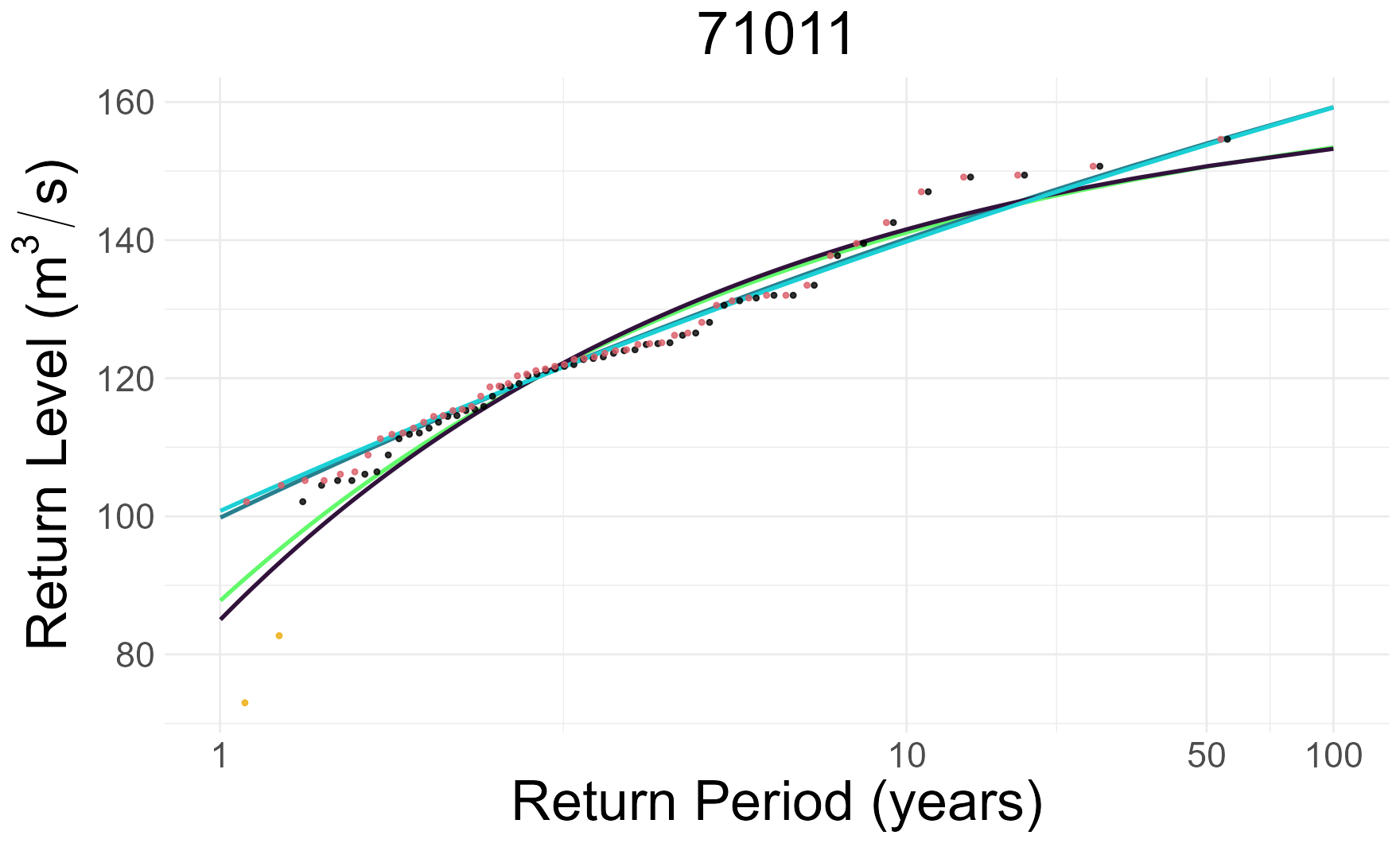}
\end{minipage}
\hfill
\begin{minipage}{0.28\textwidth}
\includegraphics[width=\textwidth]{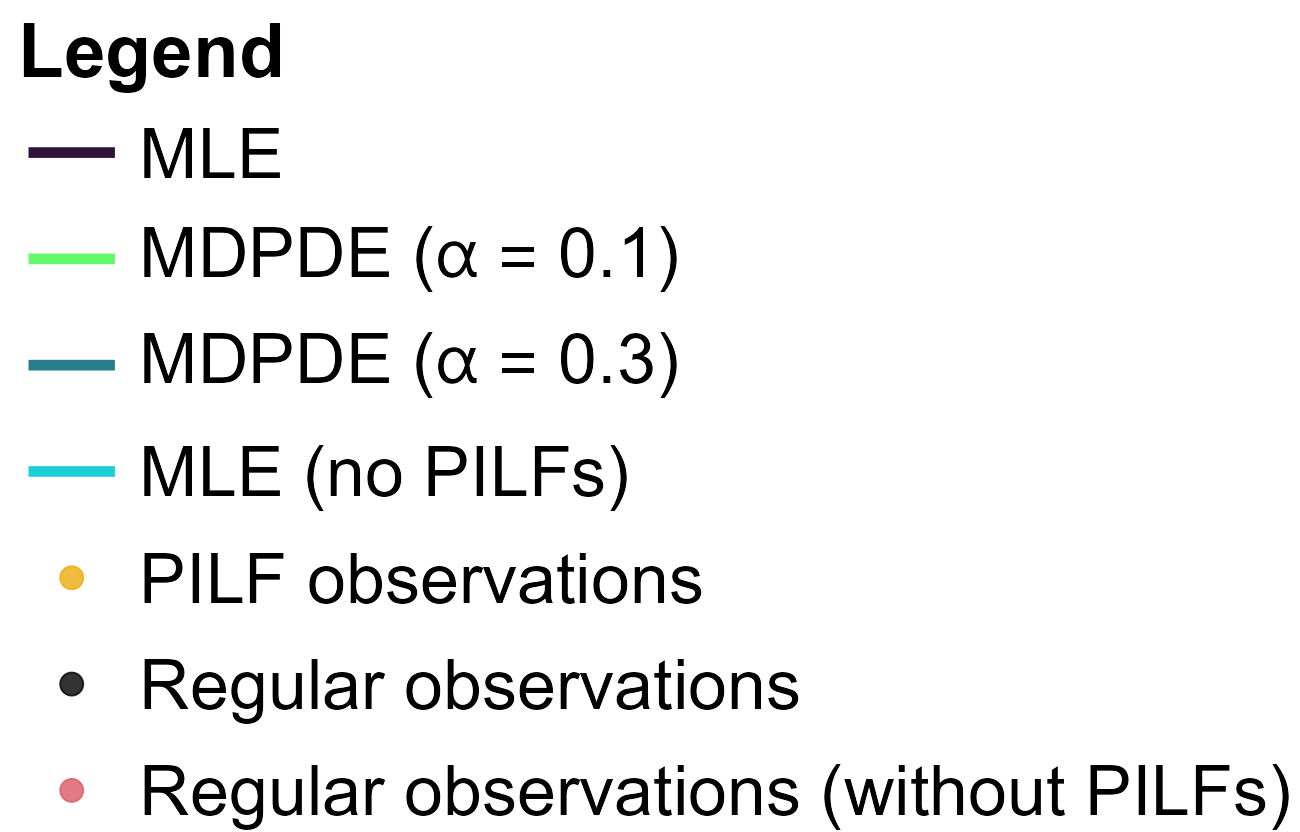}
\end{minipage}
  \caption{Return level curves at the four gauging stations 
  . The return period axis is plotted on the scale $-\log(-\log(1 - 1/T))$, where $T \in ]1, 100]$ denotes the return period associated with each return level.\label{fig:appli_return_curve}}
\end{figure}

We consider three estimators for the parameters of the GEV distribution: the ML estimator applied to the full dataset (including PILFs), and two MDPD estimators with tuning parameters $\alpha = 0.1$ and $\alpha = 0.3$. The ML estimator computed on the dataset with PILFs removed plays the role of the reference density. The four fitted GEV densities are shown in Figure~\ref{fig:appli_density}, with corresponding QQ-plots presented in Figure~\ref{fig:appli_qqplot}. The return level curves for each estimator are provided in Figure~\ref{fig:appli_return_curve}.
Tables~\ref{table:real_wd}, \ref{table:real_shape}, and~\ref{table:real_rl} present, respectively, the Wasserstein distances to the reference distribution, the estimated shape parameters, and the estimated 100-year return levels, which also serve as metrics for comparing the methods.
Additional estimates, such as those for the location and scale parameters, are provided in Section~\ref{sec:real_appendix} of the Appendix. Note that estimates of the standard deviations of the estimators, reported in Tables~\ref{table:real_shape}, \ref{tab:loc_parameter}, and \ref{tab:scale_parameter} for each model, are derived from the asymptotic variance formula proposed in Theorem~\ref{th:asympt_norm} and can be used to construct confidence intervals for the parameters and the return levels.

\begin{table}[ht!]
\caption{Wasserstein distances at each station between the reference density, \ie, the density fitted by MLE on data with PILFs removed, and the three densities fitted on the full dataset (including PILFs) using ML, MDPD with $\alpha = 0.1$, and MDPD with $\alpha = 0.3$.\label{table:real_wd}}

\vspace{0.2cm}

\begin{center}
\begin{small}
\begin{tabular}{c|ccc}
     &  MLE & MDPDE ($\alpha = 0.1)$& MDPDE ($\alpha = 0.3)$  \\ \hline
    10001 & 2.30 & 1.96 & 1.18 \\
    50014 & 1.76 & 1.70 & 1.51 \\
    69028 & 7.40 & 7.07 & 6.03 \\
    71011 & 2.96 & 2.46 & 0.27 \\
\end{tabular}
\end{small}
\end{center}
\end{table}

First, it is well established in the hydrology literature \citep[][]{robson1999statistical} that the generalized logistic distribution is recommended for modeling annual peak flow maxima in the UK \citep[and the log-Pearson Type III in the USA,][]{england2018guidelines}. Nevertheless, the GEV distribution demonstrates a satisfactory fit to the peak flow data in our case, as shown in Figures~\ref{fig:appli_density} and~\ref{fig:appli_qqplot}, with particularly strong performance in capturing the behavior of the extreme observations. More complex models, such as nonstationary GEV models \cite{prosdocimi2015detection}, could potentially improve the fit, but this lies beyond the scope of the present study.

Following the recommendations in \cite{england2018guidelines} to remove PILFs before inference, especially when estimating return levels, we consider the outcome from the ML procedure on the dataset with the PILFs removed as the "reference method". Then, we compare the MDPD estimators against the ML estimator computed on the whole dataset, according to the reference method.

Regarding overall distributional performance, the MDPD-based fits are consistently closer to the reference distribution, in terms of Wasserstein distance (Table~\ref{table:real_wd}), than the ML-based fit on the full dataset. This similarity is also visually evident in the return level plots (Figure~\ref{fig:appli_return_curve}). For instance, at station 71011, the return curve obtained with $\alpha = 0.3$ is almost indistinguishable from the reference curve. This can be explained by the large difference between the PILFs and the rest of the data, which the MDPD estimator with $\alpha = 0.3$ effectively downweights, nearly full censoring, as the reference method. In contrast, the classical ML estimator weights all observations equally, making it more sensitive to PILFs.

In the tail region, we take the shape parameter as a natural metric for comparing estimators. MDPD estimates of the shape parameter are systematically closer to the reference values than those from ML (Table~\ref{table:real_shape}), with closeness increasing as $\alpha$ grows. As expected from the downweighting of PILFs, MDPD yields larger shape estimates than ML. Importantly, positive shape parameters are generally expected for peak flow data: at station 69028, the MDPD estimate with $\alpha = 0.3$ changes the sign from negative, for the ML, to positive, resulting in a non-negligible increase, about $30 m^3/s$, in the 100-year return level (Table~\ref{table:real_rl}). For the other stations, return level estimates from MDPD and ML are broadly similar, with slight advantages for MDPD at stations 50014 and 71011, and for ML elsewhere.

A practical advantage of MDPD over the standard “ML-without-PILFs” approach is that it avoids hard censoring and retains all observations. The traditional Grubbs–Beck PILF detection can remove up to half the data, despite the already limited sample size from using only annual maxima. Such heavy censoring reduces reliability for short return periods and makes the return curve accurate only for very long periods. By contrast, MDPD offers a good compromise between the two ML procedures (on the whole dataset and the PILFs-filtered dataset): it reduces the influence of PILFs without discarding them entirely and without preprocessing steps.

\begin{table}[hb!]
\caption{Estimated shape parameter $\xi$ and standard deviation at each station using the four different models.\label{table:real_shape}}
\vspace{0.2cm}

\begin{center}
\begin{small}
\begin{tabular}{c|cccc}
     & MLE (without PILFs) &  MLE & MDPDE ($\alpha = 0.1)$& MDPDE ($\alpha = 0.3)$  \\ \hline
    10001 & 0.02 (0.11) & -0.15 (0.09) & -0.14 (0.09) & -0.11 (0.11) \\
    50014 & 0.30 (0.14) & 0.01 (0.09) & 0.03 (0.10) & 0.08 (0.12) \\
    69028 & $0.19$ (0.18) &$-0.08$ (0.08) &$-0.06$ (0.08)&$0.01$ (0.16)\\
    71011 & -0.07 (0.09)& -0.35 (0.07)& -0.32 (0.07)& -0.10 (0.10)\\
\end{tabular}
\end{small}
\end{center}
\end{table}

\begin{table}[ht!]
\caption{Estimated 100-years return level (in $m^3/s$) at each station using the four different models.\label{table:real_rl}}
\vspace{0.2cm}

\begin{center}
\begin{small}
\begin{tabular}{c|cccc}
     & MLE (without PILFs) &  MLE & MDPDE ($\alpha = 0.1)$& MDPDE ($\alpha = 0.3)$  \\ \hline
     10001 & $103.28$ &$97.62$ &$96.81$ &$95.08$ \\
     50014 & $79.54$ &$56.52$ &$57.78$ &$60.84$ \\
     52011 & $13.38$ &$13.28$ &$13.26$ &$13.24$ \\
    69028 & $381.75$ &$301.71$ &$309.76$ &$334.37$\\
    71011 & $159.57$ &$153.35$ &$153.56$ &$159.41$
    

\end{tabular}
\end{small}
\end{center}
\end{table}

%% file: appendix.tex
\section{Additional background}
\subsection{Score and information  function of the GEV distribution}\label{appendix:score_and_inf}
We recall that the density $f(\cdot;\mu,\sigma,\xi)$ of a GEV distribution with parameters $\mu,\sigma,\xi$ is given by
\begin{equation}\label{appendix_eq:gev_density}
    f(x;\mu,\sigma,\xi) = \frac{1}{\sigma}\Big(1+\xi\Big(\frac{x-\mu}{\sigma}\Big)\Big)^{-(\xi+1)/\xi}\exp\Big(-\Big(1+\xi\Big(\frac{x-\mu}{\sigma}\Big)\Big)^{-1/\xi}\Big)\1\{x \in D_{\mu,\sigma,\xi}\},
\end{equation}
where the support of the distribution $D_{\mu,\sigma,\xi}$ is defined as
  \begin{equation*}
    D_{\mu,\sigma,\xi} =
    \begin{cases*}
      [\mu - \sigma/\xi,+\infty[, & if $\xi >0$; \\
      \mb{R}, & if $\xi =0$; \\
      ]-\infty,\mu - \sigma/\xi],   & if $\xi <0$.
    \end{cases*}
    \end{equation*}
The score function is defined as
\begin{equation*}
    S(x;\mu,\sigma,\xi) = \frac{\partial \log f(x;\mu,\sigma,\xi)}{\partial (\mu,\sigma,\xi)},
\end{equation*}
and can be written as
\begin{equation*}
    S(x;\mu,\sigma,\xi) = \Big(S_\mu(x;\mu,\sigma,\xi),S_\sigma(x;\mu,\sigma,\xi),S_\xi(x;\mu,\sigma,\xi)\Big)^\top.
\end{equation*}
The information function is defined as
\begin{equation*}
    i(x;\mu,\sigma,\xi) = -\frac{\partial^2 \log f(x;\mu,\sigma,\xi)}{\partial (\mu,\sigma,\xi)^2},
\end{equation*}
and can be written as
\begin{equation*}
    i(x;\mu,\sigma,\xi) = \begin{pmatrix}
i_{\mu\mu}(x;\mu,\sigma,\xi) & i_{\mu\sigma}(x;\mu,\sigma,\xi) & i_{\mu\xi}(x;\mu,\sigma,\xi)\\
i_{\mu\sigma}(x;\mu,\sigma,\xi) & i_{\sigma\sigma}(x;\mu,\sigma,\xi) & i_{\sigma\xi}(x;\mu,\sigma,\xi) \\
i_{\mu\xi}(x;\mu,\sigma,\xi) & i_{\sigma\xi}(x;\mu,\sigma,\xi) & i_{\xi\xi}(x;\mu,\sigma,\xi)
\end{pmatrix}.
\end{equation*}
Thus, using Equation~\eqref{appendix_eq:gev_density}, explicit formula for the score and information functions can be deduced
\begin{equation*}
    S_\mu(x;\mu,\sigma,\xi) = \frac{\partial \log f(x;\mu,\sigma,\xi)}{\partial \mu}= \Big(\frac{1}{\sigma}\Big)\Big(1+\xi\Big(\frac{x-\mu}{\sigma}\Big)\Big)^{-1}\Big[\xi+1-\Big(1+\xi\Big(\frac{x-\mu}{\sigma}\Big)\Big)^{-1/\xi}\Big];
\end{equation*}
\begin{equation*}
    S_\sigma(x;\mu,\sigma,\xi) =\frac{\partial \log f(x;\mu,\sigma,\xi)}{\partial \sigma}= -\frac{1}{\sigma} + \Big(\frac{x-\mu}{\sigma^2}\Big)\Big(1+\xi\Big(\frac{x-\mu}{\sigma}\Big)\Big)^{-1}\Big[(\xi+1)-\Big(1+\xi\Big(\frac{x-\mu}{\sigma}\Big)\Big)^{-1/\xi}\Big];
\end{equation*}
\begin{align*}
    S_\xi(x;\mu,\sigma,\xi) = \frac{\partial \log f(x;\mu,\sigma,\xi)}{\partial \xi}&= \frac{1}{\xi^2}\log\Big(1+\xi\Big(\frac{x-\mu}{\sigma}\Big)\Big)\Big[1-\Big(1+\xi\Big(\frac{x-\mu}{\sigma}\Big)\Big)^{-1/\xi}\Big]\\&+\frac{1}{\xi}\Big(1+\xi\Big(\frac{x-\mu}{\sigma}\Big)\Big)^{-1}\Big[\xi+1-\Big(1+\xi\Big(\frac{x-\mu}{\sigma}\Big)\Big)^{-1/\xi}\Big];
\end{align*}
\begin{equation*}
    i_{\mu\mu}(x;\mu,\sigma,\xi) =-\frac{\partial^2 \log f(x;\mu,\sigma,\xi)}{\partial \mu^2} =  \frac{(\xi+1)}{\sigma^2}\Big(1+\xi\Big(\frac{x-\mu}{\sigma}\Big)\Big)^{-2}\Big[\Big(1+\xi\Big(\frac{x-\mu}{\sigma}\Big)\Big)^{-1/\xi}-\xi\Big]
\end{equation*}
\begin{equation*}
    i_{\mu\sigma}(x;\mu,\sigma,\xi) =-\frac{\partial^2 \log f(x;\mu,\sigma,\xi)}{\partial \mu\partial\sigma}= \frac{1}{\sigma^2}\Big(1+\xi\Big(\frac{x-\mu}{\sigma}\Big)\Big)^{-2}\Big[\xi+1 +\Big(\frac{x-\mu}{\sigma}-1\Big)\Big(1+\xi\Big(\frac{x-\mu}{\sigma}\Big)\Big)^{-1/\xi}\Big];
\end{equation*}
\begin{align*}
    i_{\mu\xi}(x;\mu,\sigma,\xi) &= -\frac{\partial^2 \log f(x;\mu,\sigma,\xi)}{\partial \mu\partial\xi} =\frac{1}{\sigma}\Big(1+\xi\Big(\frac{x-\mu}{\sigma}\Big)\Big)^{-2}\Big(\frac{x-\mu}{\sigma}-1\Big) \\ &+\Big(\frac{1}{\sigma\xi^2}\Big)\Big(1+\xi\Big(\frac{x-\mu}{\sigma}\Big)\Big)^{-1 / \xi-2}\Big[\Big(1+\xi\Big(\frac{x-\mu}{\sigma}\Big)\Big)\log \Big(1+\xi\Big(\frac{x-\mu}{\sigma}\Big)\Big) - \xi(\xi+1) \Big(\frac{x-\mu}{\sigma}\Big)\Big];
\end{align*}
\begin{align*}
    i_{\sigma\sigma}(x;\mu,\sigma,\xi) =-\frac{\partial^2 \log f(x;\mu,\sigma,\xi)}{\partial \sigma^2} &= \frac{1}{\sigma^2}\Big(\frac{x-\mu}{\sigma}\Big)\Big(1+\xi\Big(\frac{x-\mu}{\sigma}\Big)\Big)^{-1 / \xi-2}\Big[(1-\xi)\Big(\frac{x-\mu}{\sigma}\Big)-2 \Big] \\
    &+\frac{1}{\sigma^2}\Big(\frac{x-\mu}{\sigma}\Big)\Big(1+\xi\Big(\frac{x-\mu}{\sigma}\Big)\Big)^{-2}\Big[\xi\Big(\frac{x-\mu}{\sigma}\Big)^2+2\Big(\frac{x-\mu}{\sigma}\Big)-1\Big]
\end{align*}
\begin{align*}
    i_{\sigma\xi}(x;\mu,\sigma,\xi) &=-\frac{\partial^2 \log f(x;\mu,\sigma,\xi)}{\partial \sigma\partial\xi} = \frac{1}{\sigma}\Big(1+\xi\Big(\frac{x-\mu}{\sigma}\Big)\Big)^{-2}\Big(\frac{x-\mu}{\sigma}\Big)\Big(\Big(\frac{x-\mu}{\sigma}\Big)-1\Big) \\
    &+\frac{1}{\xi^2 \sigma^2}\Big(\frac{x-\mu}{\sigma}\Big)\Big(1+\xi\Big(\frac{x-\mu}{\sigma}\Big)\Big)^{-1 / \xi-2}\Big[\Big(1+\xi\Big(\frac{x-\mu}{\sigma}\Big)\Big) \log \Big(1+\xi\Big(\frac{x-\mu}{\sigma}\Big)\Big)-\xi(\xi+1) \Big(\frac{x-\mu}{\sigma}\Big)\Big]; \\
\end{align*}

\begin{align*}
    i_{\xi\xi}&(x;\mu,\sigma,\xi) = -\frac{\partial^2 \log f(x;\mu,\sigma,\xi)}{\partial \xi^2}= -\Big(1+\xi\Big(\frac{x-\mu}{\sigma}\Big)\Big)^{-2}\Big(\frac{x-\mu}{\sigma}\Big)^2 \\
    &+ \frac{3}{\xi}\Big(\frac{x-\mu}{\sigma}\Big)^2\bigg(\Big(1+\xi\Big(\frac{x-\mu}{\sigma}\Big)\Big)^{-1/\xi-2}-1\bigg) \\
    &+ \frac{1}{\xi^2}\Big(\frac{x-\mu}{\sigma}\Big)\bigg[2\Big(1+\xi\Big(\frac{x-\mu}{\sigma}\Big)\Big)^{-1/\xi-2}-2 \Big(1+\xi\Big(\frac{x-\mu}{\sigma}\Big)\Big)^{-2}+\Big(\frac{x-\mu}{\sigma}\Big)\\
    &+2\log\Big(1+\xi\Big(\frac{x-\mu}{\sigma}\Big)\Big)\Big(1+\xi\Big(\frac{x-\mu}{\sigma}\Big)\Big)^{-1}\bigg(\Big(1+\xi\Big(\frac{x-\mu}{\sigma}\Big)\Big)^{-1/\xi}-1\bigg)\bigg] \\
    &+ 2\frac{\log\Big(1+\xi\Big(\frac{x-\mu}{\sigma}\Big)\Big)}{\xi^3}\Big(1+\xi\Big(\frac{x-\mu}{\sigma}\Big)\Big)^{-1}\bigg(\Big(1+\xi\Big(\frac{x-\mu}{\sigma}\Big)\Big)^{-1/\xi}-\Big(1+\xi\Big(\frac{x-\mu}{\sigma}\Big)\Big)^{-1}+\Big(\frac{x-\mu}{\sigma}\Big)\Big(1+\xi\Big(\frac{x-\mu}{\sigma}\Big)\Big)^{-1/\xi}\bigg) \\
    &+ \frac{\log^2\Big(1+\xi\Big(\frac{x-\mu}{\sigma}\Big)\Big)}{\xi^4}\Big(1+\xi\Big(\frac{x-\mu}{\sigma}\Big)\Big)^{-1/\xi};
\end{align*}
for all $x \in D_{\mu,\sigma,\xi}$.

\begin{remark}[Minor Contribution]
    The score and information matrices for the GEV distribution are given by \cite{prescottwalden1980}. We recall that the score and information matrices are defined as the expectations of the score and information functions, respectively. However, to the best of our knowledge, the complete specification of the function needed to compute the information matrix have not been published before. 
\end{remark}


\section{Proof of Theorem~\ref{th:asympt_norm}}\label{appendix:proof_asympt_norm}

First, we must establish the existence of a consistent sequence of MDPD estimators. For the sake of conciseness (and because the detailed proofs follow directly from the case of the GP distribution; see Section~A.4.1 of \Citealp{phdjuarez}, based on results from Chapters~13 and~14 in \Citealp{hoffman1994probability}, we omit the proof here.

Assuming the existence of a consistent sequence of MDPD estimators $(\hat{\theta}_{n,\alpha})_n := (\hat{\mu}_{n,\alpha}, \hat{\sigma}_{n,\alpha}, \hat{\xi}_{n,\alpha})_n$, we now turn to the asymptotic normality of the estimator. To this end, we rely on the following result from \cite{phdjuarez}, where $\theta_0 \in \Theta_0$ denotes the true parameter, $\mathcal{F} = \{f(x;\theta) : x \in \mathcal{X}, \theta \in \Theta_0\}$ is the parametric model, and $g$ is the true density.

\begin{corollary}[Corollary A.3.4 in \Citealp{phdjuarez}]\label{cor:juarez}
Let $\theta_0$ be in the interior of $\Theta_0$. Suppose the following four conditions hold.
\begin{enumerate}
    \item The information matrix $i(x ; \theta_0)$ is finite in a neighborhood $V$ of $\theta_0$;
    \item The integral $\int_{\mathcal{X}} f^{1+\alpha}(x ; \theta) \mu(d x)$ is twice differentiable with respect to $\theta$ and the derivative can be taken under the integral sign in a neighborhood $V$ of $\theta_0$. This means that $f^{1+\alpha}(x ; \theta)$ is also twice partially differentiable with respect to $\theta$;
    \item Let $S_j$ denote the $j$-th element of the score vector. For each $j=1, \ldots, p$
    \begin{equation*}
        \int_{\mathcal{X}} S_j^2\left(x ; \theta_0\right) f^{2 \alpha}\left(x ; \theta_0\right) g(x) \mu(d x)<\infty;
    \end{equation*}
    
    \item Let $i_{j k}(x ; \theta)$ denote the $j k$-th element of the information matrix. There exists functions $\phi_{j k}, j, k=1, \ldots, p$, such that $\left|i_{j k}\left(x ; \theta_0\right) f^\alpha\left(x ; \theta_0\right)\right| \leq \phi_{j k}(x)$ and
    \begin{equation*}
        \int_{\mathcal{X}}\left|\phi_{j k}(x)\right| g(x) \mu(d x)<\infty, \quad \text { for } G \text {-a.e. in } \mathcal{X}.
    \end{equation*}
\end{enumerate}
Then, any sequence of MDPDE $(\hat{\theta}_{\alpha, n})_n$  that is consistent for $\theta_0$ is such that
\begin{equation*}
    \sqrt{n}\left(\hat{\theta}_{n, \alpha}-\theta_0\right) \overset{d}{\longrightarrow}\mathcal{N}\left(0, J_\alpha^{-1}\left(\theta_0\right) K_\alpha\left(\theta_0\right) J_\alpha^{-1}\left(\theta_0\right)\right).
\end{equation*}
\end{corollary}

Given this result, the proof of Theorem~\ref{th:asympt_norm} reduces to verifying that the required conditions are satisfied for a GEV distribution. In this work, we operate strictly in the case where the density $g(x) = f(x; (\mu_0, \sigma_0, \xi_0))$ is the GEV density corresponding to the target parameter triplet $(\mu_0, \sigma_0, \xi_0)$  \citep[which refers to \emph{under model conditions} in][]{juarez2004robust}. Let $\alpha >0$.

We now check each condition of Corollary~\ref{cor:juarez} individually. In the following, we denote a triplet of GEV parameters by $\theta = (\mu, \sigma, \xi)$, and the target parameter by $\theta_0 = (\mu_0, \sigma_0, \xi_0)$.

Recall that we assume $\xi > -(1+\alpha)/(2+\alpha)$.
\begin{enumerate}
    \item Given the explicit formulas of the information matrix a GEV distribution (Section~\ref{appendix:score_and_inf}), it is clear that the information matrix is finite in a neighborhood of $\theta_0.$
\item For $\xi > -\alpha/(\alpha+1)$, which is guaranteed by our assumption $\xi > -(1+\alpha)/(2+\alpha)$, we have \\$\int_{D_{\mu,\sigma,\xi}} f^{1+\alpha}(x;\theta) \, dx < +\infty$, and in particular:
\begin{equation*}
    \int_{D_{\mu,\sigma,\xi}} f^{1+\alpha}(x;\mu,\sigma,\xi) \, dx = \frac{1}{\sigma^\alpha} \Big(\frac{1}{1+\alpha}\Big)^{\alpha(\xi+1)+1} \Gamma(\alpha(\xi+1)+1).
\end{equation*}
Since the Gamma function is $\mathcal{C}^\infty$ on $\mathbb{R}_{>0}$, the second condition is satisfied.

\item We treat the case $0 > \xi > -(1+\alpha)/(2+\alpha)$; the case $\xi \s 0$ follows by the exact same argument and is left to the reader. Assume $0 > \xi > -(1+\alpha)/(2+\alpha)$. Since
\begin{equation}\label{eq:infty}
    f(x;\theta) \underset{x \to -\infty}{=} O\left((-x)^{-(\xi+1)/\xi} \exp\big(-(-x)^{-1/\xi}\big)\right),
\end{equation}
we deduce that $f(x;\theta_0)^{2\alpha+1} S^2_\mu(x;\theta_0)$, $f(x;\theta_0)^{2\alpha+1} S^2_\sigma(x;\theta_0)$, and $f(x;\theta_0)^{2\alpha+1} S^2_\xi(x;\theta_0)$ are integrable in a neighborhood of $-\infty$.

In addition, since
\begin{equation}\label{eq:mu-sigmaxi}
    f(x;\theta) \underset{x \to \mu - \sigma/\xi}{=} O\left(\left(1+\xi\left(\frac{x - \mu}{\sigma}\right)\right)^{-1 - 1/\xi}\right),
\end{equation}
it follows that $f(x;\theta_0)^{2\alpha+1} S^2_\mu(x;\theta_0)$, $f(x;\theta_0)^{2\alpha+1} S^2_\sigma(x;\theta_0)$, and $f(x;\theta_0)^{2\alpha+1} S^2_\xi(x;\theta_0)$ are integrable in a neighborhood of $\mu - \sigma/\xi$ provided that $\xi > -(2\alpha+1)/(2\alpha+2)$, which is satisfied under the assumption $\xi > -(\alpha+1)/(\alpha+2)$. We conclude that condition~3. is fulfilled.

\item Again, we treat the case $0 > \xi > -(1+\alpha)/(2+\alpha)$; the case $\xi \s 0$ follows by the exact same argument and is left to the reader. Using Equations~\eqref{eq:infty} and~\eqref{eq:mu-sigmaxi}, it is straightforward to verify that each of the functions
\[
|f(x;\theta_0)^\alpha i_{\mu\mu}(x;\theta_0)|,\quad |f(x;\theta_0)^\alpha i_{\mu\sigma}(x;\theta_0)|,\quad |f(x;\theta_0)^\alpha i_{\mu\xi}(x;\theta_0)|,
\]
\[
|f(x;\theta_0)^\alpha i_{\sigma\sigma}(x;\theta_0)|,\quad |f(x;\theta_0)^\alpha i_{\sigma\xi}(x;\theta_0)|,\quad |f(x;\theta_0)^\alpha i_{\xi\xi}(x;\theta_0)|
\]
can be bounded above by functions independent of $\theta$, such that the integrals of condition 4. are finite. Therefore, condition~4. is also satisfied.
\end{enumerate}

The requirements of Corollary~\ref{cor:juarez} are fulfilled for the GEV distribution which concludes the proof of Theorem~\ref{th:asympt_norm}.

%% file: appendix_simulation.tex
\section{Additional simulation studies} \label{sec:simu_appendix}
This section provides further simulation experiments, including the proportion of non-convergence for the different methods, in the setting studied in Section~\ref{sec:simulation_study} with sample size $n = 100$ and contamination proportion $\varepsilon = 0.1$ (see Section~\ref{sec:n100e01}). We also present additional simulation studies similar to that of Section~\ref{sec:simulation_study} in various settings: $n = 50$ with $\varepsilon = 0.1$ (Section~\ref{sec:n50e01}); $n = 50$ with $\varepsilon = 0.05$ (Section~\ref{sec:n50e005}); and $n = 100$ with $\varepsilon = 0.2$ (Section~\ref{sec:n100e02}).

We further compare our results with classical robust estimators, namely the \emph{Radius-Minimax Estimator} (RMXE) and the \emph{Optimal MSE Estimator} (OMSE), using the functions provided in the \texttt{R} packages \texttt{ROptEst} and \texttt{RobExtremes} \citep[][]{horbenko2018package,kohl2019package}. The comparison is restricted to the case where the true shape parameter is positive, since these packages only implement methods for positive shape parameters.

\subsection{Simulation study : $n=100,\varepsilon = 0.1$}\label{sec:n100e01}

\begin{figure}[ht!]
\centering
  \includegraphics[width=.47\textwidth]{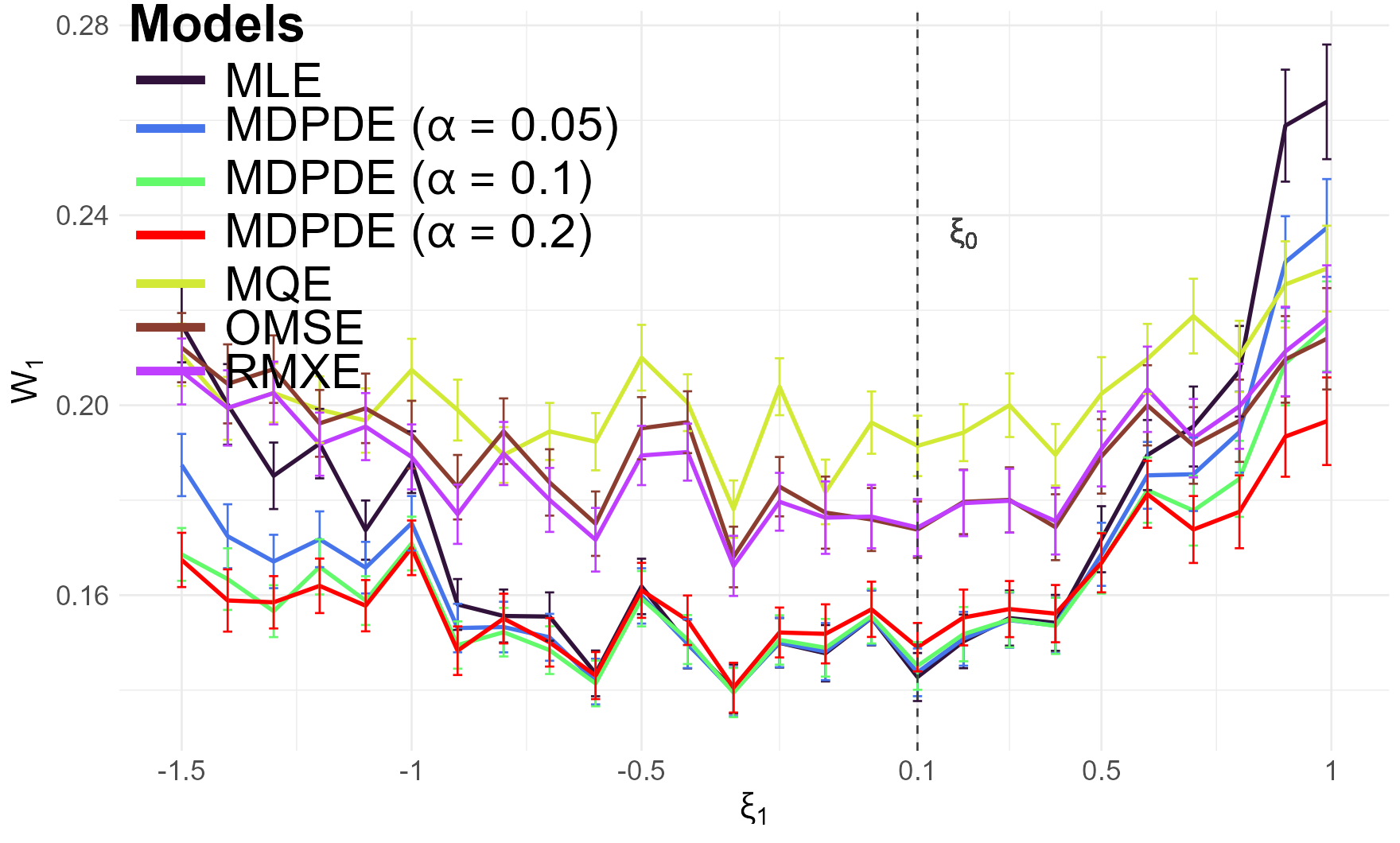}
  \hspace{0.7cm}
  \includegraphics[width=.47\textwidth]{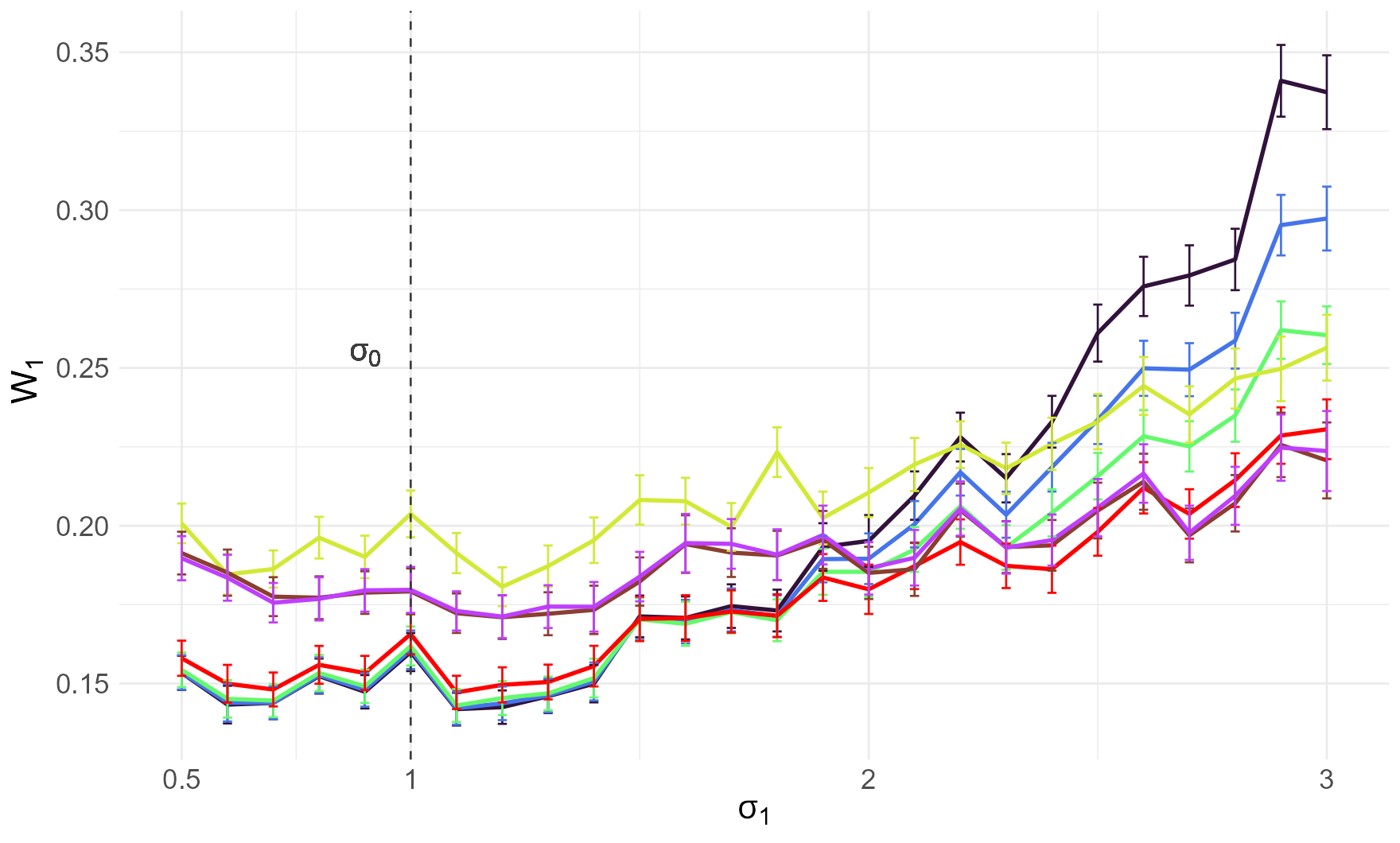}
  \caption{Average Wasserstein distance over 200 replications (with standard errors) across various contaminated models. In the left panel, the shape parameter $\xi_1$ varies while the location $\mu_0$ and scale $\sigma_0$ are fixed. In the right panel, the scale parameter $\sigma_1$ varies while $\mu_0$ and the shape $\xi_0$ remain fixed. Each sample has size $n = 100$, with contamination proportion $\varepsilon = 0.1$. The true model parameters are $\mu_0 = 0$, $\sigma_0 = 1$, and $\xi_0 = 0.1$.}
\end{figure}

Table~\ref{tab:missing_value} provides the number of samples in which either convergence fails, either estimation yields non-plausible results, for each considered configuration.



\begin{table}[ht!]
\caption{Number of samples in which either convergence fails (for the MLE or the MDPDE) or the estimation procedures return non-plausible values (i.e., $\hat{\mu}_0 < -2$, $\hat{\mu}_0 > 2$, or $\hat{\sigma}_0 > 2$). The models vary either in the shape parameter $\xi_{1}\in\{-1.5,-1.4,\dots,0.9,0.99\}$ or in the scale parameter $\sigma_{1}\in\{0.5,0.6,\dots,3\}$, out of 200 replications per contaminated model, yielding 5200 total samples for each case. All results are based on samples of size $n=100$ with contamination proportion $\varepsilon=0.1$.\label{tab:missing_value}}
\vspace{0.2cm}

\begin{center}
\begin{small}
\resizebox{\textwidth}{!}{
\begin{tabular}{ll|ccccccc}
    & & MLE & MDPDE ($\alpha = 0.05)$& MDPDE ($\alpha = 0.1)$ & MDPDE ($\alpha = 0.2)$ & MQE & OMSE & RMXE \\ \hline
    \multirowcell{2}{$\xi_0  =-0.1$}&\textit{varying} $\xi_1$ & 0 & 2 & 8 & 4 & 23 & $\times$ & $\times$\\
    &\textit{varying} $\sigma_1$ & 2 & 4 & 4 & 4 & 7 & $\times$ & $\times$\\
    \hline
    \multirowcell{2}{$\xi_0  =0$}&\textit{varying} $\xi_1$& 2 & 2 & 3 & 3 & 32 & $\times$ & $\times$ \\
    &\textit{varying} $\sigma_1$ & 0 & 1 & 2 & 1 & 34 & $\times$ & $\times$ \\
    \hline
    \multirowcell{2}{$\xi_0  =0.1$}& \textit{varying} $\xi_1$ & 10 & 0 & 0 & 3 & 16 & 616 & 615\\
    &\textit{varying} $\sigma_1$ & 0 & 0 & 1 & 0 &  9 & 597 & 597 \\
\end{tabular}
}
\end{small}
\end{center}
\end{table}

\subsection{Simulation study : $n=50,\varepsilon = 0.1$\label{sec:n50e01}}

\begin{figure}[ht!]
\centering
  \includegraphics[width=.47\textwidth]{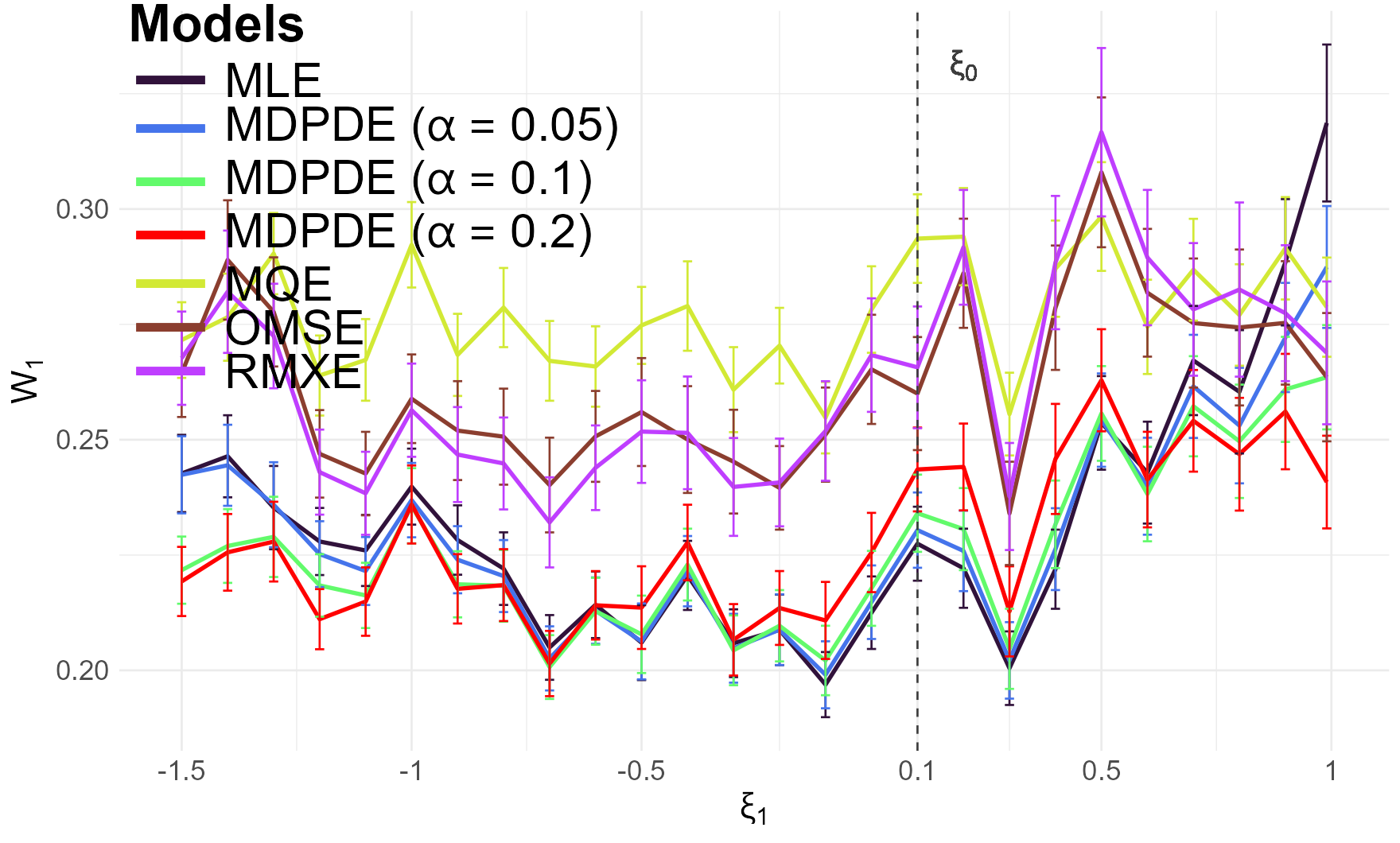}
  \hspace{0.7cm}
  \includegraphics[width=.47\textwidth]{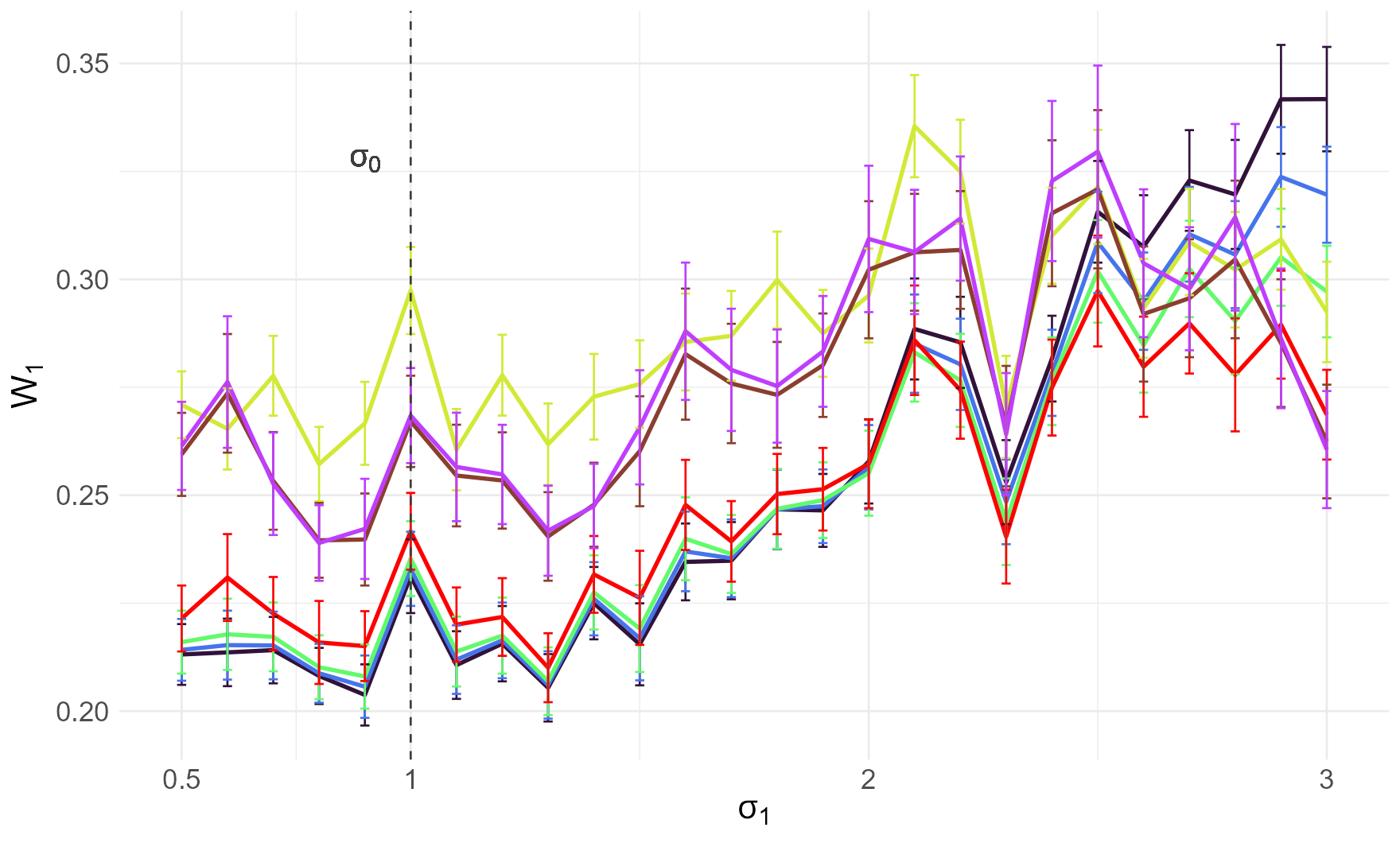}
  \caption{Average Wasserstein distance over 200 replications (with standard errors) across various contaminated models. In the left panel, the shape parameter $\xi_1$ varies while the location $\mu_0$ and scale $\sigma_0$ are fixed. In the right panel, the scale parameter $\sigma_1$ varies while $\mu_0$ and the shape $\xi_0$ remain fixed. Each sample has size $n = 50$, with contamination proportion $\varepsilon = 0.1$. The true model parameters are $\mu_0 = 0$, $\sigma_0 = 1$, and $\xi_0 = 0.1$.}
\end{figure}

\begin{figure}[ht!]
\centering
  \includegraphics[width=.47\textwidth]{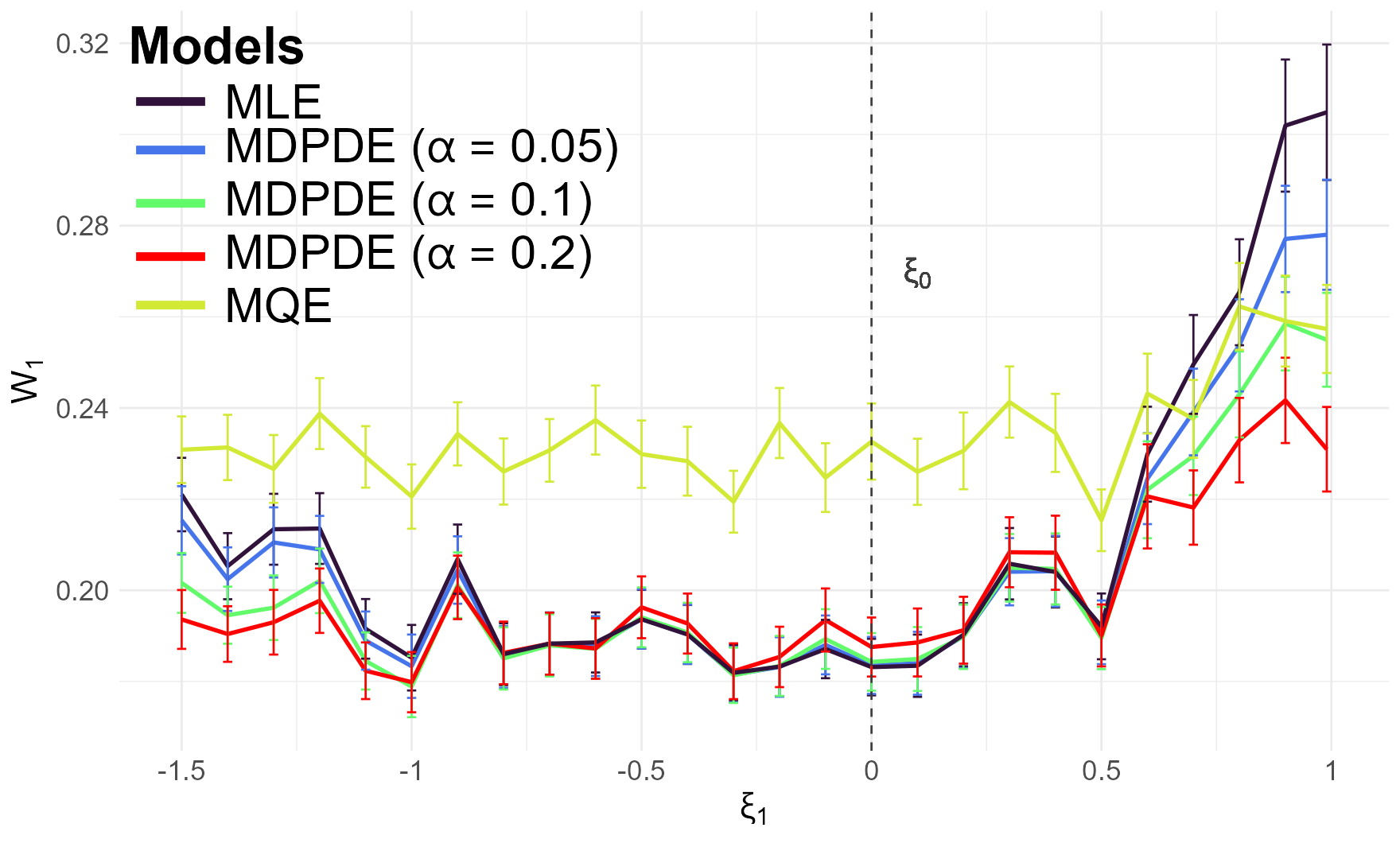}
  \hspace{0.7cm}
  \includegraphics[width=.47\textwidth]{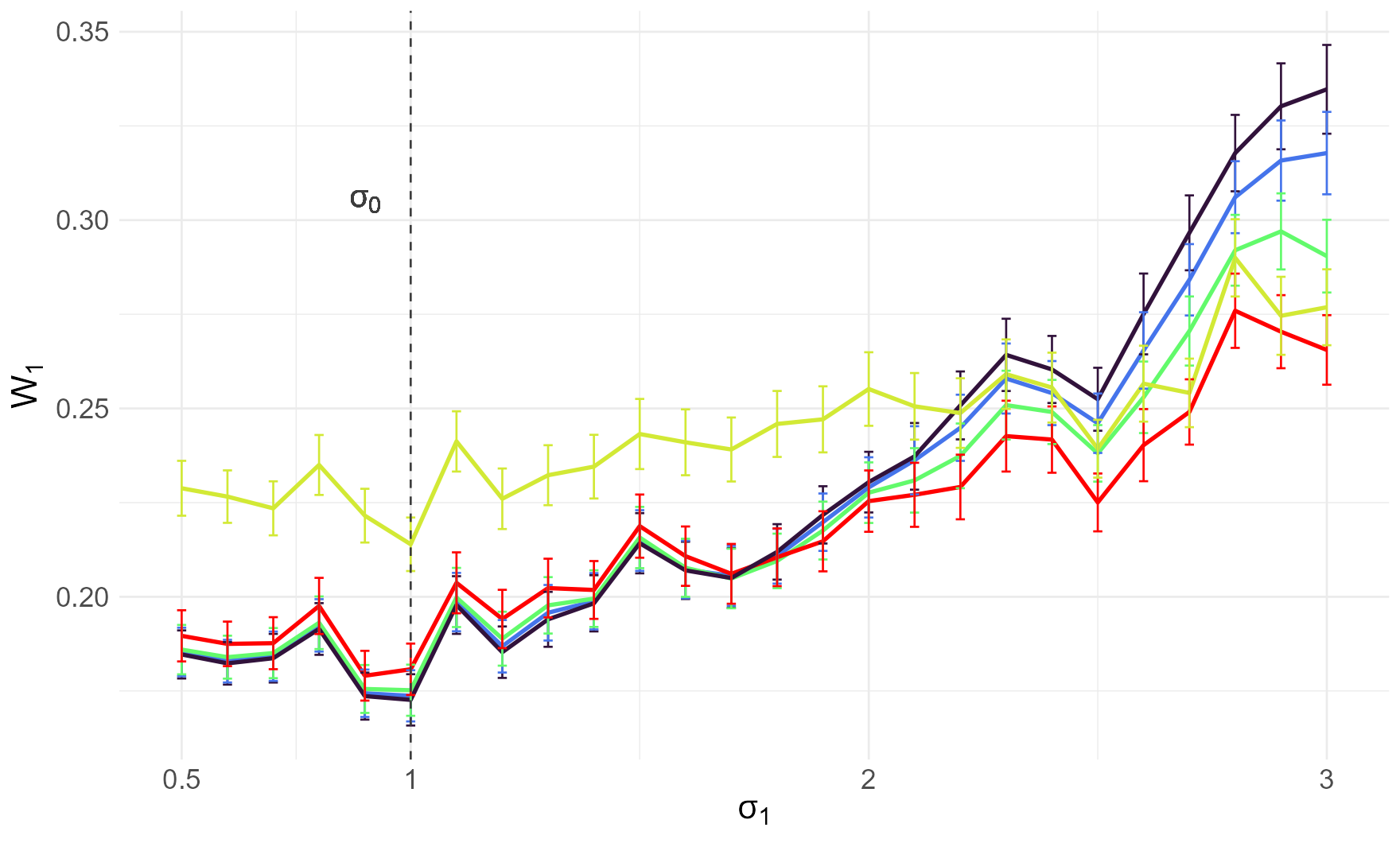}
  \caption{Average Wasserstein distance over 200 replications (with standard errors) across various contaminated models. In the left panel, the shape parameter $\xi_1$ varies while the location $\mu_0$ and scale $\sigma_0$ are fixed. In the right panel, the scale parameter $\sigma_1$ varies while $\mu_0$ and the shape $\xi_0$ remain fixed. Each sample has size $n = 50$, with contamination proportion $\varepsilon = 0.1$. The true model parameters are $\mu_0 = 0$, $\sigma_0 = 1$, and $\xi_0 = 0$.}
\end{figure}

\begin{figure}[ht!]
\centering
  \includegraphics[width=.47\textwidth]{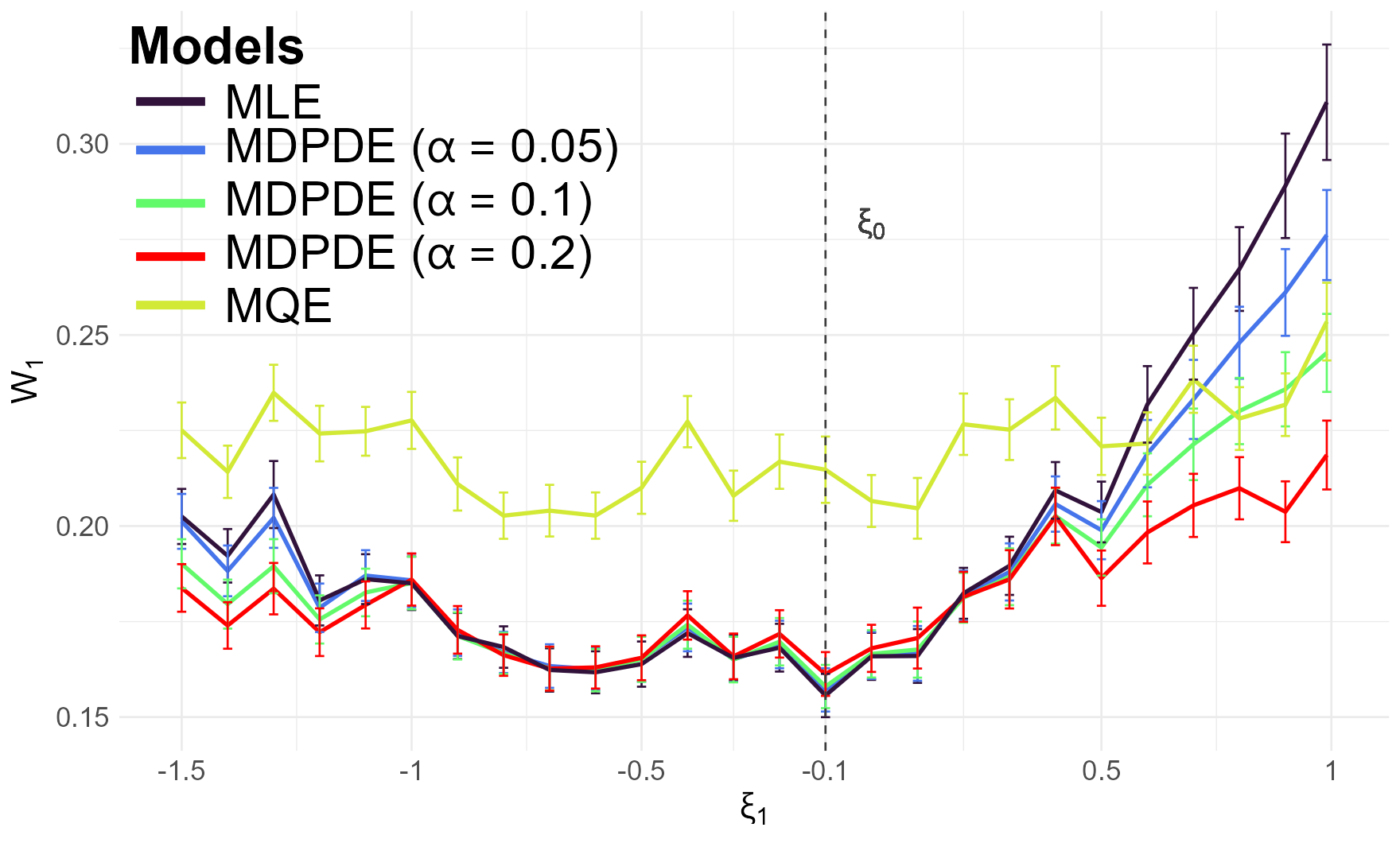}
  \hspace{0.7cm}
  \includegraphics[width=.47\textwidth]{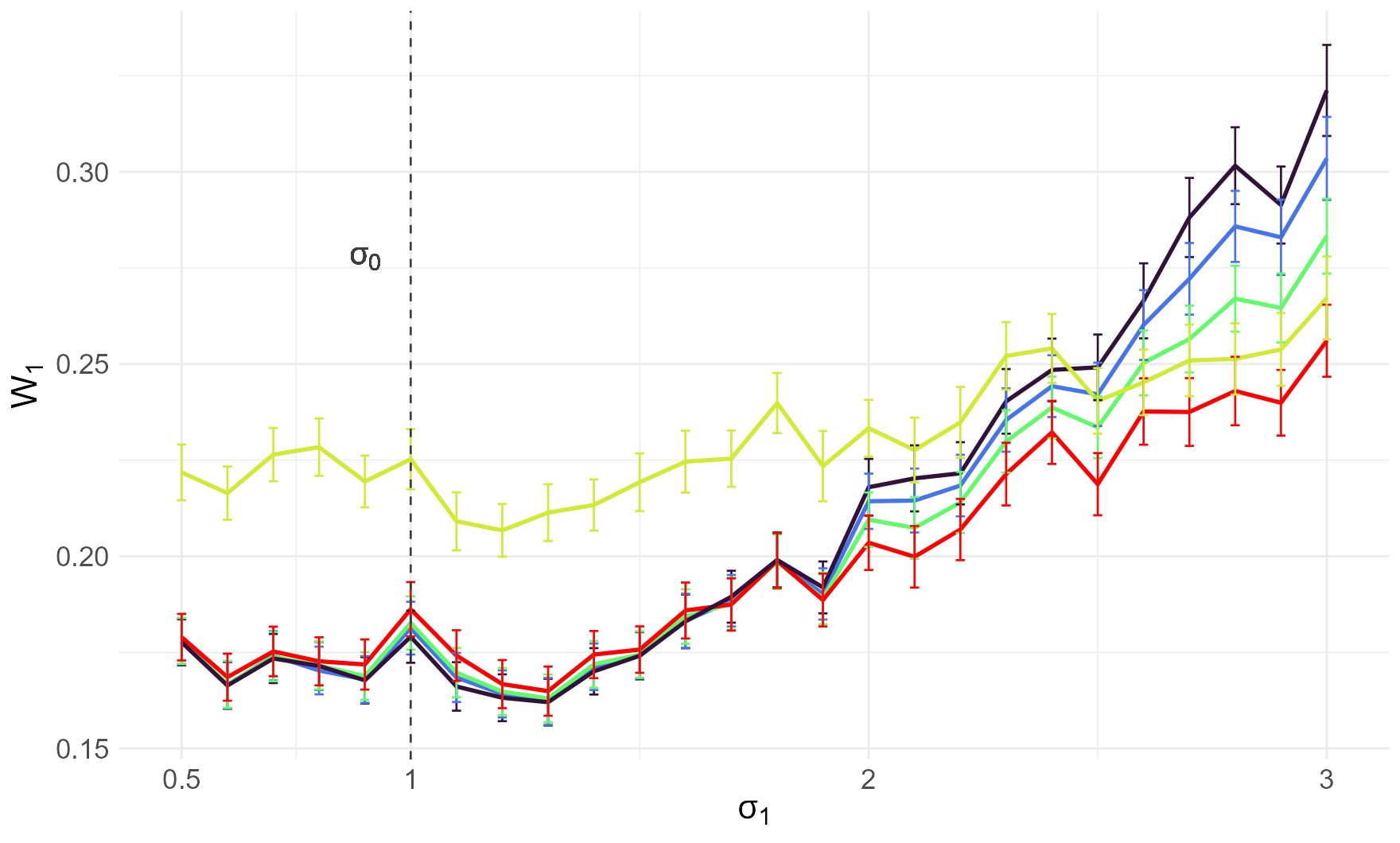}
  \caption{Average Wasserstein distance over 200 replications (with standard errors) across various contaminated models. In the left panel, the shape parameter $\xi_1$ varies while the location $\mu_0$ and scale $\sigma_0$ are fixed. In the right panel, the scale parameter $\sigma_1$ varies while $\mu_0$ and the shape $\xi_0$ remain fixed. Each sample has size $n = 50$, with contamination proportion $\varepsilon = 0.1$. The true model parameters are $\mu_0 = 0$, $\sigma_0 = 1$, and $\xi_0 = -0.1$.}
\end{figure}

\begin{table}[ht!]
\caption{Number of samples in which either convergence fails (for the MLE or the MDPDE) or the estimation procedures return non-plausible values (i.e., $\hat{\mu}_0 < -2$, $\hat{\mu}_0 > 2$, or $\hat{\sigma}_0 > 2$). The models vary either in the shape parameter $\xi_{1}\in\{-1.5,-1.4,\dots,0.9,0.99\}$ or in the scale parameter $\sigma_{1}\in\{0.5,0.6,\dots,3\}$, out of 200 replications per contaminated model, yielding 5200 total samples for each case. All results are based on samples of size $n=50$ with contamination proportion $\varepsilon=0.1$.\label{tab:missing_value_n50}}
\vspace{0.2cm}

\begin{center}
\begin{small}
\resizebox{\textwidth}{!}{
\begin{tabular}{ll|ccccccc}
    & & MLE & MDPDE ($\alpha = 0.05)$& MDPDE ($\alpha = 0.1)$ & MDPDE ($\alpha = 0.2)$ & MQE & OMSE & RMXE \\ \hline
    \multirowcell{2}{$\xi_0  =-0.1$}&\textit{varying} $\xi_1$ & 5 & 1 & 7 & 4 & 13 & $\times$ & $\times$\\
    &\textit{varying} $\sigma_1$ & 1 & 1 & 4 & 3 & 7 & $\times$ & $\times$\\
    \hline
    \multirowcell{2}{$\xi_0  =0$}&\textit{varying} $\xi_1$& 9 & 1 & 0 & 1 & 14 & $\times$ & $\times$ \\
    &\textit{varying} $\sigma_1$ & 5 & 1 & 0 & 0 & 17 & $\times$ & $\times$ \\
    \hline
    \multirowcell{2}{$\xi_0  =0.1$}& \textit{varying} $\xi_1$ & 22 & 5 & 1 & 2 & 14 & 1341 & 1338 \\
    &\textit{varying} $\sigma_1$ & 5 & 2 & 0 & 2 &  14 & 1231 & 1227 \\
\end{tabular}
}
\end{small}
\end{center}
\end{table}

\clearpage
\subsection{Simulation study : $n=50, \varepsilon = 0.05$\label{sec:n50e005}}

\begin{figure}[ht!]
\centering
  \includegraphics[width=.47\textwidth]{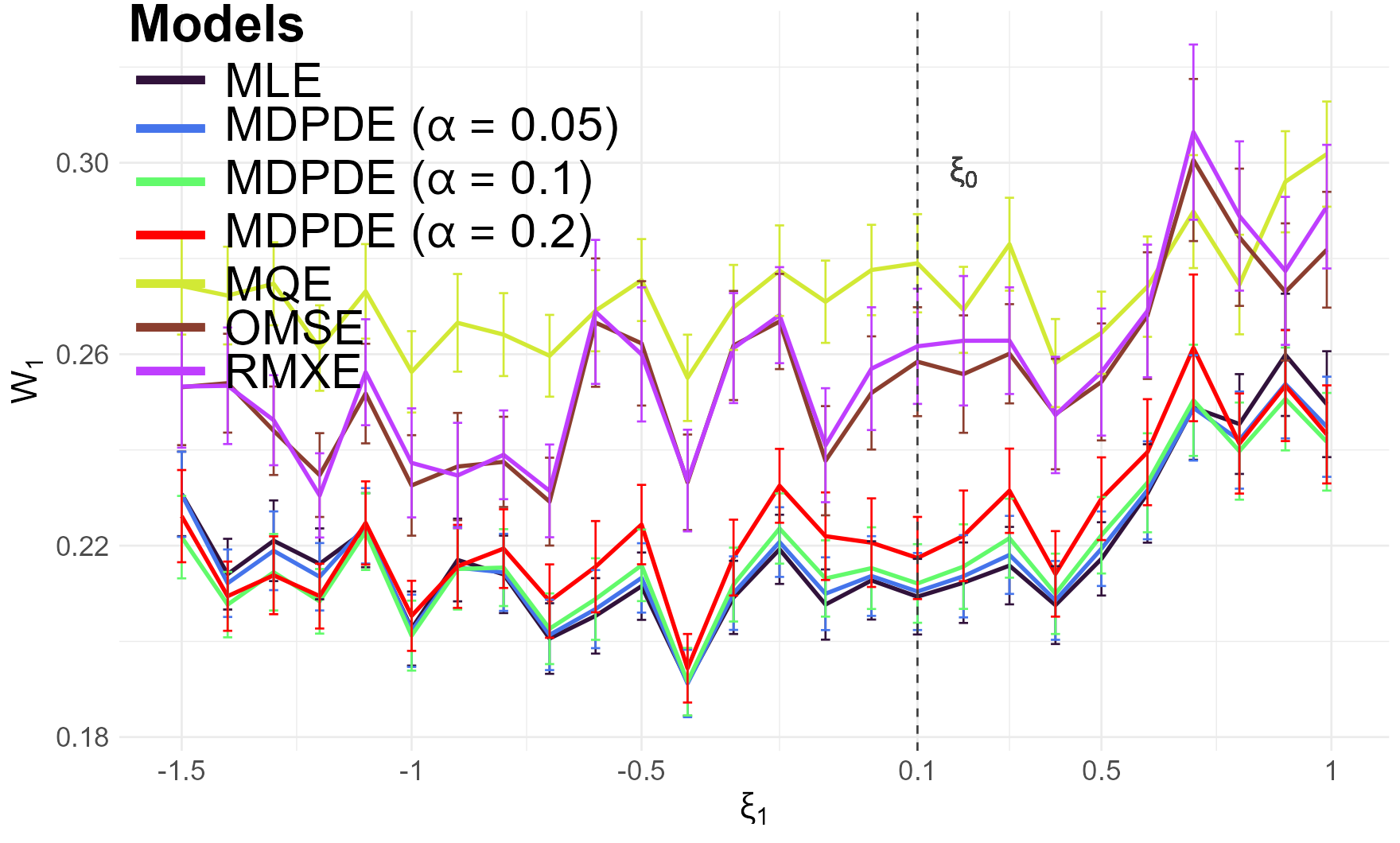}
  \hspace{0.7cm}
  \includegraphics[width=.47\textwidth]{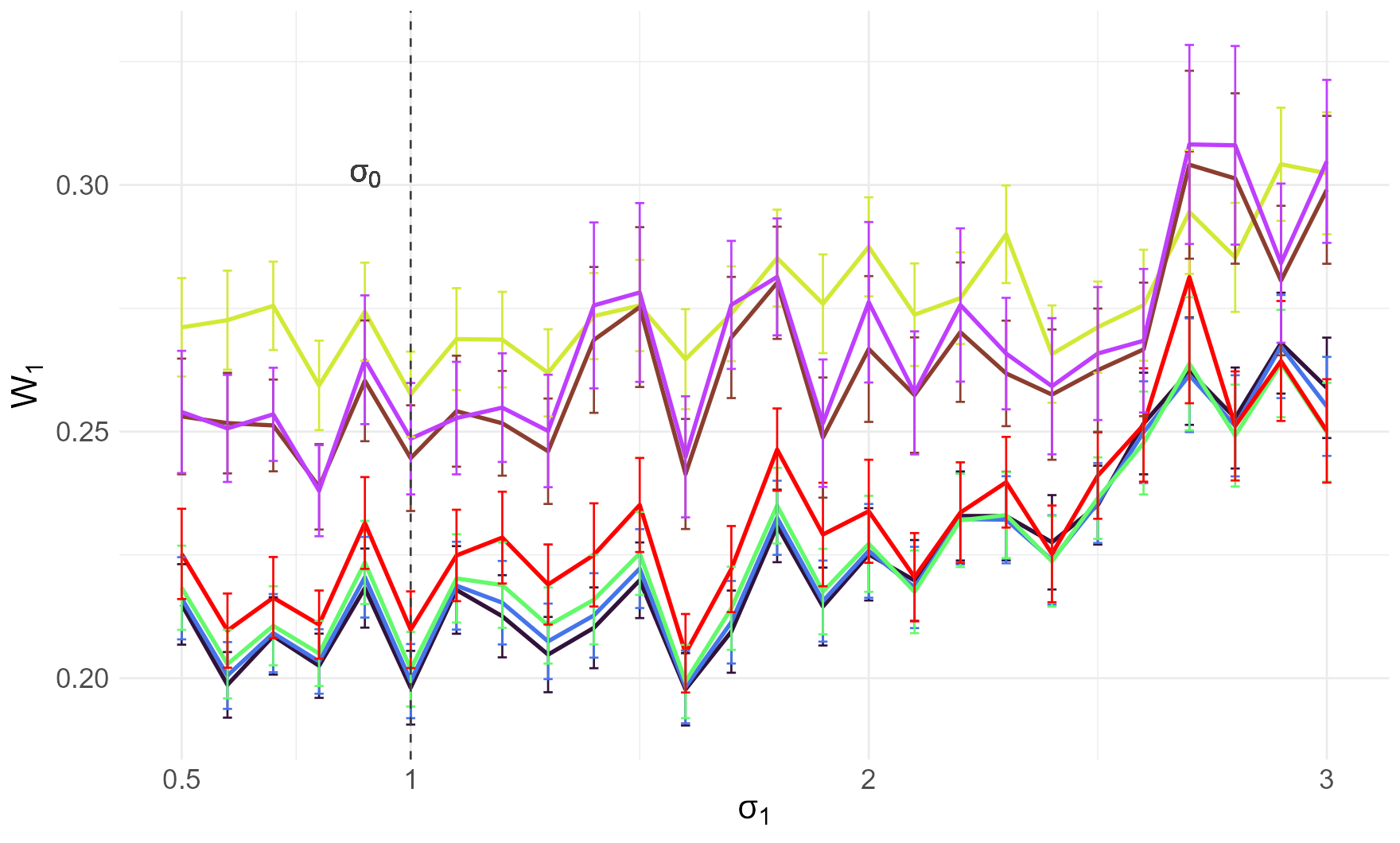}
  \caption{Average Wasserstein distance over 200 replications (with standard errors) across various contaminated models. In the left panel, the shape parameter $\xi_1$ varies while the location $\mu_0$ and scale $\sigma_0$ are fixed. In the right panel, the scale parameter $\sigma_1$ varies while $\mu_0$ and the shape $\xi_0$ remain fixed. Each sample has size $n = 50$, with contamination proportion $\varepsilon = 0.05$. The true model parameters are $\mu_0 = 0$, $\sigma_0 = 1$, and $\xi_0 = 0.1$.}
\end{figure}

\begin{figure}[ht!]
\centering
  \includegraphics[width=.47\textwidth]{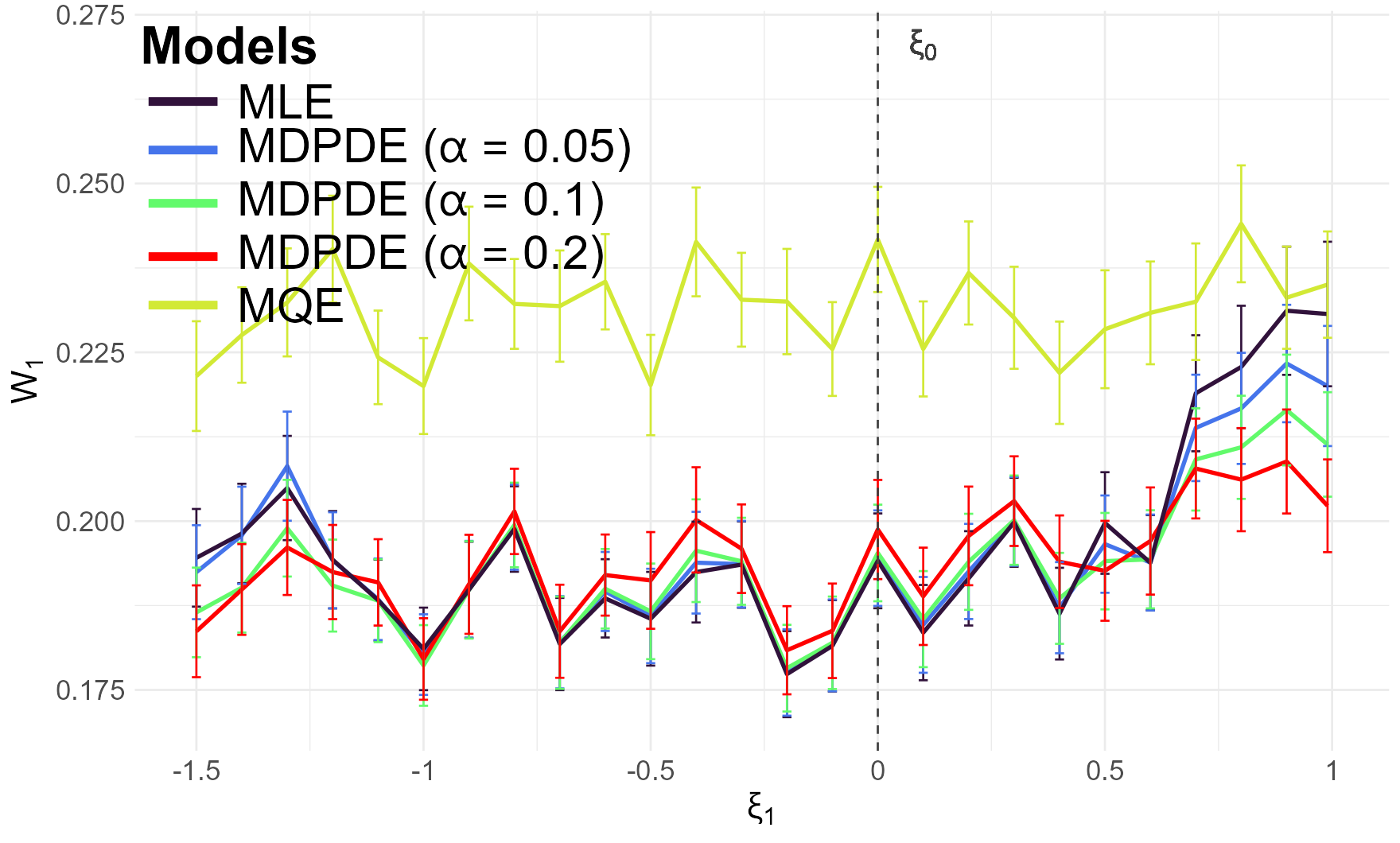}
  \hspace{0.7cm}
  \includegraphics[width=.47\textwidth]{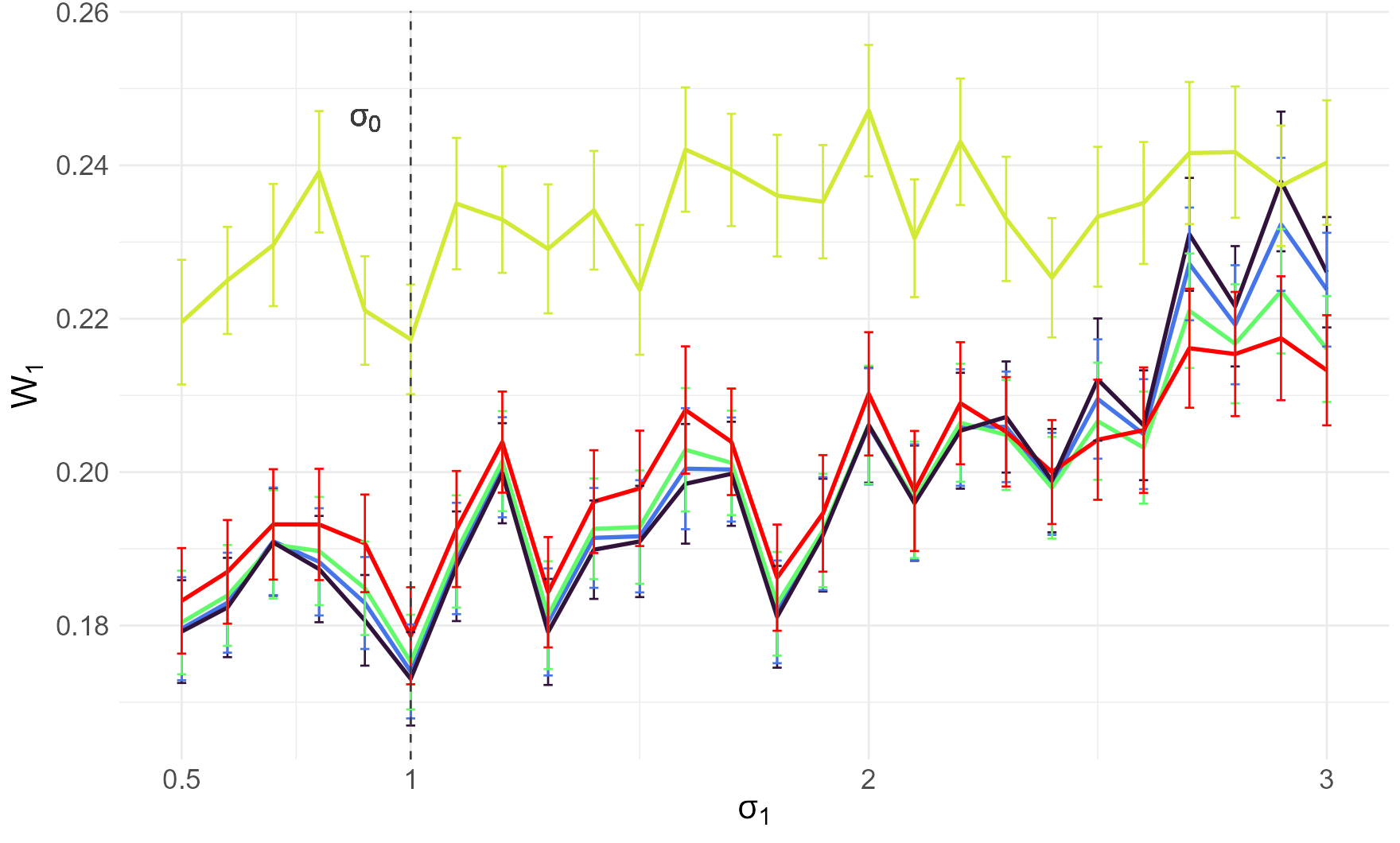}
  \caption{Average Wasserstein distance over 200 replications (with standard errors) across various contaminated models. In the left panel, the shape parameter $\xi_1$ varies while the location $\mu_0$ and scale $\sigma_0$ are fixed. In the right panel, the scale parameter $\sigma_1$ varies while $\mu_0$ and the shape $\xi_0$ remain fixed. Each sample has size $n = 50$, with contamination proportion $\varepsilon = 0.05$. The true model parameters are $\mu_0 = 0$, $\sigma_0 = 1$, and $\xi_0 = 0$.}
\end{figure}

\begin{figure}[ht!]
\centering
  \includegraphics[width=.47\textwidth]{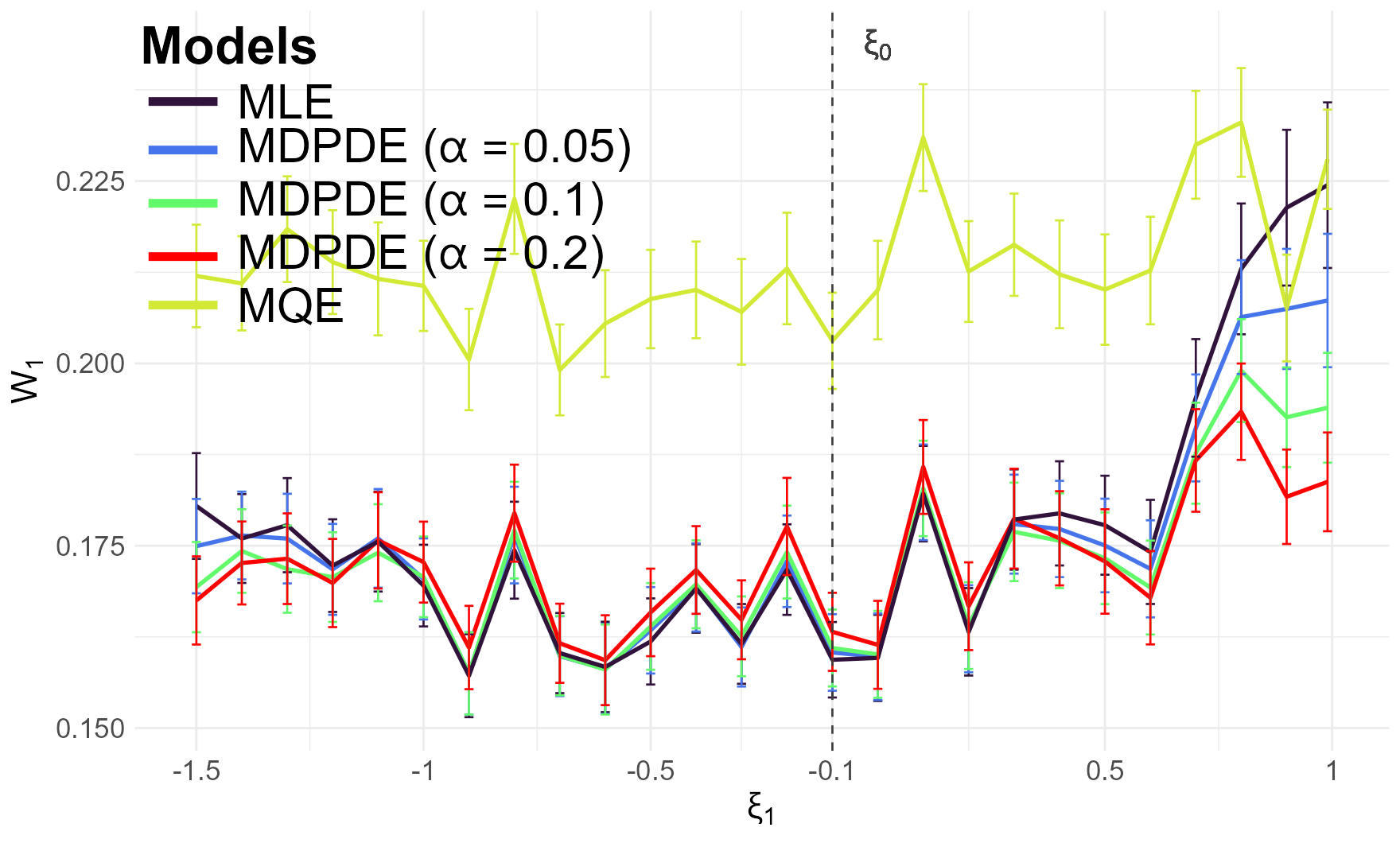}
  \hspace{0.7cm}
  \includegraphics[width=.47\textwidth]{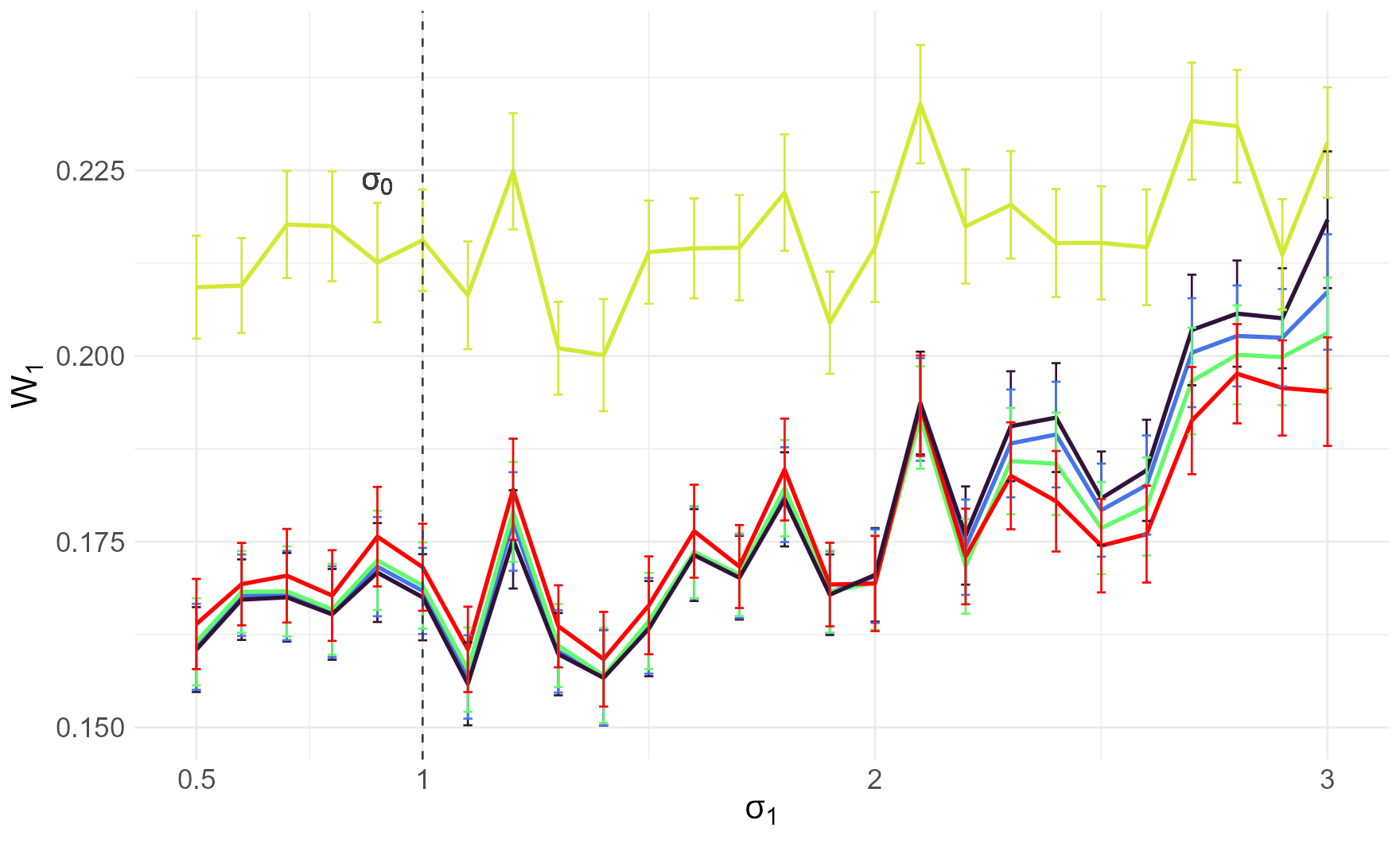}
  \caption{Average Wasserstein distance over 200 replications (with standard errors) across various contaminated models. In the left panel, the shape parameter $\xi_1$ varies while the location $\mu_0$ and scale $\sigma_0$ are fixed. In the right panel, the scale parameter $\sigma_1$ varies while $\mu_0$ and the shape $\xi_0$ remain fixed. Each sample has size $n = 50$, with contamination proportion $\varepsilon = 0.05$. The true model parameters are $\mu_0 = 0$, $\sigma_0 = 1$, and $\xi_0 = -0.1$.}
\end{figure}

\begin{table}[ht!]
\caption{Number of samples in which either convergence fails (for the MLE or the MDPDE) or the estimation procedures return non-plausible values (i.e., $\hat{\mu}_0 < -2$, $\hat{\mu}_0 > 2$, or $\hat{\sigma}_0 > 2$). The models vary either in the shape parameter $\xi_{1}\in\{-1.5,-1.4,\dots,0.9,0.99\}$ or in the scale parameter $\sigma_{1}\in\{0.5,0.6,\dots,3\}$, out of 200 replications per contaminated model, yielding 5200 total samples for each case. All results are based on samples of size $n=50$ with contamination proportion $\varepsilon=0.05$.\label{tab:missing_value_n50_e005}}
\vspace{0.2cm}

\begin{center}
\begin{small}
\resizebox{\textwidth}{!}{
\begin{tabular}{ll|ccccccc}
    & & MLE & MDPDE ($\alpha = 0.05)$& MDPDE ($\alpha = 0.1)$ & MDPDE ($\alpha = 0.2)$ & MQE & OMSE & RMXE \\ \hline
    \multirowcell{2}{$\xi_0  =-0.1$}&\textit{varying} $\xi_1$ & 2 & 3 & 2 & 5 & 23 & $\times$ & $\times$\\
    &\textit{varying} $\sigma_1$ & 1 & 1 & 1 & 1 & 9 & $\times$ & $\times$\\
    \hline
    \multirowcell{2}{$\xi_0  =0$}&\textit{varying} $\xi_1$& 8 & 2 & 4 & 1 & 22 & $\times$ & $\times$ \\
    &\textit{varying} $\sigma_1$ & 5 & 1 & 3 & 4 & 31 & $\times$ & $\times$ \\
    \hline
    \multirowcell{2}{$\xi_0  =0.1$}& \textit{varying} $\xi_1$ & 5 & 0 & 1 & 2 &  14 & 1204 & 1197\\
    &\textit{varying} $\sigma_1$ & 0 & 1 & 0 & 1 & 9 & 1174 & 1169 \\
\end{tabular}
}
\end{small}
\end{center}
\end{table}

\clearpage

\subsection{Simulation study : $n=100,\varepsilon = 0.2$} \label{sec:n100e02}

\begin{figure}[ht!]
\centering
  \includegraphics[width=.47\textwidth]{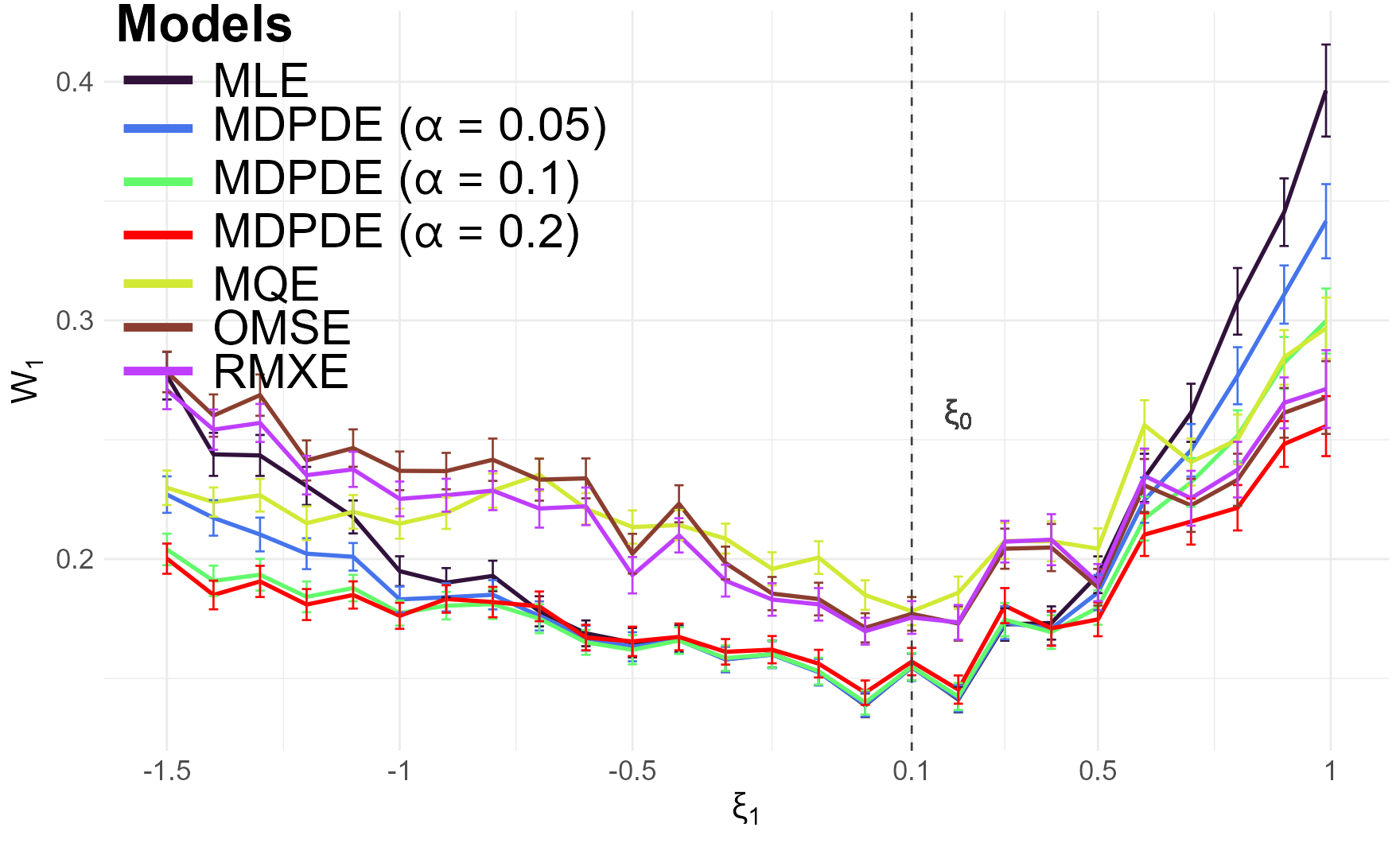}
  \hspace{0.7cm}
  \includegraphics[width=.47\textwidth]{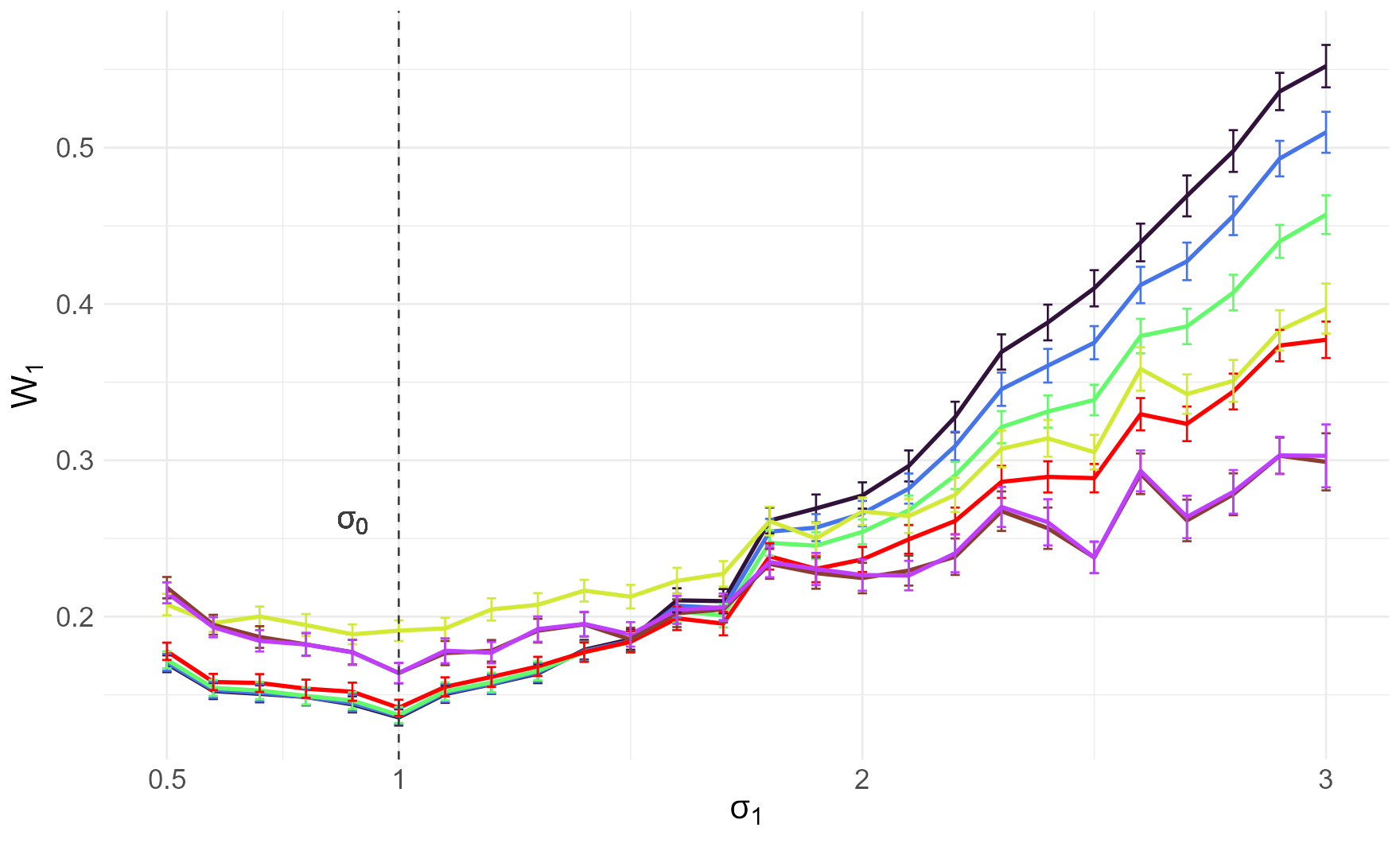}
  \caption{Average Wasserstein distance over 200 replications (with standard errors) across various contaminated models. In the left panel, the shape parameter $\xi_1$ varies while the location $\mu_0$ and scale $\sigma_0$ are fixed. In the right panel, the scale parameter $\sigma_1$ varies while $\mu_0$ and the shape $\xi_0$ remain fixed. Each sample has size $n = 100$, with contamination proportion $\varepsilon = 0.2$. The true model parameters are $\mu_0 = 0$, $\sigma_0 = 1$, and $\xi_0 = 0.1$.}
\end{figure}

\begin{figure}[ht!]
\centering
  \includegraphics[width=.47\textwidth]{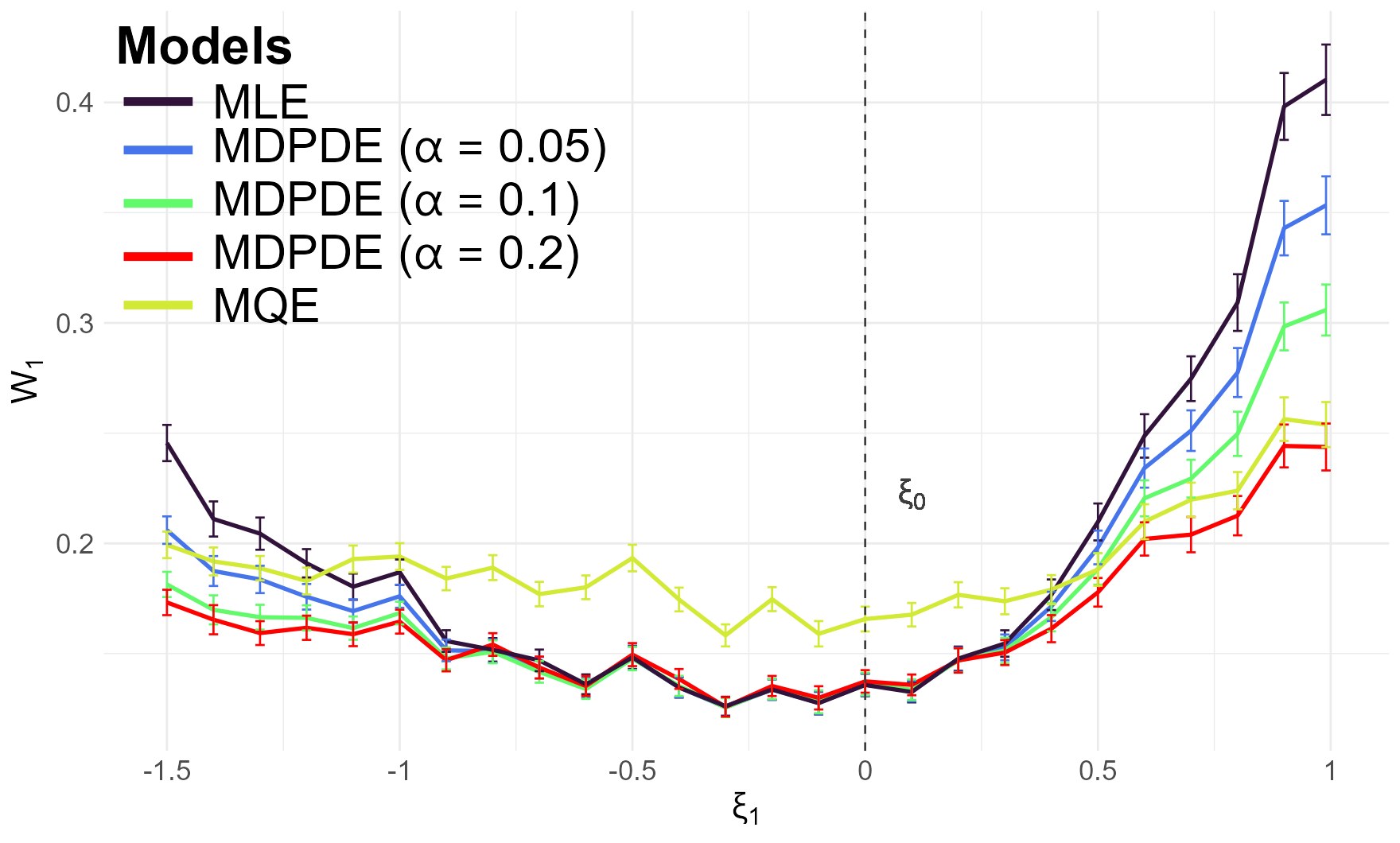}
  \hspace{0.7cm}
  \includegraphics[width=.47\textwidth]{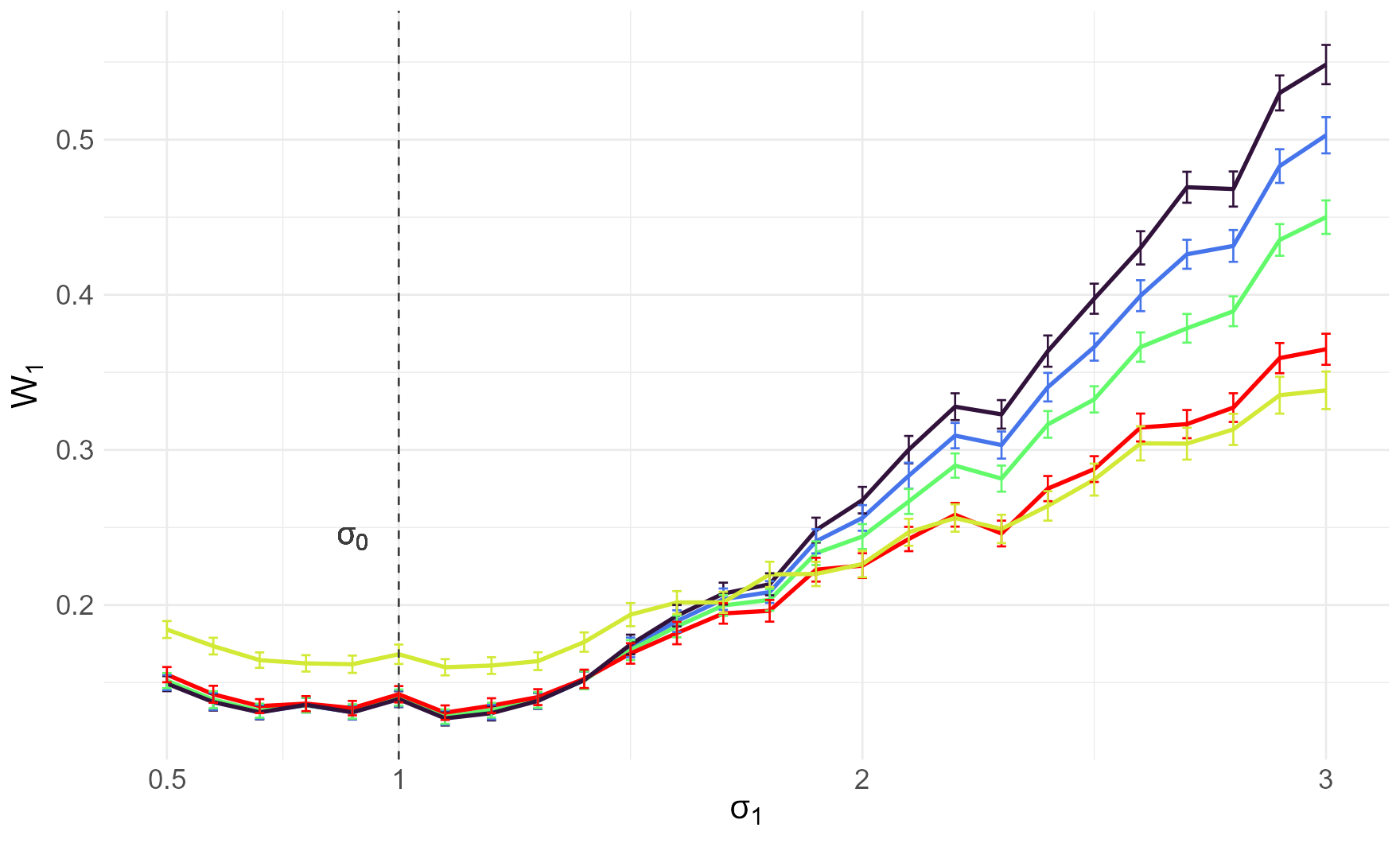}
  \caption{Average Wasserstein distance over 200 replications (with standard errors) across various contaminated models. In the left panel, the shape parameter $\xi_1$ varies while the location $\mu_0$ and scale $\sigma_0$ are fixed. In the right panel, the scale parameter $\sigma_1$ varies while $\mu_0$ and the shape $\xi_0$ remain fixed. Each sample has size $n = 100$, with contamination proportion $\varepsilon = 0.2$. The true model parameters are $\mu_0 = 0$, $\sigma_0 = 1$, and $\xi_0 = 0$.}
\end{figure}

\begin{figure}[ht!]
\centering
  \includegraphics[width=.47\textwidth]{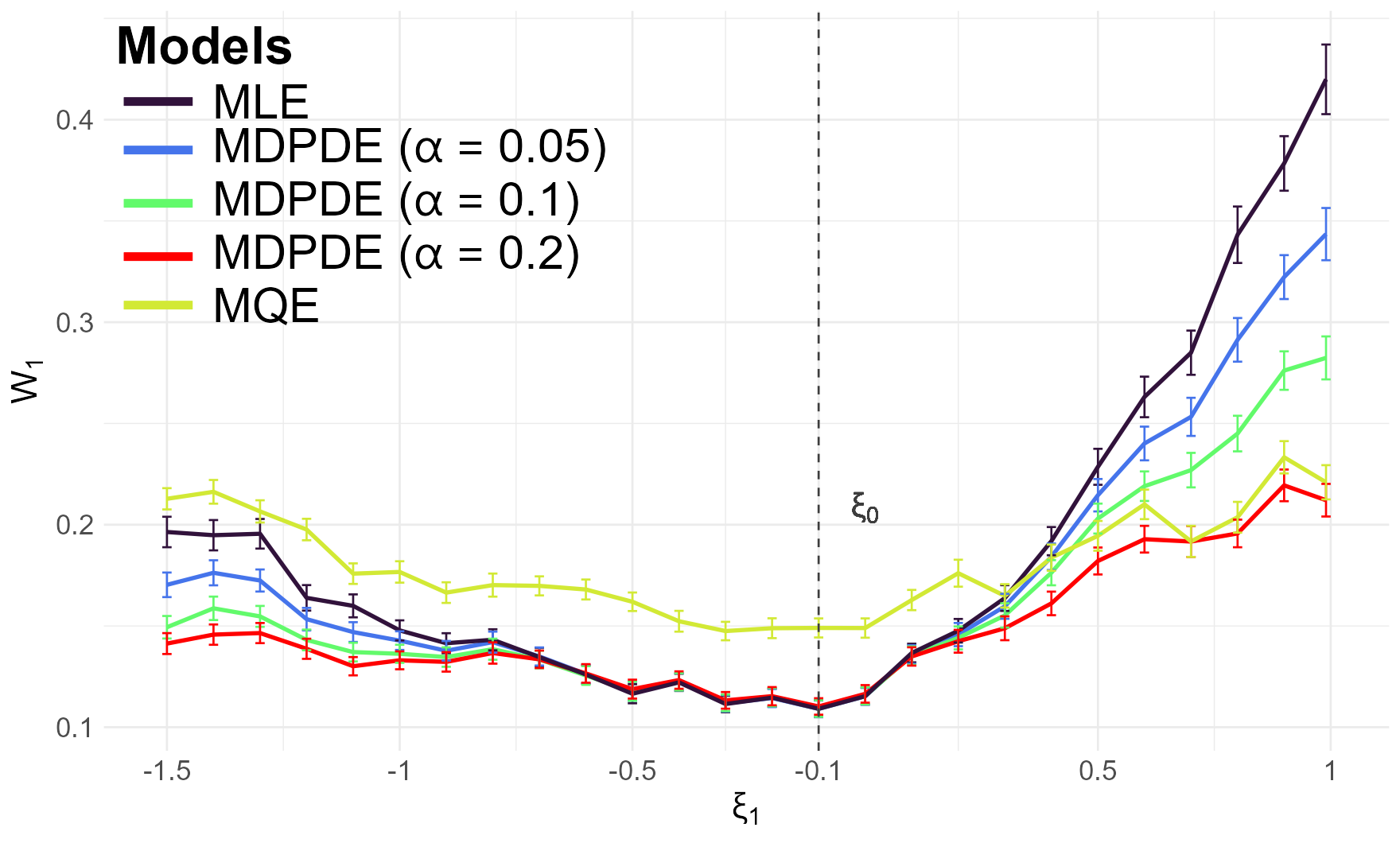}
  \hspace{0.7cm}
  \includegraphics[width=.47\textwidth]{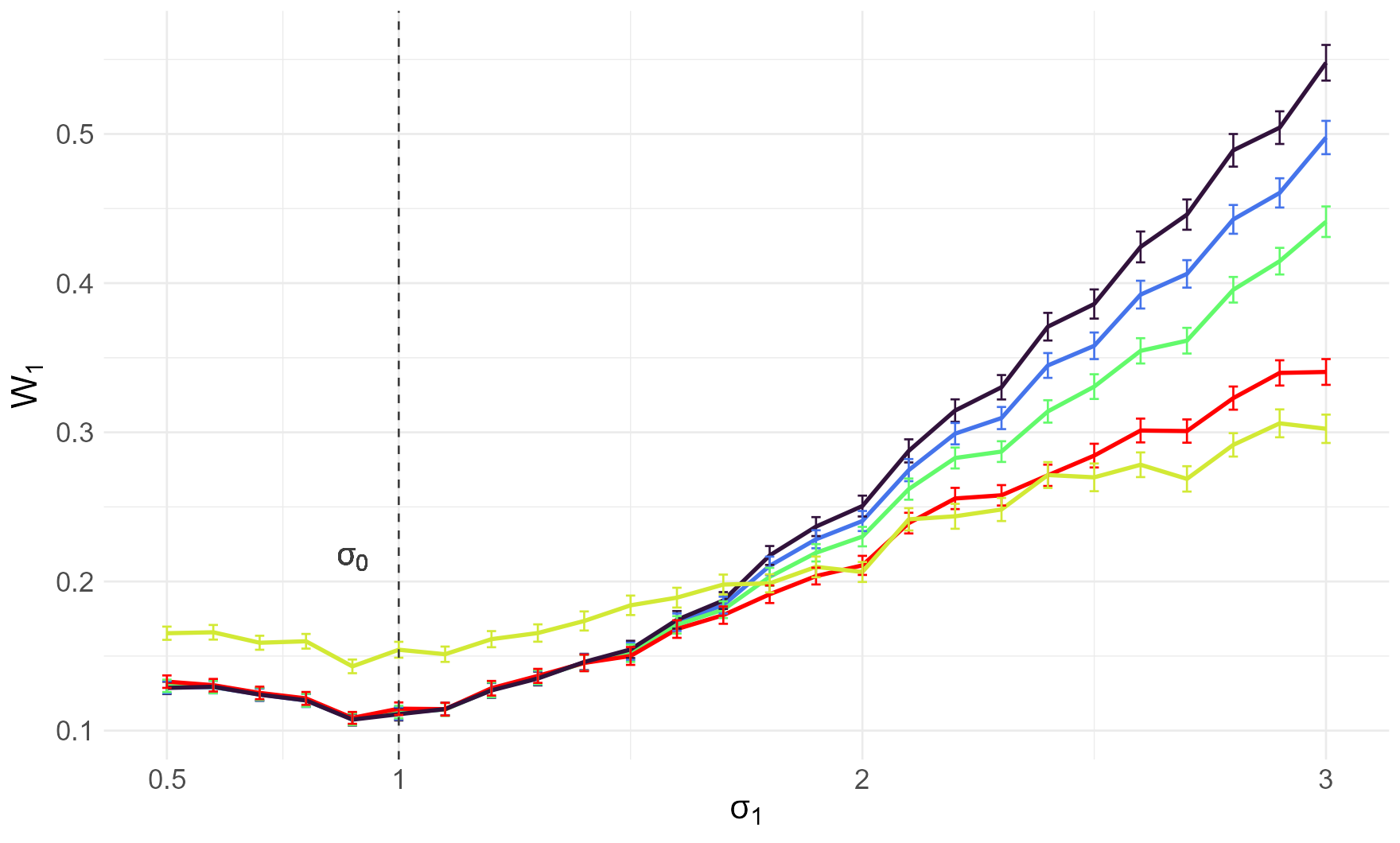}
  \caption{Average Wasserstein distance over 200 replications (with standard errors) across various contaminated models. In the left panel, the shape parameter $\xi_1$ varies while the location $\mu_0$ and scale $\sigma_0$ are fixed. In the right panel, the scale parameter $\sigma_1$ varies while $\mu_0$ and the shape $\xi_0$ remain fixed. Each sample has size $n = 100$, with contamination proportion $\varepsilon = 0.2$. The true model parameters are $\mu_0 = 0$, $\sigma_0 = 1$, and $\xi_0 = -0.1$.}
\end{figure}

\begin{table}[ht!]
\caption{Number of samples in which either convergence fails (for the MLE or the MDPDE) or the estimation procedures return non-plausible values (i.e., $\hat{\mu}_0 < -2$, $\hat{\mu}_0 > 2$, or $\hat{\sigma}_0 > 2$). The models vary either in the shape parameter $\xi_{1}\in\{-1.5,-1.4,\dots,0.9,0.99\}$ or in the scale parameter $\sigma_{1}\in\{0.5,0.6,\dots,3\}$, out of 200 replications per contaminated model, yielding 5200 total samples for each case. All results are based on samples of size $n=100$ with contamination proportion $\varepsilon=0.2$.\label{tab:missing_value_eps02}}
\vspace{0.2cm}

\begin{center}
\begin{small}
\resizebox{\textwidth}{!}{
\begin{tabular}{ll|ccccccc}
    & & MLE & MDPDE ($\alpha = 0.05)$& MDPDE ($\alpha = 0.1)$ & MDPDE ($\alpha = 0.2)$ & MQE & OMSE & RMXE \\ \hline
    \multirowcell{2}{$\xi_0  =-0.1$}&\textit{varying} $\xi_1$ & 2 & 4 & 1 & 6 & 27 & $\times$ & $\times$\\
    &\textit{varying} $\sigma_1$ & 2 & 3 & 3 & 2 & 11 & $\times$ & $\times$\\
    \hline
    \multirowcell{2}{$\xi_0  =0$}&\textit{varying} $\xi_1$& 5 & 0 & 2 & 2 & 26 & $\times$ & $\times$ \\
    &\textit{varying} $\sigma_1$ & 1 & 1 & 2 & 1 & 22 & $\times$ & $\times$ \\
    \hline
    \multirowcell{2}{$\xi_0  =0.1$}& \textit{varying} $\xi_1$ & 32 & 2 & 0 & 4 & 23 & 780 & 779\\
    &\textit{varying} $\sigma_1$ & 22 & 5 & 4 & 2 &  23 & 638 & 638 \\
\end{tabular}
}
\end{small}
\end{center}
\end{table}
\clearpage

%% file: appendix_appli.tex
\section{Appendix: application}\label{sec:real_appendix}

\begin{table}[ht!]
\caption{Estimated location parameter $\mu$ and standard deviation at each station using the four different models.\label{tab:loc_parameter}}
\vspace{0.2cm}

\begin{center}
\begin{small}
\begin{tabular}{c|cccc}
     & MLE (without PILFs) &  MLE & MDPDE ($\alpha = 0.1)$& MDPDE ($\alpha = 0.3)$  \\ \hline
     10001 & $46.62$ $(2.00)$ &$45.22$ $(2.56)$ &$45.38$ $(2.24)$&$45.91$ $(2.48)$ \\
    50014 & $16.59$ (0.99) &$15.79$ (1.26) &$15.72$ (1.25)&$15.66$ (1.22) \\
    69028 & $133.73$ (5.25) &$131.02$ (5.98) &$130.36$ (5.97) &$129.54$ (6.23)\\
    71011 & $117.86$ (1.61) &$116.85$ (2.29) &$116.96$ (2.18) &$117.76$ (1.67) \\
    

\end{tabular}
\end{small}
\end{center}
\end{table}

\begin{table}[ht!]
\caption{Estimated scale parameter $\sigma$ and standard deviation at each station using the four different models.\label{tab:scale_parameter}}
\vspace{0.2cm}

\begin{center}
\begin{small}
\begin{tabular}{c|cccc}
     & MLE (without PILFs) &  MLE & MDPDE ($\alpha = 0.1)$& MDPDE ($\alpha = 0.3)$  \\ \hline
     10001 & $11.87$ $(1.47)$ &$15.76$ $(1.79)$&$15.19$ $(1.66)$&$13.53$ $(1.74)$\\
    50014 & $6.28$ (0.84) &$8.59$ (0.92)&$8.48$ (0.92)&$8.18$ (0.95)\\
    69028 & $33.98$ (5.38)&$44.8$ (4.42) &$44.43$ (4.39) &$43.26$ (5.58)\\
    71011 & $10.67$ (1.14) &$15.95$ (1.63)&$15.1$ (1.55) &$11.2$ (1.24)\\
    

\end{tabular}
\end{small}
\end{center}
\end{table}



%% file: main.bbl
\begin{thebibliography}{}

\bibitem[Alonso et~al., 2014]{alonso2014comparing}
Alonso, A.~M., de~Zea~Bermudez, P., and Scotto, M.~G. (2014).
\newblock Comparing generalized pareto models fitted to extreme observations:
  an application to the largest temperatures in spain.
\newblock {\em Stochastic environmental research and risk assessment},
  28(5):1221--1233.

\bibitem[Asquith et~al., 2021]{MGBTpackage}
Asquith, W.~H., England, J.~F., and Herrmann, G.~R. (2021).
\newblock {\em MGBT---Multiple Grubbs-Beck Low-Outlier Test}.
\newblock U.S. Geological Survey, Reston, Va.
\newblock R package version 1.0.7.

\bibitem[Auld et~al., 2023]{AuldTemp}
Auld, G., Hegerl, G.~C., and Papastathopoulos, I. (2023).
\newblock {Changes in the distribution of annual maximum temperatures in
  Europe}.
\newblock {\em Advances in Statistical Climatology, Meteorology and
  Oceanography}, 9:45--66.

\bibitem[Barth et~al., 2017]{Barth2017mixed}
Barth, N.~A., Villarini, G., Nayak, M.~A., and White, K. (2017).
\newblock Mixed populations and annual flood frequency estimates in the western
  {United} {States}: The role of atmospheric rivers.
\newblock {\em Water Resources Research}, 53(1):257--269.

\bibitem[Basak et~al., 2021]{basak2021optimal}
Basak, S., Basu, A., and Jones, M.~C. (2021).
\newblock On the ‘optimal’ density power divergence tuning parameter.
\newblock {\em Journal of Applied Statistics}, 48(3):536--556.

\bibitem[Basu et~al., 1998]{basu1998robust}
Basu, A., Harris, I.~R., Hjort, N.~L., and Jones, M. (1998).
\newblock Robust and efficient estimation by minimising a density power
  divergence.
\newblock {\em Biometrika}, 85(3):549--559.

\bibitem[Belzile et~al., 2023]{belzile2023modeler}
Belzile, L.~R., Dutang, C., Northrop, P.~J., and Opitz, T. (2023).
\newblock A modeler’s guide to extreme value software.
\newblock {\em Extremes}, 26(4):595--638.

\bibitem[B{\"u}cher and Segers, 2017]{bucher2017maximum}
B{\"u}cher, A. and Segers, J. (2017).
\newblock On the maximum likelihood estimator for the {Generalized}
  {Extreme}-{Value} distribution.
\newblock {\em Extremes}, 20:839--872.

\bibitem[Castellarin et~al., 2012]{castellarin2012review}
Castellarin, A., Kohnov{\'a}, S., Ga{\'a}l, L., Fleig, A., Salinas, J.,
  Toumazis, A., Kjeldsen, T., and Macdonald, N. (2012).
\newblock Review of {Applied}-{Statistical} {Methods} {For} {Flood}-{Frequency}
  {Analysis} in {Europe}.
\newblock Technical report, The Centre for Ecology \& Hydrology.

\bibitem[Cohn et~al., 2013]{cohn2013generalized}
Cohn, T.~A., England, J., Berenbrock, C., Mason, R., Stedinger, J., and
  Lamontagne, J. (2013).
\newblock A generalized {Grubbs}-{Beck} test statistic for detecting multiple
  potentially influential low outliers in flood series.
\newblock {\em Water Resources Research}, 49(8):5047--5058.

\bibitem[Coles, 2001]{coles2001introduction}
Coles, S. (2001).
\newblock {\em {An Introduction to Statistical Modeling of Extreme Values}},
  volume 208.
\newblock Springer.

\bibitem[de~Zea~Bermudez and Kotz, 2010a]{de2010parameter}
de~Zea~Bermudez, P. and Kotz, S. (2010a).
\newblock {Parameter estimation of the generalized Pareto distribution—Part
  I}.
\newblock {\em Journal of Statistical Planning and Inference},
  140(6):1353--1373.

\bibitem[de~Zea~Bermudez and Kotz, 2010b]{de2010parameter2}
de~Zea~Bermudez, P. and Kotz, S. (2010b).
\newblock {Parameter estimation of the generalized Pareto distribution—Part
  II}.
\newblock {\em Journal of Statistical Planning and Inference},
  140(6):1374--1388.

\bibitem[Dombry, 2015]{dombryGEV}
Dombry, C. (2015).
\newblock Existence and consistency of the maximum likelihood estimators for
  the extreme value index within the block maxima framework.
\newblock {\em Bernoulli}, 21(1):420--436.

\bibitem[Dombry and Ferreira, 2019]{dombry2019maximum}
Dombry, C. and Ferreira, A. (2019).
\newblock Maximum likelihood estimators based on the block maxima method.
\newblock {\em Bernoulli}, 25(3):1690--1723.

\bibitem[Dupuis and Field, 1998]{dupuis1998robust}
Dupuis, D. and Field, C. (1998).
\newblock Robust estimation of extremes.
\newblock {\em Canadian Journal of Statistics}, 26(2):199--215.

\bibitem[Eastoe and Tawn, 2009]{eastoe2009modelling}
Eastoe, E.~F. and Tawn, J.~A. (2009).
\newblock Modelling non-stationary extremes with application to surface level
  ozone.
\newblock {\em Journal of the Royal Statistical Society Series C: Applied
  Statistics}, 58(1):25--45.

\bibitem[England~Jr et~al., 2018]{england2018guidelines}
England~Jr, J.~F., Cohn, T.~A., Faber, B.~A., Stedinger, J.~R., Thomas~Jr,
  W.~O., Veilleux, A.~G., Kiang, J.~E., and Mason~Jr, R.~R. (2018).
\newblock {Guidelines for determining flood flow frequency—Bulletin 17C}.
\newblock Technical report, US Geological Survey.

\bibitem[Feng et~al., 2007]{feng2007modeling}
Feng, S., Nadarajah, S., and Hu, Q. (2007).
\newblock {Modeling annual extreme precipitation in china using the generalized
  extreme value distribution}.
\newblock {\em Journal of the Meteorological Society of Japan. Ser. II},
  85(5):599--613.

\bibitem[Fisher and Tippett, 1928]{fisher1928limiting}
Fisher, R.~A. and Tippett, L. H.~C. (1928).
\newblock Limiting forms of the frequency distribution of the largest or
  smallest member of a sample.
\newblock In {\em Mathematical proceedings of the Cambridge philosophical
  society}, volume 24-2, pages 180--190. Cambridge University Press.

\bibitem[Friederichs and Thorarinsdottir, 2012]{friederichs2012forecast}
Friederichs, P. and Thorarinsdottir, T.~L. (2012).
\newblock Forecast verification for extreme value distributions with an
  application to probabilistic peak wind prediction.
\newblock {\em Environmetrics}, 23(7):579--594.

\bibitem[Gaetan et~al., 2025]{gaetan2025precip}
Gaetan, C., Opitz, T., and Toulemonde, G. (2025).
\newblock Statistical modeling of extreme precipitation.
\newblock In {\em Handbook on Statistics of Extremes}. Chapman and Hall.

\bibitem[Gnedenko, 1943]{gnedenko1943distribution}
Gnedenko, B. (1943).
\newblock Sur la distribution limite du terme maximum d'une série aleatoire.
\newblock {\em Annals of mathematics}, 44(3):423--453.

\bibitem[Grubbs and Beck, 1972]{grubbs1972extension}
Grubbs, F.~E. and Beck, G. (1972).
\newblock {Extension of Sample Sizes and Percentage Points for Significance
  Tests of Outlying Observations}.
\newblock {\em Technometrics}, 14(4):847--854.

\bibitem[Hampel et~al., 1986]{hampel86}
Hampel, F., Ronchetti, E., Rousseeuw, P., and Stahel, W. (1986).
\newblock {\em Robust Statistics: The Approach Based on Influence Functions}.
\newblock John Wiley \& Sons.

\bibitem[Hampel, 1974]{hampel1974influence}
Hampel, F.~R. (1974).
\newblock {The Influence Curve and Its Role in Robust Estimation }.
\newblock {\em Journal of the American Statistical Association},
  69(346):383--393.

\bibitem[Hoffman-Jorgensen, 1994]{hoffman1994probability}
Hoffman-Jorgensen, J. (1994).
\newblock {\em Probability with a View towards Statistics, volume II}.
\newblock New York: Chapman and Hall.

\bibitem[Hong and Kim, 2001]{hong2001automatic}
Hong, C. and Kim, Y. (2001).
\newblock {Automatic Selection of the Turning Parametter in the Minimum Density
  Power Divergence Estimation}.
\newblock {\em Journal of the Korean Statistical Society}, 30(3):453--465.

\bibitem[Horbenko et~al., 2018]{horbenko2018package}
Horbenko, N., Spangl, B., Desmettre, S., Massini, E., and Pupashenko, D.
  (2018).
\newblock {Package ‘RobExtremes’}.

\bibitem[Hosking and Wallis, 1987]{hosking1987parameter}
Hosking, J.~R. and Wallis, J.~R. (1987).
\newblock {Parameter and Quantile Estimation for the Generalized Pareto
  Distribution}.
\newblock {\em Technometrics}, 29(3):339--349.

\bibitem[Hosking et~al., 1985]{hosking1985estimation}
Hosking, J. R.~M., Wallis, J.~R., and Wood, E.~F. (1985).
\newblock {Estimation of the Generalized Extreme-Value Distribution by the
  Method of Probability-Weighted Moments}.
\newblock {\em Technometrics}, 27(3):251--261.

\bibitem[Huber and Ronchetti, 2011]{huber2011robust}
Huber, P.~J. and Ronchetti, E.~M. (2011).
\newblock {\em {Robust Statistics}}.
\newblock John Wiley \& Sons.

\bibitem[Jayaweera et~al., 2025]{jayaweera2025evidence}
Jayaweera, L., Wasko, C., and Nathan, R. (2025).
\newblock {Evidence for Non‐Stationarity in the GEV Shape Parameter When
  Modeling Extreme Rainfall}.
\newblock {\em Water Resources Research}, 61(5):e2023WR036426.

\bibitem[Jenkinson, 1955]{jenkinson1955frequency}
Jenkinson, A.~F. (1955).
\newblock The frequency distribution of the annual maximum (or minimum) values
  of meteorological elements.
\newblock {\em Quarterly Journal of the Royal meteorological society},
  81(348):158--171.

\bibitem[Ju{\'a}rez, 2003]{phdjuarez}
Ju{\'a}rez, S. (2003).
\newblock {\em {Robust and Efficient Estimation for the Generalized Pareto
  Distribution}}.
\newblock PhD thesis, Statistical Science Department, Southern Methodits
  University.

\bibitem[Ju{\'a}rez and Schucany, 2004]{juarez2004robust}
Ju{\'a}rez, S.~F. and Schucany, W.~R. (2004).
\newblock {Robust and Efficient Estimation for the Generalized Pareto
  Distribution}.
\newblock {\em Extremes}, 7:237--251.

\bibitem[Kohl et~al., 2019]{kohl2019package}
Kohl, M., Pupashenko, M., Kroisandt, G., Ruckdeschel, P., and Kohl, M.~M.
  (2019).
\newblock {Package ‘ROptEst’}.

\bibitem[Lamontagne et~al., 2016]{lamontagne2016robust}
Lamontagne, J., Stedinger, J., Yu, X., Whealton, C., and Xu, Z. (2016).
\newblock {Robust flood frequency analysis: Performance of EMA with multiple
  Grubbs-Beck outlier tests}.
\newblock {\em Water Resources Research}, 52(4):3068--3084.

\bibitem[Lin et~al., 2024]{lin2024multi}
Lin, S., Kong, A., and Azencott, R. (2024).
\newblock {Multi-Quantile Estimators for the parameters of Generalized Extreme
  Value distribution}.
\newblock {\em arXiv preprint arXiv:2412.04640}.

\bibitem[Morrison and Smith, 2002]{morrison2002stochastic}
Morrison, J.~E. and Smith, J.~A. (2002).
\newblock Stochastic modeling of flood peaks using the generalized extreme
  value distribution.
\newblock {\em {Water Resources Research}}, 38(12):41--1.

\bibitem[{National River Flow Archive}, 2024]{nrfa}
{National River Flow Archive} (2024).
\newblock Peak flow dataset - v13.

\bibitem[Negahban, 2025]{negahban2025framework}
Negahban, A. (2025).
\newblock {A framework for comparing stochastic simulation models against
  multidimensional data using the Wasserstein distance}.
\newblock {\em Journal of Simulation}, pages 1--14.

\bibitem[Nerantzaki and Papalexiou, 2022]{nerantzaki2022assessing}
Nerantzaki, S.~D. and Papalexiou, S.~M. (2022).
\newblock Assessing extremes in hydroclimatology: A review on probabilistic
  methods.
\newblock {\em Journal of Hydrology}, 605:127302.

\bibitem[P.~Prescott, 1980]{prescottwalden1980}
P.~Prescott, A. T.~W. (1980).
\newblock Maximum likelihood estimation of the parameters of the generalized
  extreme-value distribution.
\newblock {\em Biometrika}, 67(1):723–724.

\bibitem[Panaretos and Zemel, 2019]{panaretos2019statistical}
Panaretos, V.~M. and Zemel, Y. (2019).
\newblock {Statistical aspects of Wasserstein distances}.
\newblock {\em Annual review of statistics and its application}, 6(1):405--431.

\bibitem[Papalexiou and Koutsoyiannis, 2013]{papalexiou2013battle}
Papalexiou, S.~M. and Koutsoyiannis, D. (2013).
\newblock Battle of extreme value distributions: A global survey on extreme
  daily rainfall.
\newblock {\em Water Resources Research}, 49(1):187--201.

\bibitem[Peng and Welsh, 2001]{peng2001robust}
Peng, L. and Welsh, A. (2001).
\newblock {Robust Estimation of the Generalized Pareto Distribution}.
\newblock {\em Extremes}, 4:53--65.

\bibitem[Piwowar and Ku{\'z}mi{\'n}ski, 2023]{piwowar2023drought}
Piwowar, A. and Ku{\'z}mi{\'n}ski, {\L}. (2023).
\newblock Drought risk probabilistic models based on extreme value theory.
\newblock {\em Environmental Science and Pollution Research},
  30(42):95945--95958.

\bibitem[Prosdocimi et~al., 2015]{prosdocimi2015detection}
Prosdocimi, I., Kjeldsen, T., and Miller, J. (2015).
\newblock Detection and attribution of urbanization effect on flood extremes
  using nonstationary flood-frequency models.
\newblock {\em Water resources research}, 51(6):4244--4262.

\bibitem[Resnick, 2008]{resnick2008extreme}
Resnick, S.~I. (2008).
\newblock {\em Extreme Values, Regular Variation and Point Processes},
  volume~4.
\newblock Springer Science \& Business Media.

\bibitem[Robson and Reed, 1999]{robson1999statistical}
Robson, A. and Reed, D. (1999).
\newblock {\em Statistical Procedures for Flood Frequency Estimation: Flood
  Estimation Handbook (Vol 3)}.
\newblock Institute of Hydrology.

\bibitem[Singh et~al., 2005]{Singh2005nonid}
Singh, V., Wang, S., and Zhang, L. (2005).
\newblock Frequency analysis of nonidentically distributed hydrologic flood
  data.
\newblock {\em Journal of Hydrology}, 307(1):175--195.

\bibitem[Smith, 1985]{smith1985maximum}
Smith, R.~L. (1985).
\newblock {Maximum Likelihood Estimation in a Class of Nonregular Cases}.
\newblock {\em Biometrika}, 72(1):67--90.

\bibitem[Soukissian and Tsalis, 2015]{soukissian2015effect}
Soukissian, T.~H. and Tsalis, C. (2015).
\newblock The effect of the generalized extreme value distribution parameter
  estimation methods in extreme wind speed prediction.
\newblock {\em Natural Hazards}, 78(3):1777--1809.

\bibitem[Stein, 2017]{stein2017should}
Stein, M. (2017).
\newblock Should annual maximum temperatures follow a generalized extreme value
  distribution?
\newblock {\em Biometrika}, 104(1):1--16.

\bibitem[Stephenson, 2016]{stephenson2016bayesian}
Stephenson, A. (2016).
\newblock Bayesian inference for extreme value modelling.
\newblock {\em Extreme Value Modeling and Risk Analysis: Methods and
  Applications}, pages 257--280.

\bibitem[Sugasawa and Yonekura, 2021]{sugasawa2021selection}
Sugasawa, S. and Yonekura, S. (2021).
\newblock {On Selection Criteria for the Tuning Parameter in Robust
  Divergence}.
\newblock {\em Entropy}, 23(9):1147.

\bibitem[Vandewalle et~al., 2007]{vandewalle2007robust}
Vandewalle, B., Beirlant, J., Christmann, A., and Hubert, M. (2007).
\newblock {A robust estimator for the tail index of Pareto-type distributions}.
\newblock {\em Computational Statistics \& Data Analysis}, 51(12):6252--6268.

\bibitem[Warwick and Jones, 2005]{warwick2005choosing}
Warwick, J. and Jones, M. (2005).
\newblock Choosing a robustness tuning parameter.
\newblock {\em Journal of Statistical Computation and Simulation},
  75(7):581--588.

\bibitem[Yoon et~al., 2010]{yoon2010full}
Yoon, S., Cho, W., Heo, J.-H., and Kim, C.~E. (2010).
\newblock {A full Bayesian approach to generalized maximum likelihood
  estimation of generalized extreme value distribution}.
\newblock {\em Stochastic environmental research and risk assessment},
  24(5):761--770.

\end{thebibliography}
